\documentclass[12pt]{article}
\pdfoutput=1 % if your are submitting a pdflatex (i.e. if you have
\usepackage{jheppub} % for details on the use of the package, please
\usepackage[T1]{fontenc} % if needed
\usepackage{amsmath,physics}
\usepackage{comment}
\usepackage[percent]{overpic}
\usepackage[toc,page]{appendix}
\usepackage{etoolbox}
\usepackage{placeins}

\let\Oldsection\section
\renewcommand{\section}{\FloatBarrier\Oldsection}

\let\Oldsubsection\subsection
\renewcommand{\subsection}{\FloatBarrier\Oldsubsection}

\let\Oldsubsubsection\subsubsection
\renewcommand{\subsubsection}{\FloatBarrier\Oldsubsubsection}
\BeforeBeginEnvironment{appendices}{\clearpage}
\AfterEndEnvironment{appendices}{\clearpage}
\AfterEndEnvironment{appendices}{\pretocmd{\section}{\clearpage}{}{}}{}
\AfterEndEnvironment{appendices}{\pretocmd{\subsubsection}{\clearpage}{}{}}{}

\usepackage{graphicx,float,caption,subcaption,xcolor}
\usepackage{hyperref}
\hypersetup{
    colorlinks=true,
    linkcolor=blue,
    filecolor=magenta,      
    urlcolor=cyan,
    pdftitle={An Entropic Lens on Stabilizer States},
    pdfpagemode=FullScreen,
    }

\graphicspath{{./img/}} 
\usepackage{multirow}

\usepackage{bbm}% Added by cindy for that nice \mathbbm{1}

\newcommand{\Hil}{\mathcal{H}}
\def\ket#1{{|{#1}\rangle}} 

\title{An Entropic Lens on Stabilizer States}

\author[a]{Cynthia Keeler,}
\author[a]{William Munizzi,}
\author[b]{and Jason Pollack}

\affiliation[a]{Department of Physics, Arizona State University,
Tempe, AZ 85281, USA}
\affiliation[b]{Quantum Information Center, Department of Computer Science, The University of Texas at Austin,
2317 Speedway, Austin, TX 78712, USA}

\emailAdd{keelerc@asu.edu}
\emailAdd{wmunizzi@asu.edu}
\emailAdd{jasonpollack@gmail.com}

\abstract{The $n$-qubit stabilizer states are those left invariant by a $2^n$-element subset of the Pauli group. The Clifford group is the group of unitaries which take stabilizer states to stabilizer states; a physically--motivated generating set, the Hadamard, phase, and CNOT gates which comprise the Clifford gates, imposes a graph structure on the set of stabilizers. We explicitly construct these structures, the ``reachability graphs,'' at $n\le5$. When we consider only a subset of the Clifford gates, the reachability graphs separate into multiple, often complicated, connected components. Seeking an understanding of the entropic structure of the stabilizer states, which is ultimately built up by CNOT gate applications on two qubits, we are motivated to consider the restricted subgraphs built from the Hadamard and CNOT gates acting on only two of the $n$ qubits. We show how the two subgraphs already present at two qubits are embedded into more complicated subgraphs at three and four qubits. We argue that no additional types of subgraph appear beyond four qubits, but that the entropic structures within the subgraphs can grow progressively more complicated as the qubit number increases. Starting at four qubits, some of the stabilizer states have entropy vectors which are not allowed by holographic entropy inequalities. We comment on the nature of the transition between holographic and non-holographic states within the stabilizer reachability graphs.}

\notoc

\begin{document} 
\maketitle
\flushbottom

\section{Introduction}

Are all quantum states in a Hilbert space created equal? Viewed at the most abstract level, every pure quantum state can be rotated to any other by a unitary change of basis. But when the Hilbert space is endowed with some structure, different states play different roles with respect to that structure.  One natural structure is given by specifying a 
\emph{factorization} of the Hilbert space: an isomorphism between the abstract $\Hil$ and the tensor product Hilbert space $\bigotimes_{i=1}^N \Hil_i.$ For finite Hilbert spaces of composite dimension, the factorization associates to each pure state $\ket{\Psi}\in\Hil$ an \emph{entropy vector}, the collection of von Neumann entropies of the $2^N-1$ reduced density matrices formed by tracing out each possible tensor product formed from factors $\Hil_i$.

The entropy vectors provide a classification of the states in a tensor product Hilbert space, but because every state has an associated entropy vector they do not, by themselves, pick out any states as special. One way to accomplish this is by fixing one or more preferred operators acting on the Hilbert space. As a familiar example, choosing a particular Hermitian operator acting on a Hilbert space to be the Hamiltonian, the generator of time translations, specifies a basis of energy eigenstates, and every state in the Hilbert space can then be expanded in the energy basis. Most states are superpositions of more than one energy eigenstate, but not all: fixing a Hamiltonian picks out the basis of eigenstates of the Hamiltonian, the energy eigenstates themselves, which are the states where the Hamiltonian acts trivially, as scalar multiplication. The particular scalar is just given by the eigenvalue of the energy eigenstate, and if we just want to find the set of energy eigenstates this is unimportant: we could instead say that the energy eigenstates are the states where the projector onto the energy eigenspaces acts as the identity.

A similar procedure applies when instead of specifying a single Hermitian operator we pick a (multiplicative) group of Hermitian operators. Given the group $G\in L(\Hil)$, we can classify every state $\ket{\Psi}\in\Hil$ by the number of group elements that act trivially on this state: the dimension of the stabilizer subgroup $G_\ket{\Psi}$. Almost every state will have $\mathrm{dim}\:G_\ket{\Psi}=1$: the only group element that acts trivially is the identity operator. But some will have more, and the states that have the largest stabilizer subgroups relative to $G$ are called the \emph{stabilizer states} (which we will define more precisely below). For example, if the Hilbert space is that of a qubit, the stabilizer states of the Pauli group generated by $\langle X,Y,Z \rangle$ are the six states stabilized (up to sign) by the identity and one additional Pauli operator. Stabilizer states, especially with respect to the Pauli group on $n$ qubits, play an important role in the theory of quantum error correction, and in the fundamentals of quantum computing \cite{Gottesman:1997zz,Gottesman:1998hu,aaronson2004improved,knill2004fault,bravyi2005universal}. But here they emerged directly as ``generalized eigenstates,'' the states which play nicely with a specified group of operators. 

This paper initiates a research program aimed at combining these two classifications of states in Hilbert space. We seek an understanding of the stabilizer states, picked out by their interaction with a specified group of operators, in terms of their entropic structure, given by the underlying factorization of the Hilbert space. In particular, we will focus on the stabilizer states with respect to the Pauli group acting on $n$ qubits. In this setting, any stabilizer state can be reached by starting with any other stabilizer state and applying a unitary quantum circuit comprised of the Clifford gates: two one-qubit gates, the Hadamard and phase gates, and a single two-qubit gate, the controlled NOT gate. Since unitary operations on a single tensor factor do not change the entropy vector, it is already clear that moving between entropy vectors can only be accomplished via a CNOT gate. But not all such gate applications alter the entropic structure, and the picture we will find will be far richer than simply counting CNOT gates.

Besides understanding the general entropic structure of stabilizer states, we are additionally motivated by the connections between stabilizer states and holography. Stabilizer error-correcting codes satisfy a complementary recovery property which implies, and is equivalent to, an operator-algebraic version of the Ryu-Takayanagi formula \cite{Ryu:2006bv} relating entropies of states in the boundary/physical Hilbert space to the expectation value of an ``area'' operator acting on the bulk/logical Hilbert space \cite{Harlow:2016vwg,Pollack:2021yij}. A holographic bulk geometry can be discretized into a graph \cite{Bao:2015bfa}; in the limit of large bond dimension a tensor network built from random stabilizer tensors saturates the RT formula with probability one \cite{Hayden:2016cfa,Nezami:2016zni}.

Hence all holographic entropy vectors can be represented by stabilizer states, but the converse is not true: the holographic \emph{entropy cone} consisting of space of all allowed holographic entropy vectors is contained within the stabilizer entropy cone, and this containment is strict starting at three regions. Because the holographic entropy cone is well--characterized---explicitly at up to five regions \cite{HernandezCuenca:2019wgh}, and implicitly at arbitrary finite region number by the methods pioneered in \cite{Bao:2015bfa}---but the stabilizer entropy cone, and even the larger quantum entropy cone of all quantum states, are poorly understood, it is of great interest to understand in more detail how the holographic states are embedded into the larger space of stabilizer states. The stabilizer graph constructions presented in this paper allow this question to be attacked.

\subsection{Summary of Results}

\begin{figure}[!htb]
     \centering
     \begin{subfigure}{0.43\textwidth}
         \centering
         \includegraphics[width=\textwidth]{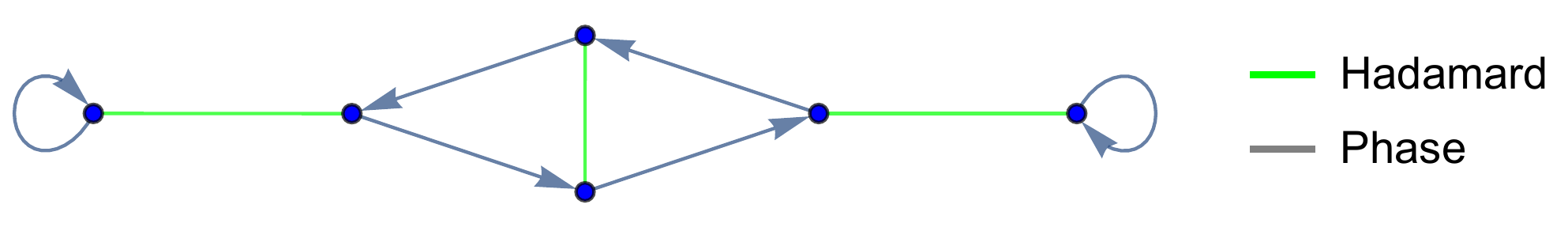}
         \caption{}%1-Qubit reachability diagram displaying all single-qubit operations.}
         \label{fig:OneQubitReachabilityGraph}
     \end{subfigure}
     \hfill
     \begin{subfigure}{0.47\textwidth}
         \centering
         \includegraphics[width=\textwidth]{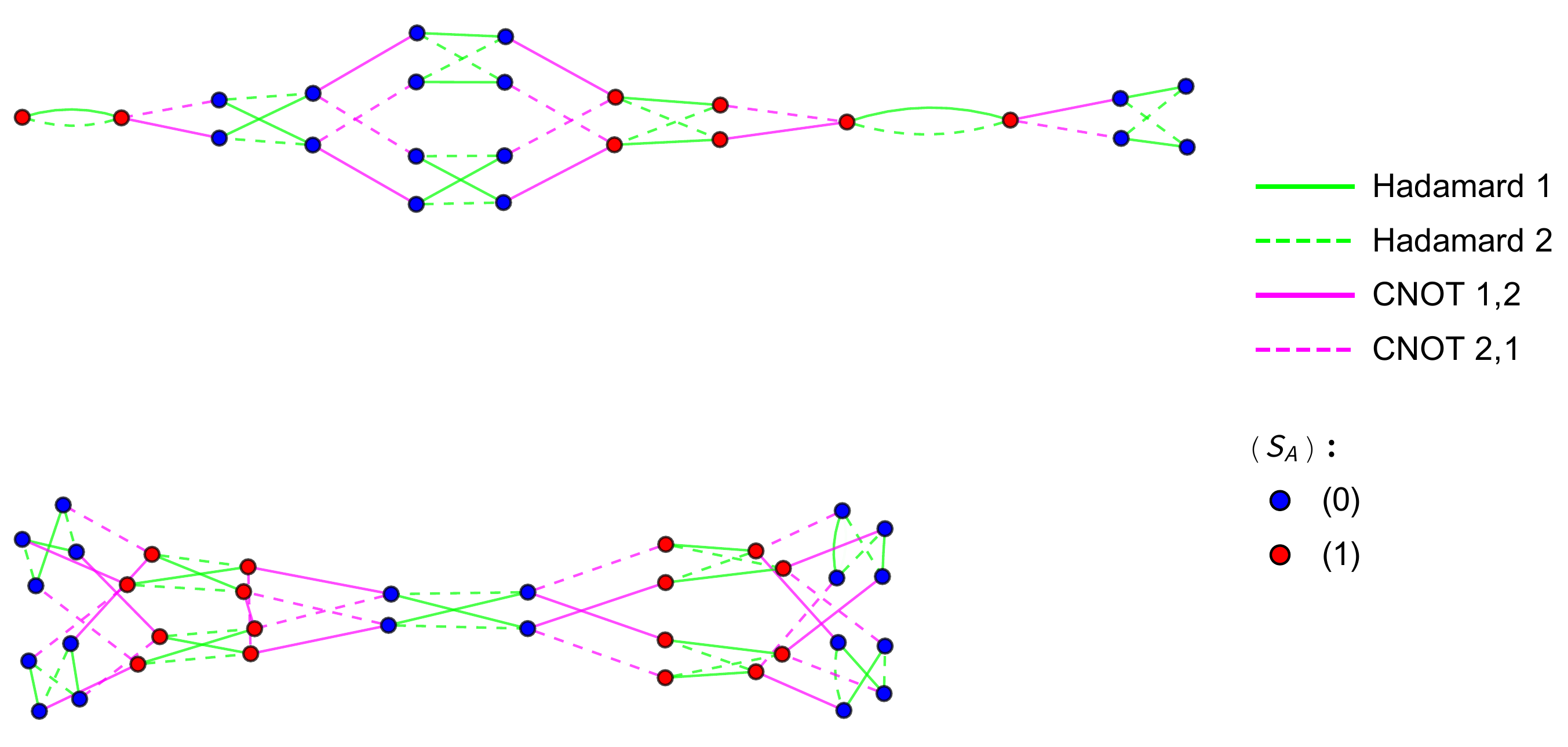}
         \caption{}%2-Qubit restricted graph consisting of subgraph structures $g_{24}$ and $g_{36}$.}
         \label{fig:TwoQubitG24G36}
     \end{subfigure}
     \hfill
     \begin{subfigure}{0.69\textwidth}
         \centering
         \includegraphics[width=\textwidth]{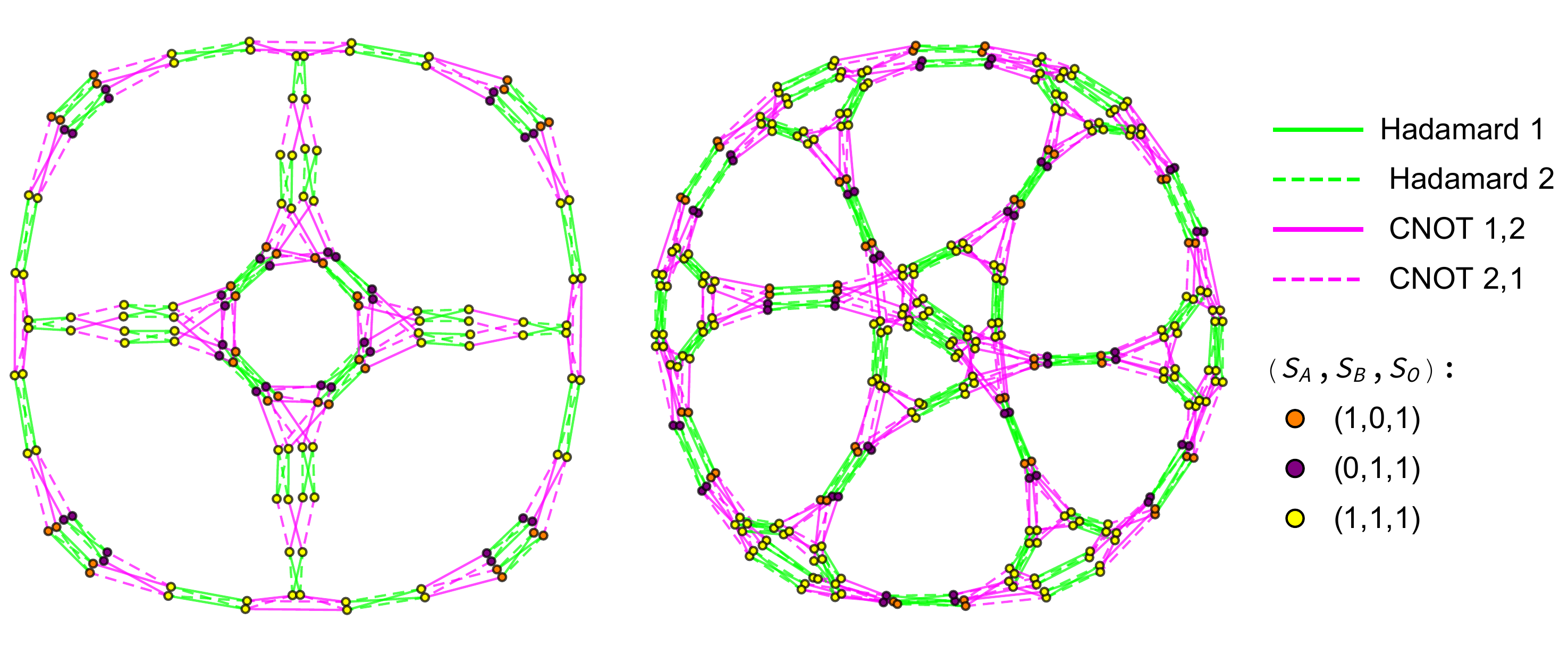}
         \caption{}%Two new subgraph structures occur at three qubits, $g_{144}$ and $g_{288}$.}
         \label{fig:ThreeQubitNewStructures}
     \end{subfigure}
     \hfill
     \begin{subfigure}{0.8\textwidth}
         \centering
         \includegraphics[width=\textwidth]{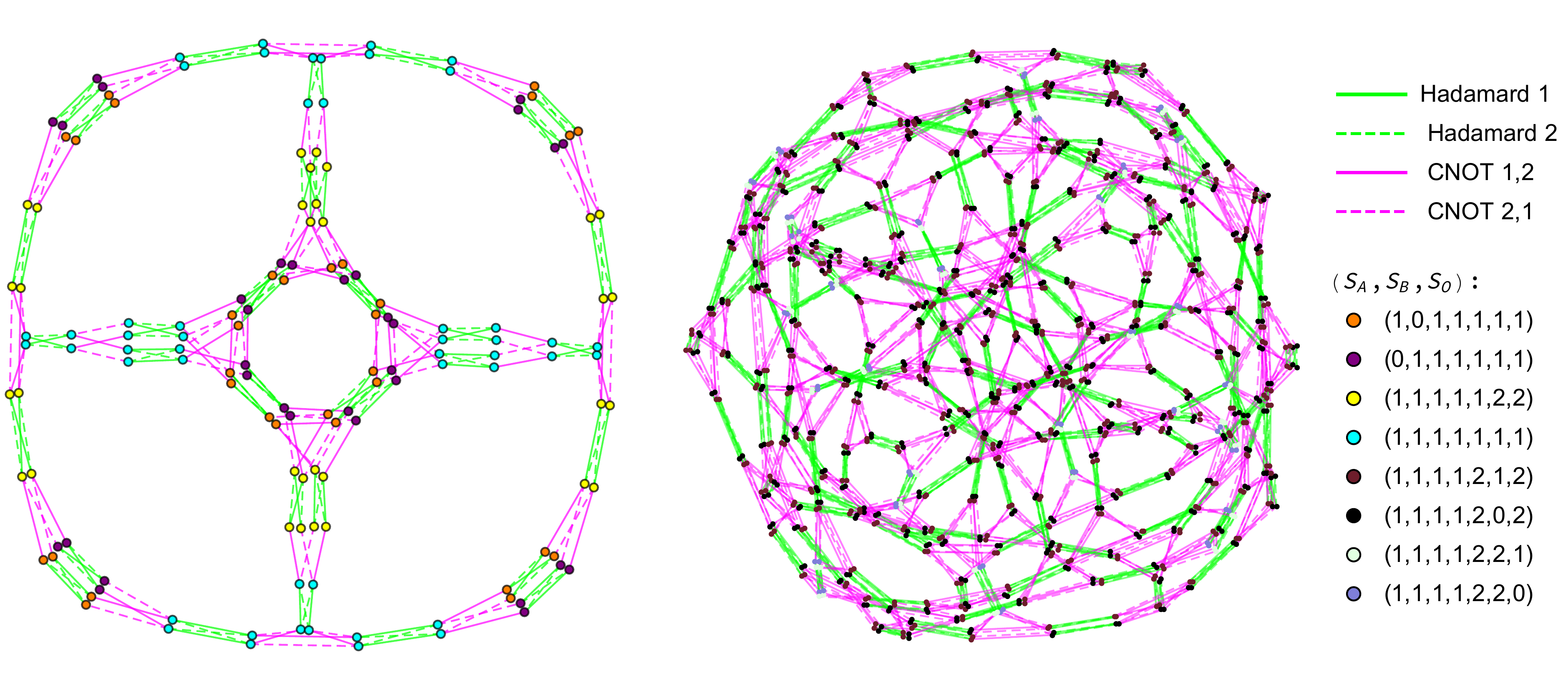}
         \caption{}%At four qubits we witness the arrival of a new subgraph $g_{1152}$, as well as an increase of entropy vectors on $g_{144}$. No new shapes arise at five (or more) qubits.}
         \label{fig:4-Qubit}
     \end{subfigure}
        \caption{The collection of subgraph structures up through five qubits. (a) 1-Qubit reachability diagram displaying all single-qubit operations. (b) 2-Qubit restricted graph consisting of subgraph structures $g_{24}$ and $g_{36}$. (c) Two new subgraph structures occur at three qubits, $g_{144}$ and $g_{288}$. (d) At four qubits we witness the arrival of a new subgraph $g_{1152}$, as well as an increase of entropy vectors on $g_{144}$. No new subgraph structures emerge beyond $g_{1152}$, but the number of different entropy vectors present in each subgraphs continues to increase for higher qubit number. At four qubits we witness the first instance of non-holographic states, appearing on subgraph $g_{144}$.}
        \label{fig:AllStructures}
\end{figure}

Figure \ref{fig:AllStructures} presents a preview of our results. All of the objects in the figure are ``reachability graphs,'' which display the structure of a (subset of) the stabilizer states. Each vertex is a particular stabilizer state, colored by its entropy vector. The edges connecting them correspond to the action of a particular Clifford gate. The first panel presents the full reachability graph at one qubit, showing the action of the two one-qubit Clifford gates, Hadamard and phase, on the six one-qubit stabilizer states. The remaining panels show reachability graphs at higher qubits, constructed from a subset of the Clifford gates consisting of the Hadamard and CNOT gates applied to two particular qubits; we will argue below that this restricted set of gates is sufficient to capture the changes in entropic structure as we move between different stabilizer states. The second panel presents the two subgraphs found in the restricted two-qubit reachability graph: one with 24 vertices and one with 36. 

At three qubits, the restricted graph has 16 connected components, made up of copies of the 24- and 36-vertex subgraphs, as well as two additional structures, with 144 and 288 vertices, respectively, which are presented in the third panel. We will show how these more complicated structures, as well as an 1152-vertex subgraph that appears at four qubits, can be constructed in a simple way by certain ``lifts'' of states in the structures found at two qubits to stabilizer states in a larger Hilbert space. By understanding the lifts, we will also understand how to map entropy vectors at lower qubit numbers to entropy vectors at higher qubit numbers: for example, although only three distinct entropy vectors appear in the 144-vertex subgraphs at three qubits, at four qubits there are versions of this subgraph with four distinct entropy vectors, as seen in the final panel of the Figure.

At four qubits, the stabilizer states can have any of eighteen different entropy vectors (see Table \ref{tab:FourQubitEntropyVectors}), and one of these entropy vectors, shown in blue in the fourth panel of Figure \ref{fig:AllStructures}, is not holographic. The reachability graphs allow us to see how, at four and five qubits, applications of Clifford gates can move states out of, and back into, the holographic entropy cone. We see, for example, that moving from the inner octagonal structure to the outer ring cannot be accomplished without passing through non-holographic states.

\subsection{Structure of the Paper}

The remainder of this paper is organized as follows. Section \ref{sec:review} reviews the notions of stabilizer states and the entropy cone, which are of fundamental importance for the rest of the paper. Section \ref{sec:two} initiates the study of the intersection of these two concepts by presenting the two-qubit stabilizer graph colored by entropy vector. We illustrate that additional insight can be gained into the graph structure by considering the \emph{restricted graphs} generated by only a subset of the Clifford gates; in particular, we highlight the special role of the restricted graph generated by only Hadamard and CNOT gates.

Section \ref{sec:three} extends the discussion to the three-qubit stabilizer graph. We show that much of the three-qubit graph can be understood as the natural extension of the two-qubit graph, but that nontrivial new structures appear in addition. We argue that because the Clifford gates act on at most two qubits it is natural to consider the restricted graphs generated by gates that act on any two of the three qubits. In Section \ref{sec:general}, we pass to a discussion of the situation for higher qubit numbers, illuminated by some direct results at four and five qubits. We show that no fundamentally new objects appear as we go to higher qubit number, but the existing objects develop progressively more complicated entropic structures. At higher qubit numbers, entropy vectors appear which cannot be represented by holographic states, and we comment on how they fit into the stabilizer graph. Finally, in Section \ref{sec:discussion} we discuss and conclude. Additional results and graphs are presented in appendices.

\section{Reminder: Stabilizer States and the Entropy Cone}\label{sec:review}

\subsection{Review of Stabilizer States}\label{sec:StabilizerReview}

The Pauli matrices
\begin{equation}
    I=\begin{pmatrix}1&0\\0&1\end{pmatrix}, \,\, \sigma_X=\begin{pmatrix}0&1\\1&0\end{pmatrix}, \,\,
    \sigma_Y=\begin{pmatrix}0&-i\\i&0\end{pmatrix}, \,\,
    \sigma_Z=\begin{pmatrix}1&0\\0&-1\end{pmatrix},
\end{equation}
are a set of four Hermitian and unitary matrices with eigenvalues $\pm 1$, which, given a fixed basis \{$\ket{0},\ket{1}\}$, can be interpreted as operators acting on the Hilbert space $\mathbb{C}^2$ of a single qubit. The (nontrivial) Pauli operators $\{\sigma_X,\sigma_Y,\sigma_Z\}$ generate the full algebra of linear operators $L(\mathbb{C}^2)$. More importantly for our purposes, they generate a 16-element multiplicative matrix group, the Pauli group on one qubit 
\begin{equation}
    \Pi_1\equiv\langle\sigma_X,\sigma_Y,\sigma_Z\rangle
    = c \{I, \sigma_X, \sigma_Y, \sigma_Z\}, 
    c\in\{\pm 1,\pm i\}.
\end{equation} 

Recall that for a pure state $\ket{\Psi}\in \Hil$ and a group of operators $G\subset L(\Hil)$, the stabilizer group of $\ket{\Psi}$ is defined by
\begin{equation}
    G_\ket{\Psi} \equiv \{g\in G\: | \:g\ket{\Psi}=\ket{\Psi}\}.
\end{equation}
That is, the stabilizer group of $\ket{\Psi}$ consists of the operators in $G$ for which $\ket{\Psi}$ is an eigenvector with eigenvalue one.  By inspection, the Pauli group on one qubit $\Pi_1$ has one element, $I$, with two unit eigenvalues, which therefore stabilizes all states in $\mathbb{C}^2$; one element with two negative eigenvalues, $-I$, which stabilizes no states; eight elements with purely imaginary eigenvalues, which also stabilize no states; and six elements with one unit eigenvalue, $\{\pm \sigma_X,\pm \sigma_Y, \pm \sigma_Z\}$, which therefore each stabilize one state in $\mathbb{C}^2$. Hence there are six states in the Hilbert space, the \emph{one-qubit stabilizer states}, which are stabilized by a two-element subgroup:
\begin{equation}\label{OneQubitStabilizerStates}
    S_1 \equiv \left\{ \ket{0}, \ket{1}, \ket{\pm}\equiv\frac{1}{\sqrt{2}}(\ket{0}\pm \ket{1}), \ket{\pm i}\equiv\frac{1}{\sqrt{2}}(\ket{0}\pm i\ket{1})  \right\},
\end{equation}
and all other states in the Hilbert space are stabilized by only the identity. For example, $\ket{+}$ is stabilized by $I$ and $\sigma_X$, while $\ket{1}$ is stabilized by $I$ and $-\sigma_Z$.

If we fix a state $\ket{\Psi}\in S_1$, then there are a limited number of operations we can do that will map $\ket{\Psi}$ to some other $\ket{\Psi^\prime}\in S_1$. In particular, we can ask what unitary operators $U\in L(\mathbb{C}^2)$ are guaranteed to map \emph{any} state in $S_1$ back into $S_1$. Since we defined the stabilizer states by their property of having the largest stabilizer group with respect to $\Pi_1$, we can equivalently ask which unitaries $U$ \emph{normalize} the Pauli group, i.e.\ take group elements to group elements under conjugation by U. Clearly every element of the Pauli group itself has this property, but in general there is a larger group of unitaries which do as well. The group of unitaries which normalize the Pauli group is called the Clifford group,
\begin{equation}
    C_1=\left\{ U\in L(\mathbb{C}^2) \: | \: 
    UgU^\dagger\: \forall g \in \Pi_1\right\}.
\end{equation}
It suffices to check that a unitary takes $\sigma_X$ and $\sigma_Z$ to elements of $C_1$. For example, the Hadamard \cite{sylvester1867lx,hadamard1893resolution} and phase gates,
\begin{equation}
    H\equiv\frac{1}{\sqrt{2}}\begin{pmatrix}1&1\\1&-1\end{pmatrix}, \qquad P\equiv\begin{pmatrix}1&0\\0&e^{\frac{i \pi}{2}}\end{pmatrix}=\begin{pmatrix}1&0\\0&i\end{pmatrix},
\end{equation}
are both elements of the Clifford group, since $H\sigma_XH^\dagger=\sigma_Z$, $H\sigma_ZH^\dagger=\sigma_X$, and $P\sigma_XP^\dagger=\sigma_Y$, $P\sigma_ZP^\dagger=\sigma_Z$. In fact, these two gates suffice to generate the Clifford group, $C_1=\langle H, P \rangle$. We see, in particular, that because $PP=\sigma_Z$, we can easily construct the Paulis themselves out of $H$ and $P$.

	\begin{figure}[t]
		\begin{center}
		\begin{overpic}[width=0.9\textwidth]{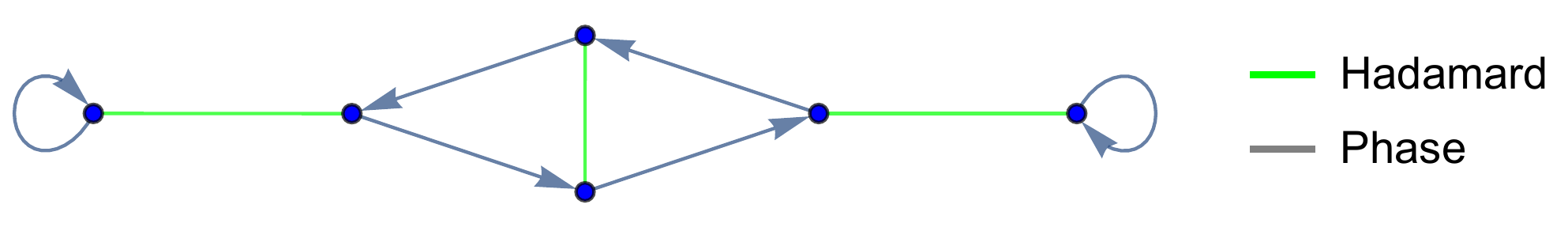}
		\put (5,2.5) {$\ket{0}$}
		\put (66,2.5) {$\ket{1}$}
		\put (20.5,3.5) {$\ket{+}$}
		\put (50.5,9.5) {$\ket{-}$}
		\put (38.8,0) {$\ket{i}$}
		\put (32.8,14.2) {$\ket{-i}$}
        \end{overpic}
		\caption{Complete one-qubit reachability diagram. Because the Hadamard gate is its own inverse we have depicted edges corresponding to Hadamard gate applications as undirected, but gates corresponding to phase gates as directed.
		\label{OneQubitReachabilityDiagram}}
	\end{center}
	\end{figure}

Given the set of stabilizer states $S_1$ and a set of generators of the Clifford group $C_1$, which we call \emph{Clifford gates} \cite{Gottesman:1997zz}, we can arrange the states into a \emph{reachability graph}, with each vertex labeled by a stabilizer state and each edge between vertices labeled by the Clifford gate which maps one vertex to the other. (The Hadamard gate is its own inverse, so it can be represented by an undirected edge, but the phase gate is not, so it is represented by a directed edge.) The reachability graph for $S_1$ is shown in Figure \ref{OneQubitReachabilityDiagram}. Note that the reachability graph depends on a particular choice of generators of $C_1$, e.g. the Clifford gates. The significance of choosing $H$ and $P$, in particular, as our generators is that both operators have physical significance and can be implemented experimentally (relatively) easily.

Note that the reachability graph consists of a single component: as implied by the definition of the Clifford group, we can use the Clifford gates to get (in some number of steps) from any initial stabilizer state, e.g. $\ket{0}$, to any other stabilizer state. Furthermore, the graph contains many cycles: trivial ones, where a Clifford gate acts as the identity on a particular stabilizer state, but also longer ones. For example, because $H^2=P^4=I$, every edge corresponding to a Hadamard application is part of a cycle of length two, and every edge corresponding to a nontrivial phase application is part of a cycle of length four, such as the diamond representing a nontrivial cycle consisting of four phase gates we see at the center of Figure \ref{OneQubitReachabilityDiagram}. More interestingly, we have cycles consisting of more than one type of gate, such as the triangles with two phase gates and one Hadamard.

To better understand these cycles, we can apply two basic group-theoretic results. First, Lagrange's theorem says that the order of any subgroup $H$ of a finite group $G$ gives an integer partition of that group \cite{Alperin1995}: explicitly,
	\begin{equation}\label{LagrangeTheorem}
		|G| = [G:H]\cdot |H|, \hspace{.5cm} \forall H \leq G,
	\end{equation}
with $[G:H]$ the index of $H$ in $G$. Second, the orbit-stabilizer theorem says that, when $H$ is a stabilizer subgroup of $G$ with respect to some state $x\in X$, i.e.\ $H=G_x$, there exists a bipartition between the orbit of x in G, $|G \cdot x|$, and the set of cosets of the stabilizer subgroup in $G$, $G/H$. Hence these two objects have the same dimension:
\begin{equation}
    |G \cdot x|=\frac{|G|}{|H|} = [G:H]\label{eq:orbit_stabilizer},
\end{equation}
where in the last equality we have used \eqref{LagrangeTheorem}.

The Clifford group $C_1$ is a group of operators acting on the one-qubit Hilbert space $\mathbb{C}^2$. As we have discussed, for each $\ket{\Psi} \in S_n \subset \mathbb{C}^2$, there exists a subgroup $G_\ket{\Psi}$. Hence we can apply this group-theoretic machinery to our case of interest. Substituting $\ket{\Psi}\in S_1$ for $x\in X$, $C_1$ for $G$, and $G_{\ket{\Psi}}$ for $H$ in \eqref{eq:orbit_stabilizer} gives
	\begin{equation}\label{Burnside+Lagrange}
		|C_1 \cdot \ket{\Psi}| = \frac{|C_1|}{|G_{\ket{\Psi}}|},
	\end{equation}
where $|C_1 \cdot \ket{\Psi}|$ denotes the length of the orbit of $\ket{\Psi}$. When we represent the action a of group element by a graph, the orbit length is the largest number of vertices in any connected component of the graph.

Explicitly, consider the set of one-qubit stabilizer states $S_1$ defined in \ref{OneQubitStabilizerStates}. The one-qubit Clifford group $C_1$ is constructed from the generating set $\langle H_1,P_1 \rangle$. The orbit of each $\ket{\Psi} \in S_1$ can be computed directly using Equation \eqref{Burnside+Lagrange}, with results shown in Table \ref{tab:OneQubitFullOrbits}. One can easily verify these results by comparing with the reachability diagram in Figure \ref{OneQubitReachabilityDiagram}.
\begin{table}
\begin{center}
\begin{tabular}{|c||c|c|c|}
	\hline
	$\ket{\Psi}$ & $|C_1|$ & $|C_{1_{\ket{\Psi}}}|$ & $|C_1 \cdot \ket{\Psi}|$ \\ 
	\hline
	\hline
	$\ket{0}$ & 192 & 32 &6\\
	\hline
	$\ket{1}$ & 192 & 32 &6\\
	\hline
	$\ket{+}$ & 192 & 32 &6\\
	\hline
	$\ket{-}$ & 192 & 32 &6\\
	\hline
	$\ket{i}$ & 192 & 32 &6\\
	\hline
	$\ket{-i}$ & 192 & 32 &6\\
	\hline
\end{tabular}
    \caption{Orbit lengths for each single-qubit stabilizer state under the one-qubit Clifford group $C_1$.}
    \label{tab:OneQubitFullOrbits}
    \end{center}
\end{table}

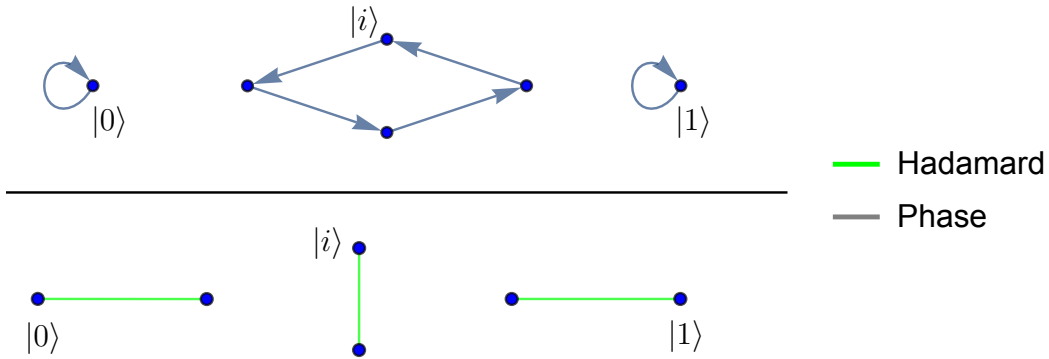
\begin{figure}
		\begin{center}
		\begin{overpic}[width=0.9\textwidth]{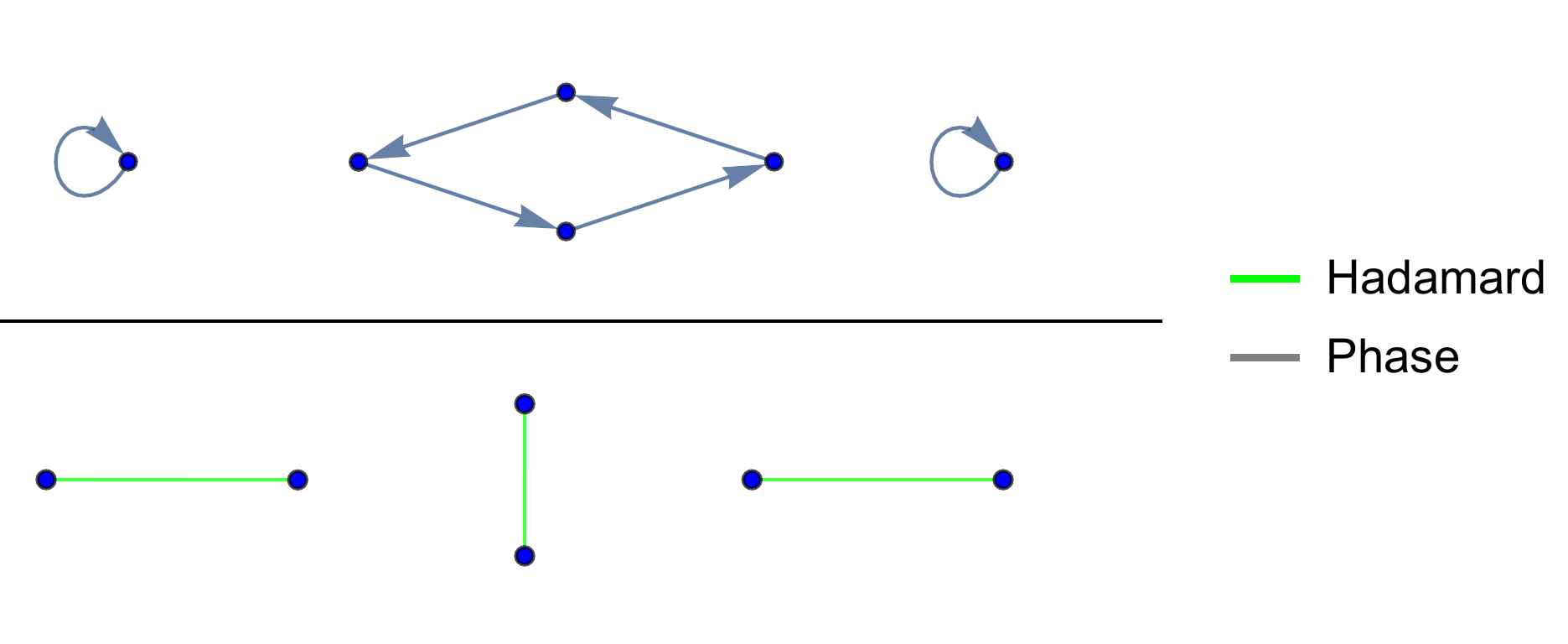}
		\put (8,26.5) {$\ket{0}$}
		\put (63.5,26.5) {$\ket{1}$}
		\put (32.5,36) {$\ket{i}$}
		\put (1.7,6) {$\ket{0}$}
		\put (62.8,6.2) {$\ket{1}$}
		\put (29,15) {$\ket{i}$}
        \end{overpic}
	\caption{The one-qubit reachability diagram restricted to only phase (Top) and only Hadamard (Bottom) operations reveals disconnected orbits of varying length for states $\ket{\Psi} \in S_1$.}
	\label{OneQubitH1OnlyP1Only}
	\end{center}
	\end{figure}

We can also use this machinery to consider the orbits of states $\ket{\Psi} \in S_1$ under subgroups of $C_1$, see Table \ref{tab:OneQubitSubgroupOrbits}. Let $G_P < C_1$ denote the subgroup generated by only the phase gate. This group contains the $4$ unique elements $P_1, P_1^2, P_1^3,$ and $P_1^4$ (recall $P_1^4 = I$). Each element acts trivially on $\ket{0}$ and $\ket{1}$, and thus these two states are stabilized by all elements of $G_P$. The remaining states $(\ket{+},\ket{-},\ket{i},\ket{-i})$ form a cycle of length $4$ under operation of $P_1$, each stabilized only by the identity $P_1^4$. These orbits manifest as subgraphs of the reachability graph as seen in Figure \ref{OneQubitH1OnlyP1Only}.

\begin{table}
\begin{center}
\begin{tabular}{|c||c|c|c|}
	\hline
	$\ket{\Psi}$ & $|G_P|$ & $|G_{P_{\ket{\Psi}}}|$ & $|G_P \cdot \ket{\Psi}|$ \\ 
	\hline
	\hline
	$\ket{0}$ & 4 & 4 & 1\\
	\hline
	$\ket{1}$ & 4 & 4 & 1\\
	\hline
	$\ket{+}$ & 4 & 1 & 4\\
	\hline
	$\ket{-}$ & 4 & 1 & 4\\
	\hline
	$\ket{i}$ & 4 & 1 & 4\\
	\hline
	$\ket{-i}$ & 4 & 1 & 4\\
	\hline
\end{tabular}
\hspace{.2cm}
\begin{tabular}{|c||c|c|c|}
	\hline
	$\ket{\Psi}$ & $|G_H|$ & $|G_{H_{\ket{\Psi}}}|$ & $|G_H \cdot \ket{\Psi}|$ \\ 
	\hline
	\hline
	$\ket{0}$ & 2 & 1 & 2\\
	\hline
	$\ket{1}$ & 2 & 1 & 2\\
	\hline
	$\ket{+}$ & 2 & 1 & 2\\
	\hline
	$\ket{-}$ & 2 & 1 & 2\\
	\hline
	$\ket{i}$ & 2 & 1 & 2\\
	\hline
	$\ket{-i}$ & 2 & 1 & 2\\
	\hline
\end{tabular}
    \caption{Orbit lengths for each single-qubit stabilizer state under subgroups $G_P<C_1$ and $G_H<C_1$, generated by only the phase gate and Hadamard gate respectively.}
    \label{tab:OneQubitSubgroupOrbits}
    \end{center}
\end{table}

Similarly, consider the subgroup $G_H<C_1$, generated by only the Hadamard gate. This group only has $2$ elements since $H_1 = H_1^{-1}$ and $H_1^2 = I$. Distinctly, the Hadamard gate stabilizes no element of $S_n$, and thus each state in $S_1$ is stabilized by only the identity $(H_1^2)$. Each $\ket{\Psi}\in S_1$ subsequently has an orbit of length $2$, as shown in Table \ref{tab:OneQubitSubgroupOrbits}, which results in the decomposition of the reachability graph into pairs of states (Figure \ref{OneQubitH1OnlyP1Only}), equivalent up to a change of basis. 

In the remainder of the paper, we will often consider splitting the reachability graph into subgraphs constructed from a subset of the Clifford gates. Ultimately, the underlying structure behind this graphical representation is precisely the partitioning of the stabilizer set by orbit length.

As the alert reader will have realized from our notation, we can extend the definitions of the Pauli group, stabilizer states, and Clifford gates to more than one qubit. The Pauli group $\Pi_n$ on $n$ qubits consists of ``Pauli strings'' acting on each of the qubits, and is generated by\footnote{Note that writing a Pauli string requires not just a factorization of the $2^n$ dimensional Hilbert space into tensor factors representing qubits (each of which has a specified basis $\{0,1\}$), but a particular \emph{ordering} of the qubits from $1$ to $n$: we write $\ket{a_1\ldots a_n}\equiv \ket{a_1}_1\otimes\ldots\ket{a_n}_n$. It should be clear that the set of length-one Pauli strings as a whole are the independent of choice of ordering, and hence so is the Pauli group $\Pi_n$ they generate. This will also be the case for the Clifford group $C_n$ and the set of all Clifford gates. However, individual gates will of course depend on the choice of ordering. We will often consider a subset of the Clifford gates which act only on the first two qubits, which again depends on the choice of ordering, but there is an equivalent subset which acts on any two specified qubits, so although the position of a given state within the graphs we will generate depends on a choice of ordering, the overall graph structures themselves will not.} the length-one Pauli strings like $I^1\otimes\ldots\otimes I^{k-1} \otimes \sigma_Z^k \otimes I^{k+1} \otimes \ldots \otimes I^n$. The stabilizer states $S_n$ are those states $\ket{\Psi}\in (\mathbb{C}^2)^{\otimes n}$ with stabilizer groups of maximal size, $\mathrm{dim}\: G_\ket{\Psi} = 2^n$.\cite{aaronson2004improved,garcia2017geometry} 
\begin{equation}
    \left|S_n\right| = 2^n \prod_{k=0}^{n-1} (2^{n-k}+1)
    = 2 (2^n +1) \left|S_{n-1}\right| \approx 2^{(.5+o(1))n^2}.
\end{equation}
We have given a closed-form expression, recursion relation (with base case $|S_1|=6$, as constructed explicitly above), and asymptotic expression.

The Clifford group $C_n$ is again the group of unitaries which normalize $\Pi_n$, which contains the Hadamard and phase gates acting on each individual qubit. However, these gates no longer suffice to generate the full Clifford group: because the gates act only on a single qubit, they cannot change the entanglement structure of a state. Yet not every stabilizer state has the same entanglement structure. One subset of the two-qubit stabilizer states consists of the tensor product of a one-qubit stabilizer state on the first qubit and another one-qubit stabilizer state on the second qubit: these states are, of course, product states. But, for example, the Bell state $\frac{1}{\sqrt{2}}(\ket{00}+\ket{11})$ is a two-qubit stabilizer state, stabilized by $\{I^1I^2,\sigma_X^1\sigma_X^2,-\sigma_Y^1\sigma_Y^2,\sigma_Z^1\sigma_Z^2\}$. Hence any set of operators generating $C_n$ must contain operators which map product states to entangled states and vice versa.

One convenient gate which accomplishes this task is the $CNOT_{i,j}$ gate, which performs a controlled $NOT$ operation on the $j$th qubit depending on the state of the $i$th qubit:
\begin{equation}
		CNOT_{1,2} \equiv \begin{pmatrix}
            1 & 0 & 0 & 0\\
            0 & 1 & 0 & 0\\
	    0 & 0 & 0 & 1\\
	    0 & 0 & 1 & 0
            \end{pmatrix} \in L(\mathbb{C}^4).
\end{equation}
To check that $CNOT_{i,j}\in C_n$, it suffices to check its action on the length-one Pauli strings consisting of $\{\sigma_X^i,\sigma_Z^i,\sigma_X^j,\sigma_Z^j\}$ tensored with identities on every other qubit. A standard calculation shows that conjugation by $CNOT_{i,j}$ maps these four strings to $\{\sigma_X^i\otimes\sigma_X^j,\sigma_Z^i,\sigma_X^j,\sigma_Z^i\otimes\sigma_Z^j\}$ tensored with identities, respectively. Hence the CNOT gates are indeed members of the Clifford group; together with the Hadamards and phase gates, they generate\footnote{In fact, we only need half of the CNOT gates, the $n(n-1)/2$ gates $CNOT_{i,j}$ with $i<j$, because we have the relation $CNOT_{j,i}=H_i H_j CNOT_{i,j} H_j H_i$. Note that this expression is not unique, because $H_i$ and $H_j$ acting on distinct qubits commute. Following convention, we will nevertheless take the Clifford gates to include all $n(n-1)$ CNOT gates; e.g.\ our two-qubit reachability graph will show the actions of both $CNOT_{1,2}$ and $CNOT_{2,1}$.\label{fn:cnot_dependence}} $C_n$, so the \emph{$n$-qubit Clifford gates} are taken to be the $n$ Hadamard gates $H_n$, the $n$ phase gates $P_n$, and the $n(n-1)$ gates $CNOT_{i,j}$ (for $i\ne j$).

We can thus construct, for the Hilbert space of $n$ qubits $\mathbb{C}^{2n}$, a reachability graph for the stabilizer states $S_n$ using the Clifford gates which generate $C_n$. We will devote the rest of the paper to studying this object, and the subgraphs formed from it by restricting to a subset of the Clifford gates, at various qubit numbers $n>1$. Before we begin this study, however, we will first categorize the various possible allowed entropic structures of the stabilizer states $S_n$.
    
\subsection{Review of the Entropy Cone}\label{sec:EntropyCone}

All stabilizer states themselves are pure states; that is, for a given stabilizer state $\ket{\psi}$, the density matrix $\rho_{\psi}$ is idempotent and thus has zero total entropy:
\begin{equation}
    \rho_{\psi}=\ket{\psi}\bra{\psi}, \quad \rho_{\psi}^2=\rho_{\psi}, \quad S(\psi)=-\tr \rho_\psi \log_2 \rho_\psi=0.
\end{equation}
For this paper, we measure entropy in \emph{bits}: every $\log$ should be interpreted as $\log_2$ throughout.  As we will see shortly, this convention results in positive integer entries for every element in the entropy vector for all stabilizer states.

Non-trivial entropic structure arises when we consider how one subset of qubits relates to its complement.  Suppose we pick a $p$-qubit subset $I$ of the $n$ qubits in a full stabilizer state.  Then, the entanglement entropy between the $p$-qubit subset $I$ and its $(n-p)$-qubit complement $\bar{I}$ is given by
\begin{equation}
    \rho_{I}=\tr_{I} \ket{\psi}\bra{\psi}, \quad S_I=-\tr \rho_I \log_2 \rho_I.
\end{equation}
Here the trace $\tr_{I}$ is taken over only the $p$ qubits in the subset $I$. The density matrix $\rho_I$ is called a reduced density matrix, and since stabilizer states are pure only their reduced density matrices have nonzero entropy.    Since the entropy for the full state is zero, we also have $S_I=S_{\bar{I}}$.

For an $n$-qubit stabilizer state, there are thus $2^{n-1}-1$ entropies.  Listing all of these entropies produces the \emph{entropy vector} for a given state. As an example, 2-qubit stabilizer states have full entropy vector $\vec{S}=(S_A,S_O,S_{AO})$.  However, since the state is pure, $S_{AO}=0$ and $S_A=S_O$. We thus write the 2-qubit entropy vector as just $\vec{S}=(S_A)$. Here, we have labelled our last qubit with $O$ to indicate it acts as a purifier for the other qubits.

Similarly, for a $3$-qubit state, we write $\vec{S}=(S_A,S_B,S_O)$, or equivalently, $\vec{S}=(S_A,S_B,S_{AB})$. For $4$ qubits we have $\vec{S}=(S_A,S_B,S_C,S_O,S_{AB},S_{AC},S_{AO})$.  We could have written $S_O=S_{ABC}$ and $S_{AO}=S_{BC}$ instead; some sources choose a different ordering for the entropy vector accordingly.

As reviewed in Section \ref{sec:StabilizerReview}, only CNOT gates can create or destroy entanglement entropy.  We can now refine this statement: the $CNOT_{i,j}$ gate can only alter entropies $S_I$ where qubit $i\in I$ but qubit $j\in \bar{I}$, or vice versa.

In addition to the equation $S_I=S_{\bar{I}}$, which holds for any pure state, entropies for subsets of qubits also obey entropy inequalities.  The full set of entropy inequalities obeyed by a given set of states defines the \emph{entropy cone} \cite{10.1109/18.641561,1193790}. The quantum entropy cone is the largest region we will discuss; any quantum state obeys the inequalities that define its boundaries. The Araki-Lieb inequality \cite{cmp/1103842506} $S_{IJ}+S_I\geq S_J$ and subadditivity $S_I+S_J\geq S_{IJ}$, where $I,J$ are disjoint sets of qubits, are both examples of inequalities
obeyed by all quantum states. The full quantum cone at arbitrary qubit number is not
known, but many classes of inequalities are \cite{681320,4215134,Linden:2004ebt,Schnitzer:2022exe}. 

Instead, we will be interested in two smaller cones: the stabilizer cone, and the holographic entropy cone. The stabilizer cone \cite{Linden:2013kal,doi:10.1063/1.4818950,HernandezCuenca:2019wgh,Bao:2020zgx,Bao:2020mqq} is defined as the smallest convex cone which contains all stabilizer states.  Since all states we study are stabilizer states, they will all lie within the stabilizer cone. Although we will not study this cone in further detail, we will use the fact that it is larger than our next cone: the holographic entropy cone.

As defined in \cite{Bao:2015bfa}, the holographic entropy cone is the smallest convex cone in entropy space which contains all quantum states that have a dual representation as a classical gravity state.  The Ryu-Takayanagi formula relates the entanglement entropies for subregions of field theoretic states to areas of extremal surfaces in their dual holographic geometries.  The geometry of these extremal surfaces constrains the allowed entropy vectors.  The first such constraint was the monogamy of mutual information%
\footnote{As discussed in \cite{Bao:2015bfa,HernandezCuenca:2019wgh}, at 6 qubits (5 regions), further entropy inequalities arise not described here.  Since we limit our detailed discussion to 5 qubits or fewer, the Araki-Lieb, subadditivity, and monogamy inequalities are sufficient to test if a state lies within the holographic entropy cone. }%
\ \cite{Hayden:2011ag}, 
\begin{equation}\label{eq:MonogamyofMutualInformation}
    S_{IJ}+S_{IK}+S_{JK}\geq S_{IJK}+S_I+S_J+S_K.
\end{equation}
Here again $I,\, J,\, K$ are disjoint sets of qubits. This inequality is not obeyed by all quantum states, nor by all stabilizer states, but it is obeyed by all states which have a dual smooth classical geometry.  As a consequence, it sets the first boundary between the stabilizer and holographic cones.  As we will review below, beginning at four qubits (or, in the holographic dual language, three regions plus a purifier), some stabilizer states cannot have a smooth holographic dual, because they do not lie within the holographic cone. One of our interests in studying the reachability diagrams is to understand what gate actions on a given state can move it from within the holographic cone to outside of it.  These gate actions then describe how to create a state whose geometry is definitely nonclassical.

\section{The Two-Qubit Stabilizer Graph}\label{sec:two}

\begin{figure}[h]
\begin{center}\includegraphics[scale=0.45]{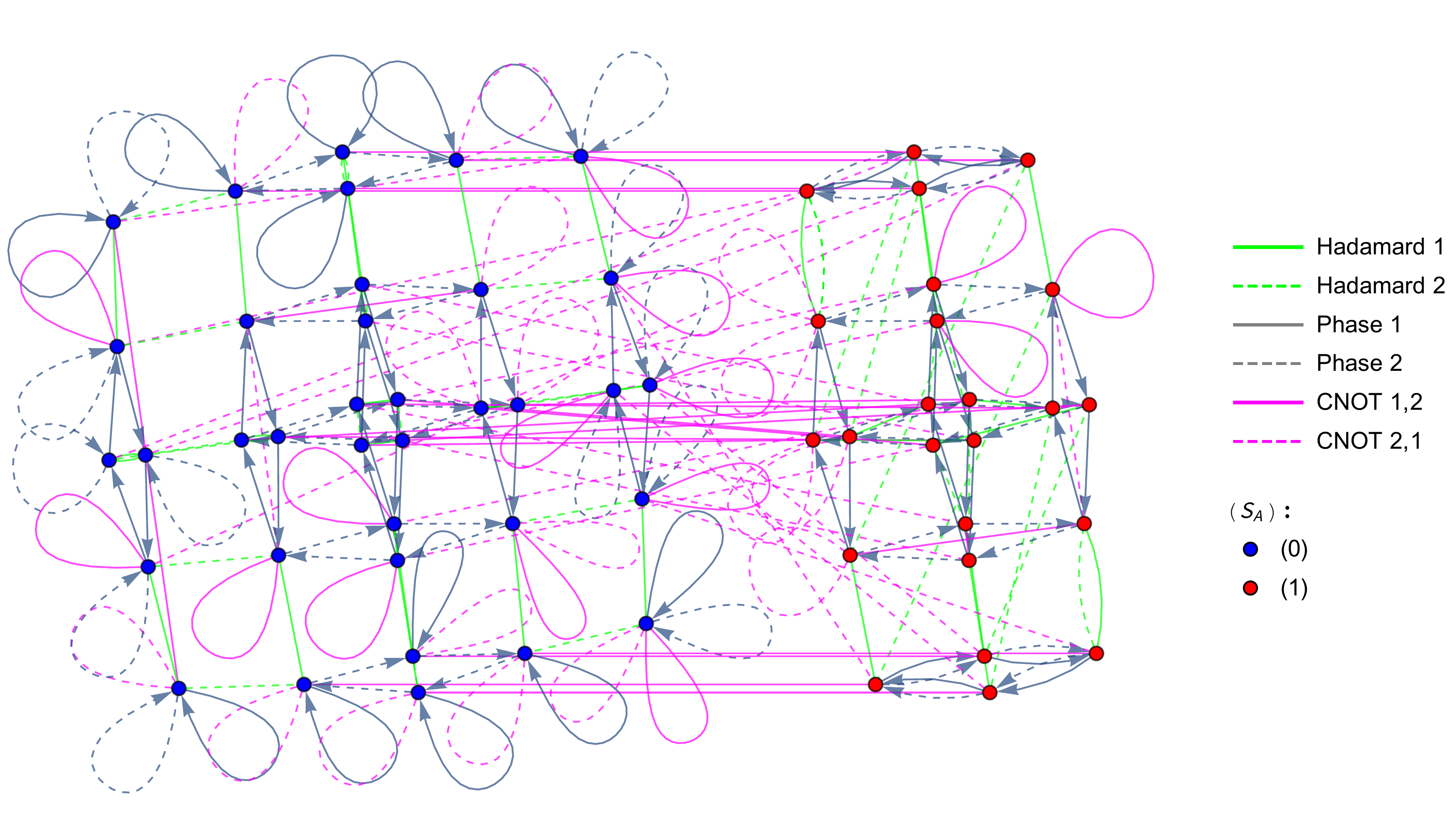}
\caption{This complete 2-qubit reachability graph depicts the full map between all stabilizer states under action of the Clifford group. Edges depicting $H_i$ and $CNOT_{i,j}$ are undirected since they are each their own inverse.  $P$ is not its own inverse since $P^4 = I$;  consequently the phase gates are represented by directed edges.  The color of the edge indicates the type of gate, and the line texture (solid vs. dashed) indicates the qubits which the gate acts on.}
\label{TwoQubitCompleteGraph}
\end{center}
\end{figure}

At two qubits, the reachability graph contains $60$ vertices, representing the full set of 2-qubit stabilizer states. These states are connected by the six Clifford gates: $H_1,\, H_2,\, P_1,\, P_2,\, CNOT_{1,2},$ and $CNOT_{2,1}$. The full graph is visible in Figure \ref{TwoQubitCompleteGraph}. 

As discussed in section \ref{sec:EntropyCone}, 2-qubit stabilizer states have a one-component reduced entropy vector $S_A$. For these states, $S_A$ is either zero or one,%
\footnote{Since we are working with qubits, we measure entropies using $\log_2$. That is, the reduced density matrix of one qubit in a maximally-entangled pair has von Neumann entropy 1.} %
so states on the reachability graph (represented by vertices) are either unentangled (blue) or form an maximally entangled pair (red). Since only a CNOT gate can alter the entropy vector, the graph has two subgraphs which are connected only by CNOT gates (pink lines).  One subgraph has all of the entangled stabilizer states, while the other has all of the unentangled ones.  At any number of qubits, removing all of the CNOT gates breaks the full graph into subcomponents. Each subcomponent has the same entropy vector throughout.

The graph here depicts every gate acting on each vertex. Since some of these gate actions act trivially on particular states, the graph contains loops. These loops thus represent gate actions which stabilize the state represented by the vertex attached to the loop. This graph also contains degenerate gate action, i.e. multiple edges that map one vertex to another as can be seen in the bottom-rightmost pair connected by $H_1$ and $H_2$. Beginning in section \ref{2qubitHCNOT}, we will suppress trivial loops since we are most interested in understanding the gates that move us between states. 

While Figure \ref{TwoQubitCompleteGraph} completely describes the connections between 2-qubit stabilizer states via the Clifford gates, the graph's complexity obscures some of the important features.  At higher qubit numbers, the full graphs quickly increase in complexity; we thus relegate their complete graphs to Appendix \ref{Two-Qubit Graphs}.  In order to further explore the structure of the 2-qubit graph, and to extend our understanding to higher qubits, we will now explore restricted graphs which only depict the action of various subsets of the Clifford gates.  

\subsection{Restricted Graphs}

Beginning at three qubits, we will further restrict our focus to the restricted graphs composed of the Hadamard and CNOT gates on only the first two qubits. To help motivate why we concentrate on this gate subset, we begin by constructing several different restricted graphs for two qubits.

We construct a restricted graph by considering only select operations of the full Clifford group. This restriction corresponds to removing edges, representing eliminated gate operations, from the complete reachability graph. Each restricted graph reveals different details about the 
connectivity and physics of the stabilizer states.

\subsubsection{Two-Qubit Hadamard}

\begin{figure}[h]
\begin{center}
\includegraphics[scale=0.8]{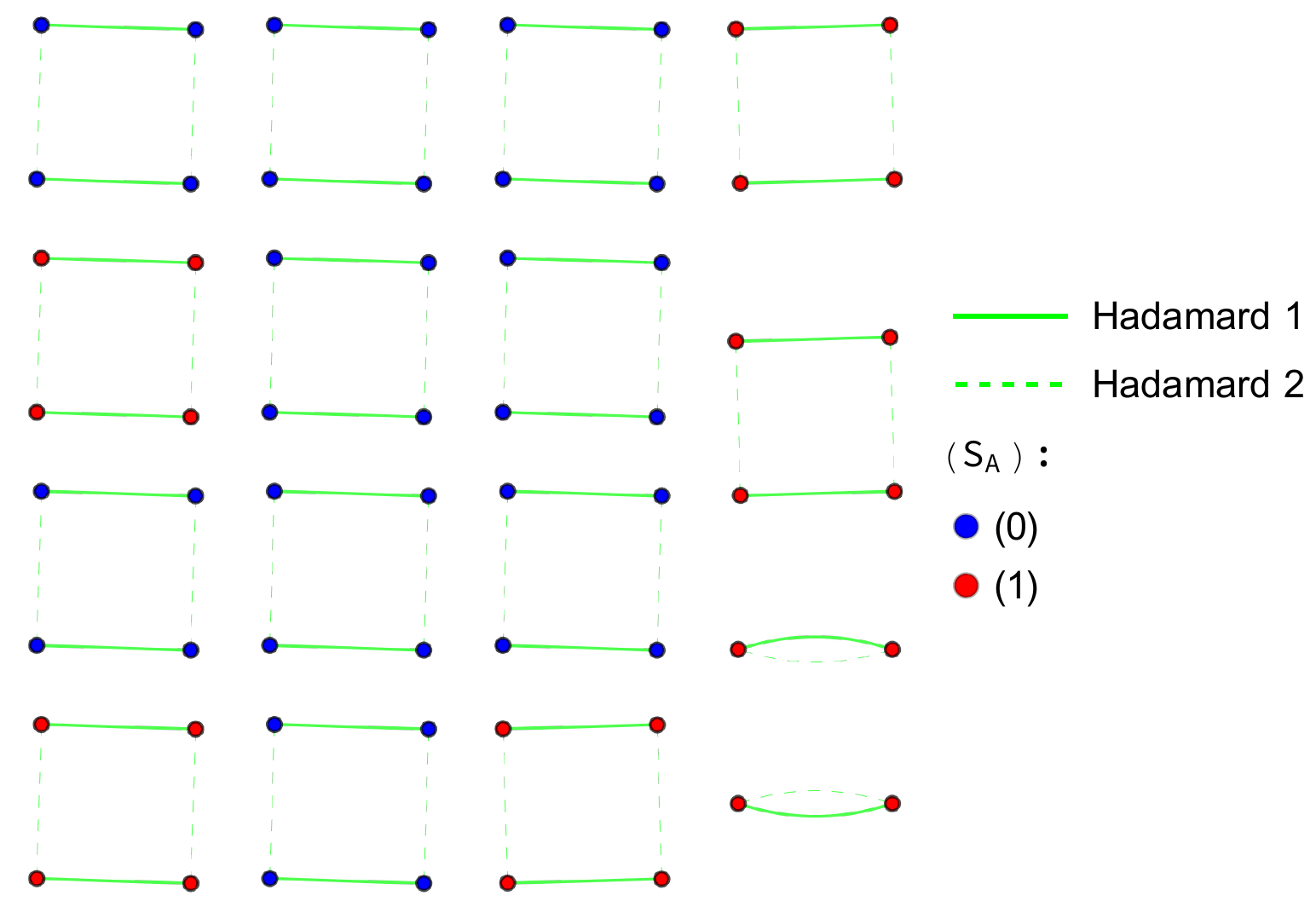}
\caption{The $14$ squares and $2$ connected pair subgraphs in the graph restricted to $H_1$ and $H_2$ at two qubits. Since $[H_1,H_2]=0$ and $H_1^2=H_2^2=\mathbbm{1}$, these subgraphs are the only allowed shapes. The four states which connect in pairs rather than squares are given in Equation \eqref{2qubitHadamardPairs}.}
\label{TwoQubitH1H2}
\end{center}
\end{figure}

We begin by considering only the two Hadamard gates on two qubits, $H_1$ and $H_2$, as in  Figure \ref{TwoQubitH1H2}. The Hadamard gate $H_i$ enacts a basis change on the $i$th qubit.  Consequently, Hadamards on different qubits commute.  Since each Hadamard gate also satisfies $H_i^2=\mathbbm{1}$, each subgraph can have at most four different states.  Indeed, the majority of the 2-qubit states organize themselves into squares, consisting of a starting state $\ket{\psi}$ and the states $H_1\ket{\psi}$, $H_2\ket{\psi}$, and $H_1H_2\ket{\psi}=H_2H_1\ket{\psi}$. Additionally, since Hadamard gates cannot change entanglement, each square has the same entropy vector throughout.%

The four states arranged in pairs in the lower right of Figure \ref{TwoQubitH1H2} are worth further note.  They are
\begin{align}\nonumber
    \ket{00}+\ket{11}, \qquad \ket{00}+\ket{01}+\ket{10}-\ket{11}&=H_1\left(\ket{00}+\ket{11}\right)=H_2\left(\ket{00}+\ket{11}\right);
    \\\label{2qubitHadamardPairs}
    \ket{01}-\ket{10}, \qquad \ket{00}-\ket{01}-\ket{10}-\ket{11}&=H_1\left(\ket{01}-\ket{10}\right)=H_2\left(\ket{01}-\ket{10}\right).
\end{align}
Since $H_1$ and $H_2$ produce the same action on each of these states, their subgraphs thus form degenerate pairs instead of full squares. That is, the four states given in \eqref{2qubitHadamardPairs} are eigenstates of $H_1\otimes H_2$; the two states given in the first line have eigenvalue $+1$, while the two states on the second line have eigenvalue $-1$.

\subsubsection{Two-Qubit Phase and Hadamard}\label{TwoQubitPandHSection}

\begin{figure}[h]
\begin{center}
\includegraphics[scale=0.42]{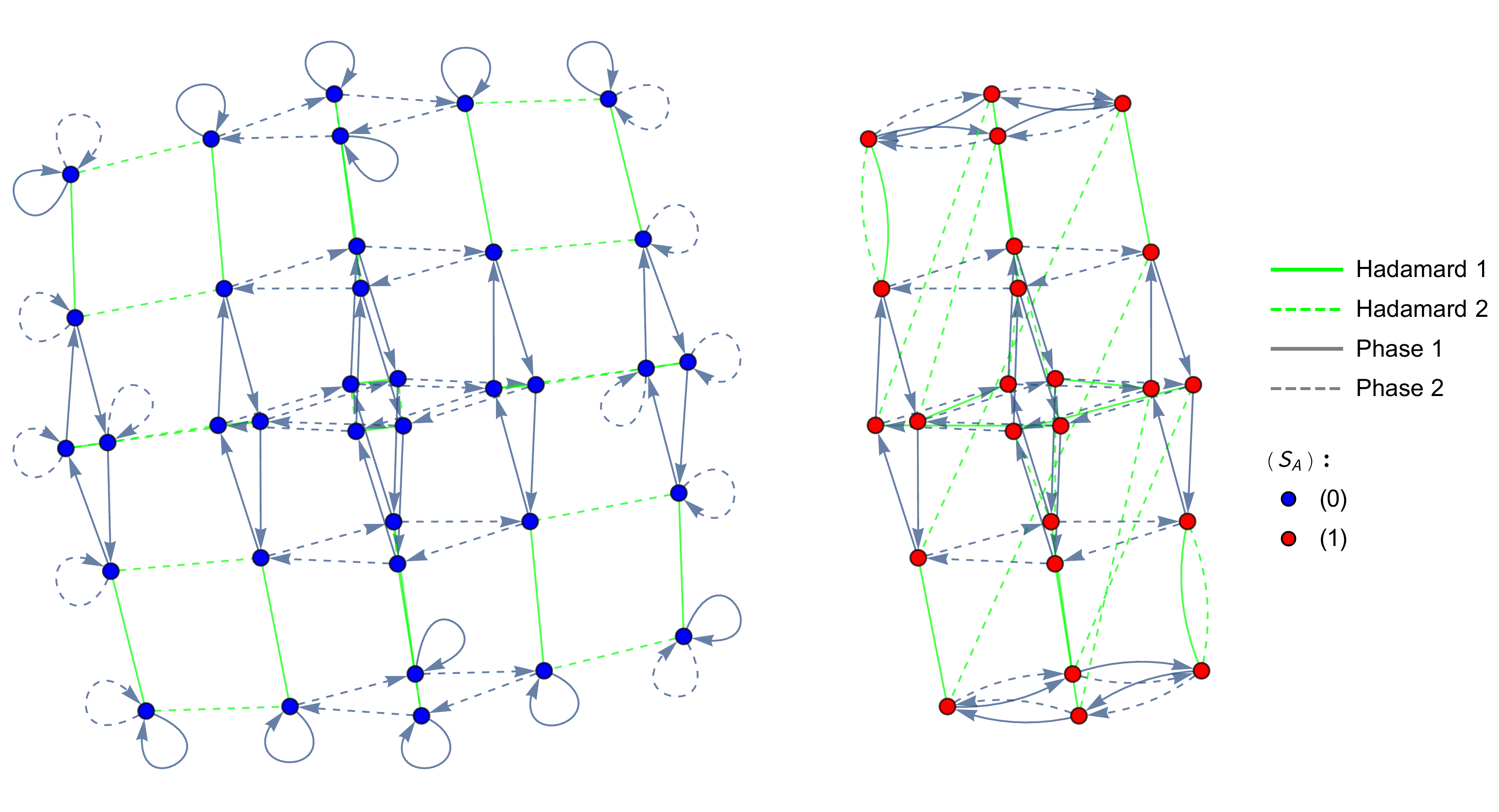}
\caption{Removing the set of CNOT operations from the full reachability graph Figure \ref{TwoQubitCompleteGraph} reveals two disconnected subgraphs.  Since only the CNOT gates can change the entropy, each subgraph has the same entropy vector for all of its states.}
\label{TwoQubitH1H2P1P2}
\end{center}
\end{figure}

Considering both the Hadamard and phase gates yields the restricted graph depicted in Figure \ref{TwoQubitH1H2P1P2}. The right subgraph of Figure \ref{TwoQubitH1H2P1P2} contains all entangled states. The left subgraph of this figure consists entirely of unentangled states. This bisection of the restricted graph is required by the removal of CNOT edges since neither Hadamard nor phase can alter the entanglement entropies.

The Hadamard boxes from Figure \ref{TwoQubitH1H2} are clearly visible in the unentangled subgraph.  For the entangled state subgraph, the two degenerate pairs are present at the upper left and lower right, while the Hadamard boxes are still present but slightly harder to visualize.  Removing the phase gates of course reproduces the Hadamard-only Figure \ref{TwoQubitH1H2}, while removing both Hadamard operations yields the phase-only restricted graph (Figure \ref{TwoQubitP1P2} in Appendix \ref{Two-Qubit Graphs}).

The four basis states  $(\ket{00},\ket{01},\ket{10},\ket{11})$, are located at the corners of the unentangled subgraph.%
\footnote{Qubits added to a system are appended to the right of the quantum register, as described in section \ref{sec:StabilizerReview}. Thus, an $n$-qubit product state is represented by $\ket{a_1\ldots a_n}\equiv \ket{a_1}_1\otimes\ldots\ket{a_n}_n$.}
Since both phase gates act trivially on basis states, each corner has two attached directed loops.  For the other sixteen states on the outside of the unentangled subgraph (four on each side), one of the phase gates is trivial while the other makes a square (since $P_i^4=\mathbbm{1}$).  As an example, the state $\ket{+0}=H_1\ket{00}$ satisfies
\begin{equation}
    P_2\ket{+0}=\ket{+0}, \quad P_1\ket{+0}=\ket{i0}, \quad P_1^4\ket{+0}=\ket{+0}.
\end{equation}
The remaining 16 states in the center of the unentangled graph are connected by 4 $P_1$ squares and 4 $P_2$ squares, arising because $P_1^4=P_2^4=\mathbbm{1}$.

In the entangled subgraph, again the 16 states in the center are connected by 4 $P_1$ squares and 4 $P_2$ squares.  The remaining 8 entangled states, 4 at the top and 4 at the bottom of the entangled subgraph in Figure \ref{TwoQubitH1H2P1P2}, are in degenerate pairs.  For these states, either $P_1\ket{\psi}=P_2^3\ket{\psi}$ or $P_1\ket{\psi}=P_2\ket{\psi}$. More precisely, $\ket{01}+\ket{10}$ is one of the states at the top of the figure, so it satisfies
\begin{equation}
    P_1(\ket{01}+\ket{10})=P_2^3(\ket{01}+\ket{10})
\end{equation}
while $\ket{00}+\ket{11}$ is at the bottom of the figure and satisfies
\begin{equation}
    P_1(\ket{00}+\ket{11})=P_2(\ket{00}+\ket{11}.
\end{equation}
Consequently, the phase squares degenerate to connected pairs on all eight of these states.

We can also see the single-qubit reachability diagram reflected here.  In the unentangled graph, removing $H_2$ results in 6 copies of the one-qubit diagram in Figure \ref{OneQubitReachabilityDiagram}, arranged vertically in Figure \ref{TwoQubitH1H2P1P2}. The leftmost copy results from tensoring the six stabilizer states on the first qubit $\ket{0},\, \ket{1}, \, \ket{\pm},\,\ket{\pm i}$ with the state $\ket{0}$ on the second qubit.  Similarly the rightmost copy includes the states
$\ket{01},\, \ket{11},\, \ket{\pm,1},\,\ket{\pm i,1}$.  The four copies in the middle, still connected by the phase gate $P_2$, are constructed similarly except with $\ket{\pm},\, \ket{\pm i}$ for the second qubit.  This tensoring is our first example of a lift, in this case from a one-qubit structure to a two-qubit structure.

In general, a lift of a $k$-qubit state $\ket{\psi}$ to $n$ qubits is a quantum channel which maps $\ket{\psi}$ into $\mathbb{C}^{2n}$ by first tensoring on a $(n-k)$-qubit state $\ket{\phi}$ and then applying an operator $\mathcal{O}$:
\begin{equation}
    \ket{\psi} \rightarrow \mathcal{O}\left(\ket{\Psi}\otimes\ket{\phi} \right): \mathcal{O}\in L(\mathbb{C}^{2n}),\:\ket{\phi}\in\mathbb{C}^{2(n-k)}.
\end{equation}
 We will always have in mind lifts which take stabilizer states to stabilizer states, so we will take $\ket{\phi}$ to be an $(n-k)$-qubit stabilizer state and $\mathcal{O}$ to be a product of $n$-qubit Clifford gates. Given these restrictions, all lifts from $\mathbb{C}^{2k}$ to $\mathbb{C}^{2n}$ are on the same footing, and indeed lifts of $\ket{\psi}$ also successfully lift any other $k$-qubit stabilizer state. What makes a particular lift useful is that it preserves some of the structure seen at $k$ qubits when we go to $n$ qubits. In the example above, the lift given by $\ket{\phi}=\ket{0}$, $\mathcal{O}=I$ mapped all six one-qubit stabilizer states to six two-qubit stabilizer states, preserving their arrangement in the one-qubit reachability graph.

The entangled subgraph is not composed of tensor products of one-qubit stabilizer states, so it is unsurprising that the one-qubit reachability diagram has become more complicated.  We still see four almost-copies of the one-qubit diagram, except the Hadamard gate which connects $\ket{\pm i}$ in Figure \ref{OneQubitReachabilityDiagram} instead connects the four almost-copies to each other. Last, removing $H_1$ instead of $H_2$ results in exactly the same structures; we have just chosen to arrange the entangled subgraph to prioritize the $H_1$ structure.

\subsubsection{Two-Qubit Phase and CNOT}

\begin{figure}[h]
\begin{center}
\includegraphics[scale=0.5]{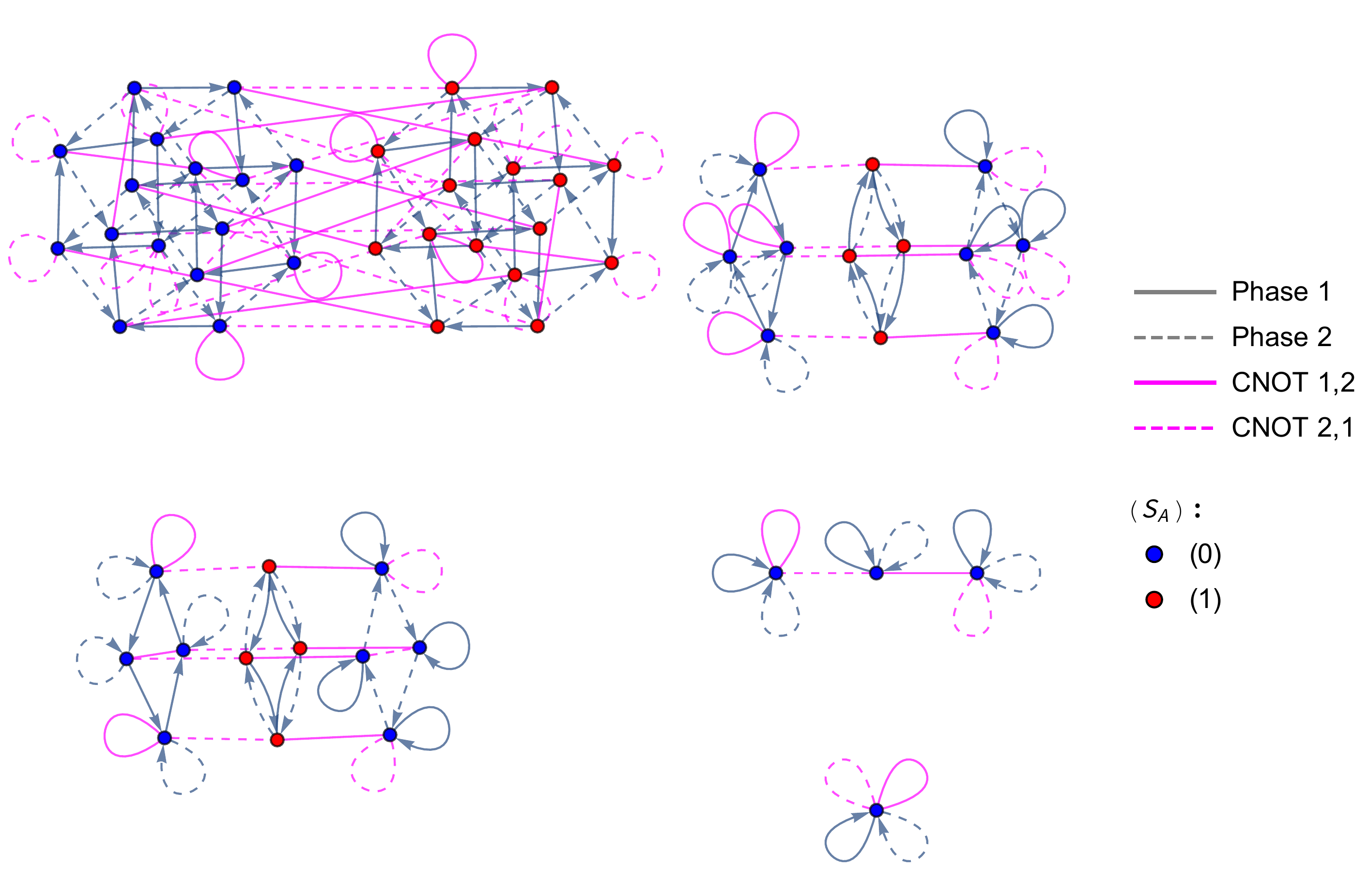}
\caption{Subgraph of 2-qubit complete reachability graph restricted the subgroup generated by CNOT and phase operations. Included structures are various attachments of the simpler structures found in the only-phase and only-CNOT restricted graphs in Figures \ref{TwoQubitP1P2} and \ref{TwoQubitCNOT12CNOT21} in Appendix \ref{Two-Qubit Graphs}.}
\label{TwoQubitP1P2CNOT12CNOT21}
\end{center}
\end{figure}

Considering only operations of the subgroup generated by phase and CNOT, we observe a reduced graph composed of five disconnected substructures (Figure \ref{TwoQubitP1P2CNOT12CNOT21}). Each substructure is inherited from a combination of phase-only and CNOT-only restricted graphs (Figures \ref{TwoQubitP1P2} and \ref{TwoQubitCNOT12CNOT21} in Appendix \ref{Two-Qubit Graphs}). 

Again, only CNOT gates can alter the entropy, so states with different entanglement are connected only via CNOT gates.  However, only some CNOT gates modify the entropy.  For example,
\begin{equation}
    CNOT_{1,2}\ket{00}=CNOT_{2,1}\ket{00}=\ket{00}.
\end{equation}
Since $P_1$ and $P_2$ also stabilize $\ket{00}$, this state is represented by the isolated vertex in Figure \ref{TwoQubitP1P2CNOT12CNOT21}. The CNOT gates permute the remaining basis states
$\ket{01},\, \ket{10},\, \ket{11}$
among each other since CNOT can only flip a bit, not introduce a superposition. As with $\ket{00}$, phase acts trivially on all the remaining basis states.

The upper right subgraph in Figure \ref{TwoQubitP1P2CNOT12CNOT21} consists of $8$ unentangled states and $4$ entangled states. The unentangled states arrange into a $P_1^4$ cycle to the left, and a $P_2^4$ cycle to the right. For the four central entangled states, $P_1\ket{\psi}=P_2\ket{\psi}$, so they are linked in one phase cycle. The entangled states and unentangled states are necessarily connected by CNOT gates.  In this subgraph, all CNOT gates either act trivially or move between entangled and unentangled states.  The CNOT gates alone connect the states into 4 lines of 3 states each.

The bottom left subgraph is similar, except the central four entangled states satisfy
$P_1\ket{\psi}=P_2^{-1}\ket{\psi}$ instead. The three top states, and the three bottom states, both have either trivial CNOT action or CNOT moves between an entangled and unentangled states. For the middle states, however, both CNOT gates act nontrivially on each state, producing a hexagon.  Explicitly, we have the cycle
\begin{equation}
    \ket{1i}=CNOT_{2,1}CNOT_{1,2}CNOT_{2,1}CNOT_{1,2}CNOT_{2,1}CNOT_{1,2}\ket{1i},
\end{equation}
where only two of the states along the way are entangled.

Finally the largest structure, located in the upper left of Figure \ref{TwoQubitP1P2CNOT12CNOT21}, contains only states stabilized by neither $P_1$ nor $P_2$. Again the CNOT gates are the only connections between the entangled and unentangled states. Considering only these CNOT gates, this subgraph contains the remaining $2$ hexagonal cycles from the CNOT-only graph (Appendix \ref{Two-Qubit Graphs}, Figure \ref{TwoQubitCNOT12CNOT21}), as well as $6$ 3-state lines of paired CNOT edges, and $2$ additional states invariant under both CNOT gates.

\subsection{Two-Qubit Hadamard and CNOT}\label{2qubitHCNOT}

\begin{figure}[h]
\begin{center}
    \begin{overpic}[width=0.8\textwidth]{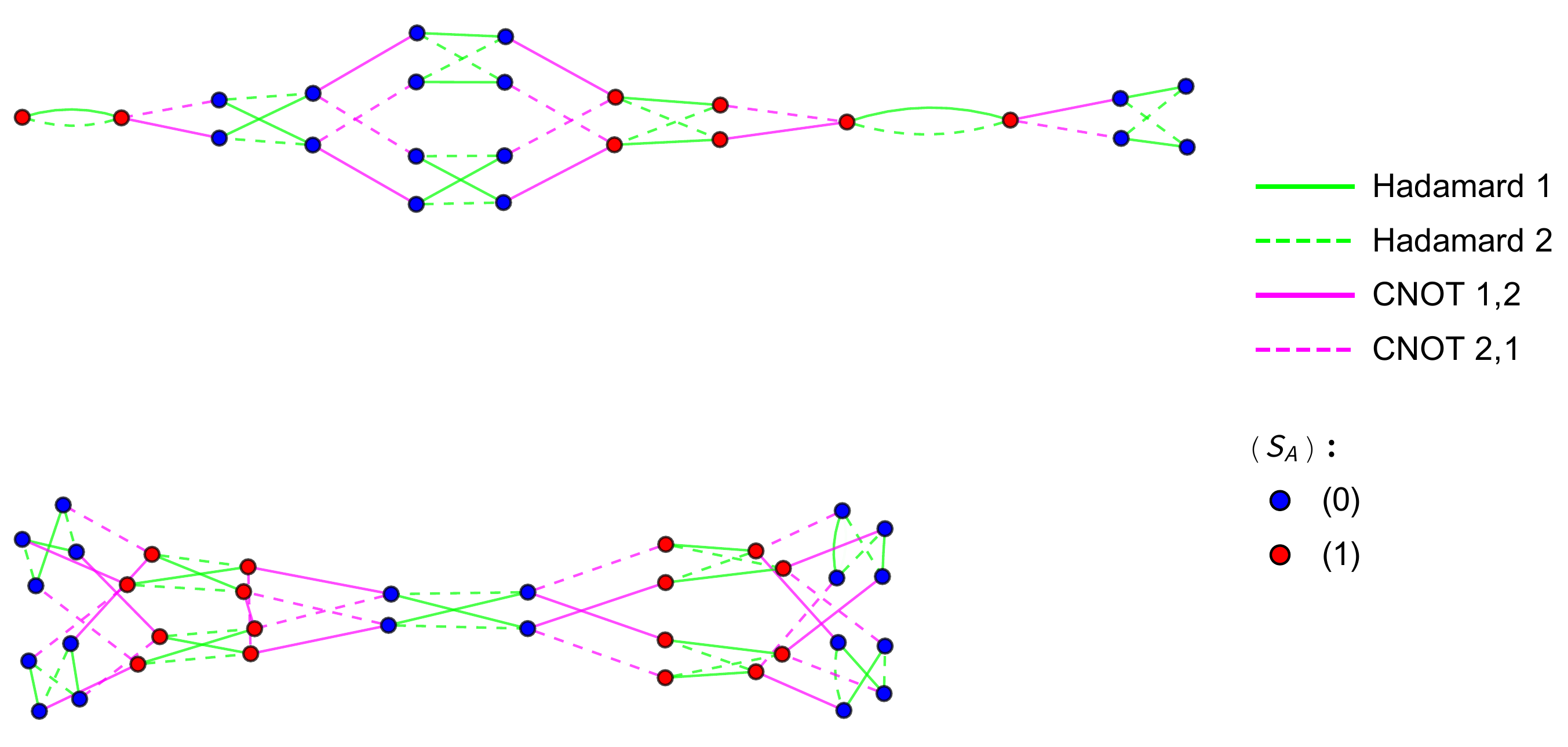}
		\put (37,33) {$g_{24}$}
		\put (27,13) {$g_{36}$}
		\put (76.6,41.4) {\footnotesize{$\leftarrow \ket{00}$}}
		\put (54.1,42) {\footnotesize{$\ket{GHZ} \downarrow$}}
		\end{overpic}
\caption{The 2-qubit subgraph restricted to $H_1,H_2,CNOT_{1,2},$ and $CNOT_{2,1}$ has two subgraphs which are connected only via phase gates. Trivial loops have been removed in this representation of the 2-qubit $H_1,H_2,CNOT_{1,2},CNOT_{2,1}$ restricted graph.}
\label{TwoQubitH1H2CNOT12CNOT21}
\end{center}
\end{figure}

For the remainder of this work, we will focus on restricted graphs generated by $H_1,\, H_2,$  $CNOT_{1,2},$ and $CNOT_{2,1}$ (see Figure \ref{TwoQubitH1H2CNOT12CNOT21} for the two-qubit version). In order to understand changes in entanglement, we need to consider a restricted graph which includes CNOT gates since they are the only gates which alter entropy.  We specifically choose to include the Hadamard gates (and exclude the phase gates) because in the Hadamard-CNOT restricted graphs, each subgraph contains states with different entropy vectors.  Additionally, at two qubits, the Hadamard-CNOT graph will have only two subgraphs; we will see echoes of these structures repeated at higher qubit number.  When we go to higher qubit number, we will continue to use only the gate set  $H_1,\,H_2,\,CNOT_{1,2},CNOT_{2,1}$ because all entropic arrangements found in stabilizer states can be built from successive bipartite entanglements.%

Beginning with this graph, and continuing below, we omit any gate whose action is the identity; thus no further trivial loops will appear. Because we have four possible gates that can act on each state, a vertex with valency $4-k$ has $k$ trivial loops. We also label the subgraph structures by the number of vertices they contain, so e.g. we use $g_{24}$ for the $24$-vertex substructure in Figure \ref{TwoQubitH1H2CNOT12CNOT21}. As noted in Footnote \ref{fn:cnot_dependence} above, we have the relation
\begin{equation}
    CNOT_{2,1}=H_1 H_2 CNOT_{1,2} H_2 H_1,
\end{equation}
which can be checked explicitly for $2$-qubit states using the figure; hence the presence of the $CNOT_{2,1}$ edges is completely fixed by the structure of the other three gates.

The subgraph $g_{24}$ contains all 2-qubit stabilizer states connected to the basis states $(\ket{00},\ket{01},\ket{10},\ket{11})$ via only Hadamard and CNOT operations. As we can see from the graph, acting a Hadamard and then a CNOT on $\ket{00}$ produces the GHZ state, which is entangled:
\begin{equation}
    CNOT_{1,2}H_1\ket{00}=\ket{GHZ}.
\end{equation}

Because the phase gate is the only Clifford gate with imaginary matrix elements, all states in $g_{24}$ can be written as superpositions of the basis states with purely real coefficients. States located in $g_{36}$, on the other hand, have relative phases $\pm i$ between different basis components. Accordingly, a phase gate is required to move between the two subgraphs.  In fact, every phase gate either acts trivially, or connects states in $g_{24}$ with states in $g_{36}$, since the CNOT and Hadamard gates are both Hermitian, but phase is not. Thus, the only way a product of CNOTs and Hadamards can have the same action as a non-Hermitian operator $O$ on a state is when the state has support only on the eigenspaces of the operator $O$ with real eigenvalues. For $O=\mathrm{phase}$, the only eigenspace with real eigenvalue has eigenvalue one. Hence the phase gate either acts as the identity or its action is non-Hermitian (and thus moves us between the $g_{24}$ and $g_{36}$ subgraphs).  We will see echoes of this structure when comparing subgraphs in the $H_1,\, H_2,\, CNOT_{1,2},\, CNOT_{2,1}$ restricted graphs at higher qubit number.

\begin{figure}[h]
\begin{center}
    \begin{overpic}[width=0.95\textwidth]{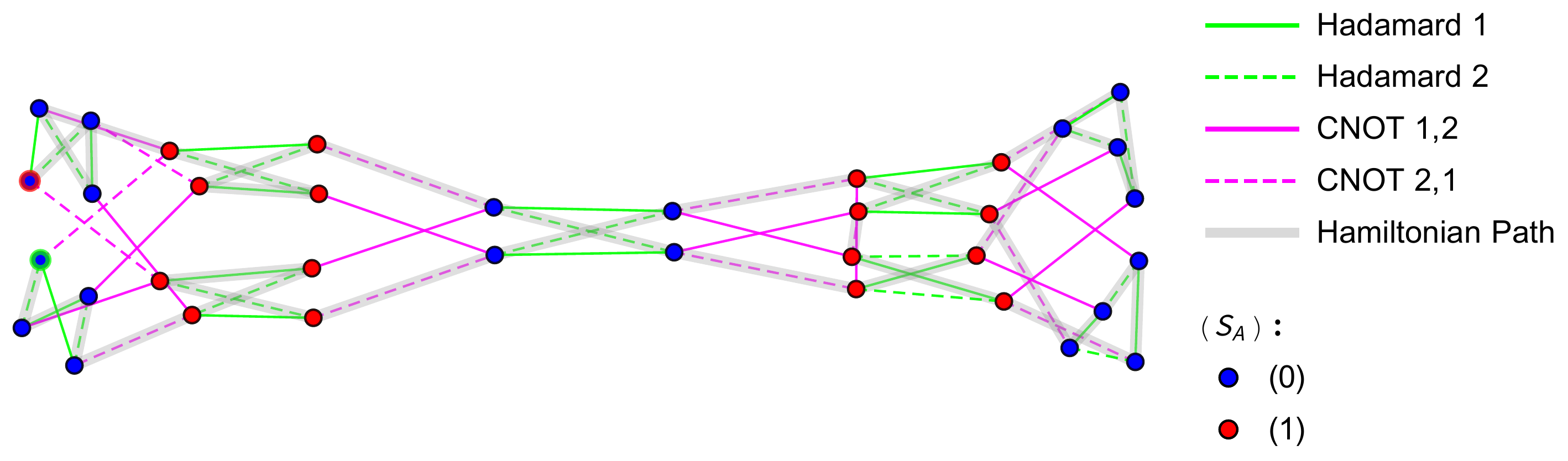}
		\put (-6.5,18) {\footnotesize{$\ket{-,-i}$}}
		\put (-3,12) {\footnotesize{$\ket{i,1}$}}
		\end{overpic}
\caption{Hamiltonian path $\mathcal{C}$ beginning on state $\ket{i,1}$ (encircled in green) and ending on state $\ket{-,-i}$ (encircled in red).}
\label{TwoQubitHamiltonianPath}
\end{center}
\end{figure}

In our analysis at higher qubits, we will also rely on Hamiltonian paths and Hamiltonian cycles.  Hamiltonian paths visit every vertex in a graph only once (and thus do not self-intersect). Hamiltonian cycles are closed loops with the same property. The subgraph $g_{24}$ has no Hamiltonian paths, and therefore no Hamiltonian cycles either.
Subgraph $g_{36}$ does have Hamiltonian paths (although again no Hamiltonian cycles). One example is shown in Figure \ref{TwoQubitHamiltonianPath}.  The specific circuit depicted is
\begin{equation}\label{g36HamiltonianCircuit}
\begin{split}
	\mathcal{C} \equiv (H_2,&H_1,H_2,CNOT_{1,2},H_2,H_1,H_2,CNOT_{1,2},H_2,CNOT_{1,2},H_2,CNOT_{1,2},\\
&H_1,H_2,H_1,CNOT_{1,2},H_1,CNOT_{1,2},H_2,CNOT_{1,2},H_2,H_1,H_2,CNOT_{1,2},\\
&H_1,CNOT_{1,2},H_2,CNOT_{1,2},H_2,H_1,H_2,CNOT_{2,1},H_2,H_1,H_2).
\end{split}
\end{equation}
When applying this circuit to the graph, the leftmost gate acts first and the rightmost gate last. 

Note that the path $\mathcal{C}$ starting on $\ket{i1}$ is not unique; other Hamiltonian paths exist on $g_{36}$.  First, $\mathcal{C}$ also traverses a Hamiltonian path starting on $\ket{-i,1}$ instead. Another example is $\mathcal{C^\top}$, which swaps qubits 1 and 2 in \eqref{g36HamiltonianCircuit}.  Two further possibilities are the inverse paths $\mathcal{C}^{-1}$ and $(\mathcal{C}^\top)^{-1}$, which simply apply the respective circuits in reverse order. In fact, there exists a Hamiltonian path on $g_{36}$ starting from $30$ (out of $36$) of its vertices.

For definiteness, we will use the Hamiltonian path $\mathcal{C}$, starting on $\ket{i1}$, in the lifting procedure introduced in Section \ref{liftTwoToThree}, where we turn to the three-qubit stabilizer graph and a detailed analysis of its reduced graphs. 

\section{The Three-Qubit Stabilizer Graph}\label{sec:three}

\begin{figure}[h]
\begin{center}
\includegraphics[scale=0.8]{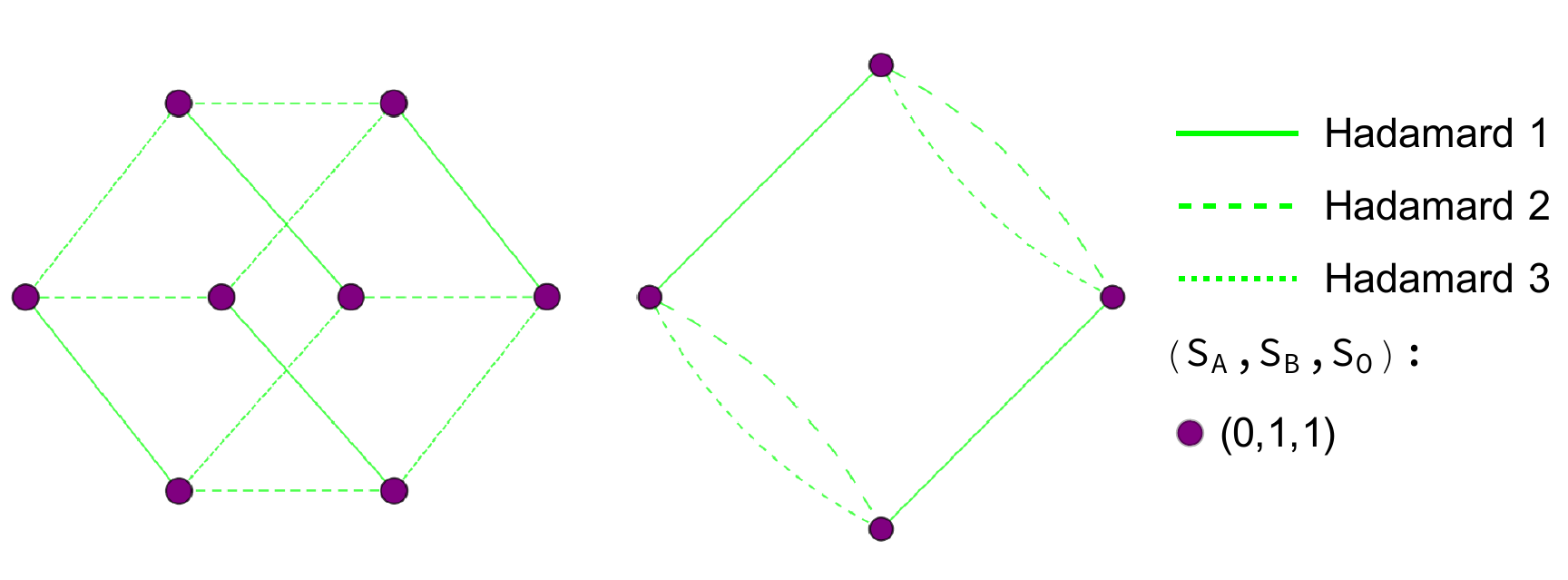}
\caption{Pictured are the two unique subgraph types that occur in the 3-qubit restricted graph of only $H_1,H_2,$ and $H_3$ operations. Degenerate pairs in the 2-qubit $H_1,H_2$ restricted graph (Figure \ref{TwoQubitH1H2}) are promoted to boxes at three qubits. Boxes from the 2-qubit $H_1,H_2$ graph are promoted to cubes by the addition of the $H_3$ gate.}
\label{ThreeQubitH1H2H3}
\end{center}
\end{figure}

At three qubits there are now $1080$ stabilizer states, and accordingly the full reachability graph, which we defer to Figure \ref{ThreeQubitCompleteGraph} in Appendix \ref{Two-Qubit Graphs}, becomes unwieldy; we thus proceed in this section immediately to the restricted graphs. For three qubits, many of the reduced graphs show features similar to two qubits.  As an example, the Hadamard-only restricted graph at three qubits contains cubes and degenerate squares instead of squares and degenerate pairs.  As before, each subgraph has only one entropy type, since Hadamard gates cannot alter the entropy vector.  Figure \ref{ThreeQubitH1H2H3} shows the structures exhibited in the $H_1,\, H_2,\, H_3$ restricted three-qubit graph.  The phase graph at three qubits similarly extends, as exhibited by the full phase graph $P_1,\, P_2,\, P_3$, shown in Appendix \ref{Two-Qubit Graphs} in Figure \ref{ThreeQubitP1P2P3Subgraphs}.%

In all of these graphs, the color of the vertex indicates the entropy vector of the associated state. As reviewed in section \ref{sec:EntropyCone} the entropy vector at three qubits is $\vec{S}=(S_A,S_B,S_{AB})$.  There are five different entropy vectors among the 3-qubit stabilizer states, as shown in Table \ref{tab:ThreeQubitEntropyVectors}.  All five entropy vectors lie within the holographic cone (that is, they satisfy the inequalities required for a holographic state, as discussed in Section \ref{sec:EntropyCone}).

Just as in the 2-qubit case, the entropy vector can only be changed by the action of a CNOT gate.  Accordingly the phase-- and Hadamard--restricted graphs do not allow us to study changes in entropy. Instead, as discussed in Section \ref{2qubitHCNOT}, we are most interested in the restricted graph which considers only $H_1,\, H_2,\, CNOT_{1,2},$ and $CNOT_{2,1}$, since it will allow us to understand the changes in entropy induced by the CNOT gates on a pair of qubits.

\subsection{Three-Qubit CNOT+Hadamard on 1 and 2 only}

We extend our analysis to three qubits, restricting the full stabilizer group to the subgroup generated by $H_1,\,H_2,\,CNOT_{1,2}$, and $CNOT_{2,1}$.  We concentrate on this gate set in order to focus on the entropic structure of qubits $1$ and $2$. The choice of qubits $1$ and $2$ is arbitrary; any pair would do. Because all stabilizer gates act on at most two qubits, and all entropic structure can be built from these bipartite interactions, our analysis of qubits $1$ and $2$ is sufficient to understand reachability for all $n$ qubits. 

\begin{figure}[h]
\begin{center}
        \begin{overpic}[width=0.94\textwidth]{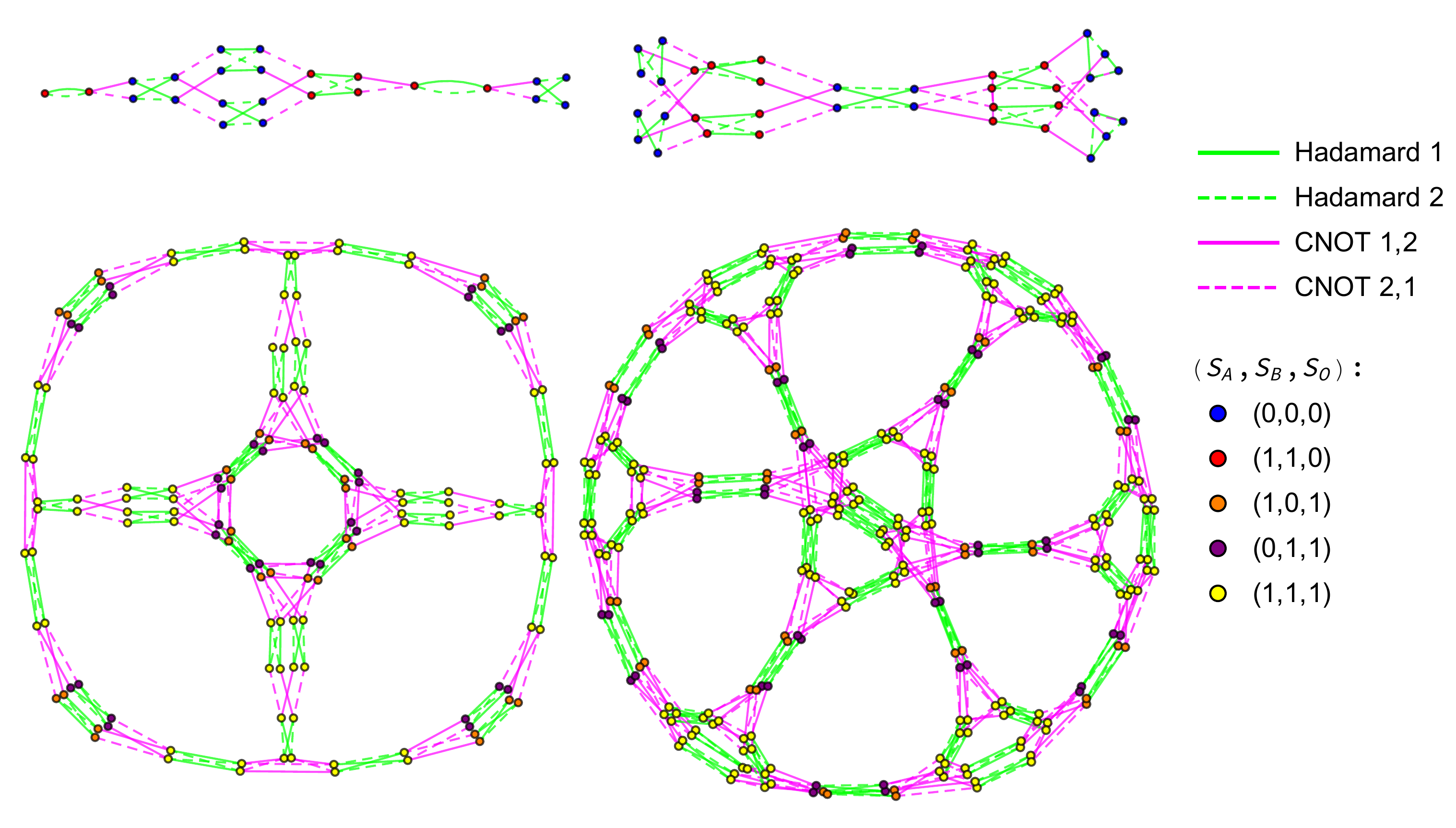}
		\put (20,54) {$g_{24}$}
		\put (58,53) {$g_{36}$}
		\put (17.1,42) {$g_{144}$}
		\put (56.6,43) {$g_{288}$}
		\end{overpic}
\caption{This figure depicts the four unique subgraphs which arise in the full 3-qubit $H_1,H_2,CNOT_{1,2},CNOT_{2,1}$ restricted graph, pictured in Figure \ref{ThreeQubitH1H2CNOT12CNOT21} of Appendix \ref{Two-Qubit Graphs}. Each subgraph is labeled by its corresponding vertex count. The full graph consists of $6$ copies of $g_{24}$ and $g_{36}$, $3$ copies of $g_{144}$, and a single copy of $g_{288}$.}
\label{ThreeQubitH1H2CNOT12CNOT21Subgraphs}
\end{center}
\end{figure}
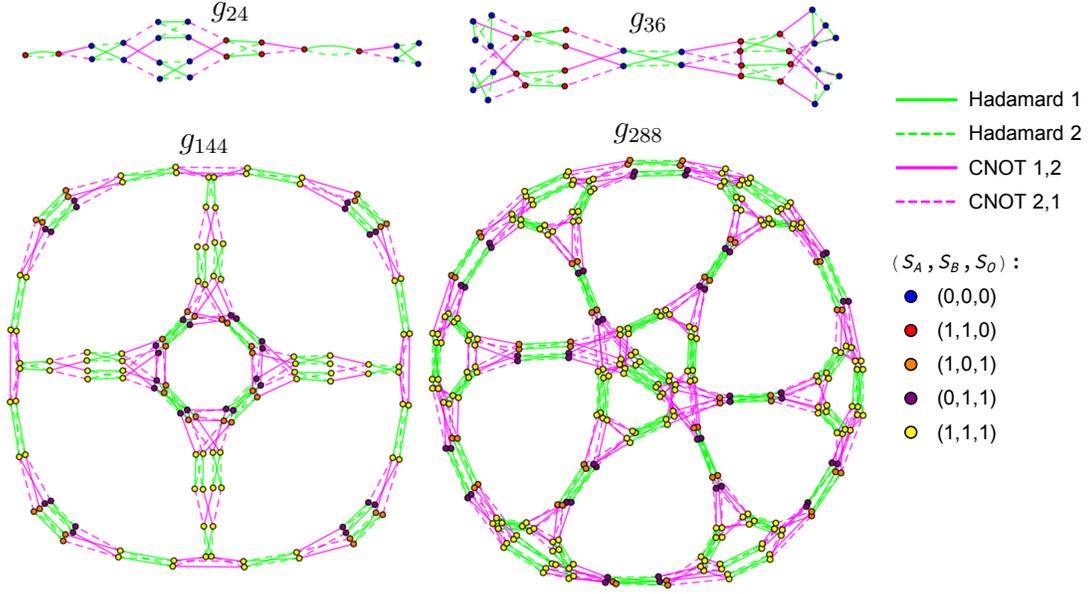

The full restricted graph for the gate set $H_1,\,H_2,\,CNOT_{1,2}, \, CNOT_{2,1}$ is shown in Figure \ref{ThreeQubitH1H2CNOT12CNOT21} of Appendix \ref{Two-Qubit Graphs}. There are only four types of subgraph, shown in Figure \ref{ThreeQubitH1H2CNOT12CNOT21Subgraphs}, which arise in the full restricted graph. The full graph consists of $6$ copies of $g_{24}$ and $g_{36}$, $3$ copies of $g_{144}$, and a single copy of $g_{288}$, where as in the previous subsection the subscript denotes the number of vertices in the subgraph. 

As mentioned above, there are only five different entropy vectors at three qubits. In Table \ref{tab:ThreeQubitEntropyVectors}, we record the number of stabilizer states with each of these entropy vectors, and also record the subgraph types those states appear in.
\begin{table}[h]
    \begin{center}
    \begin{tabular}{|c||c|c|c|} 
\hline
Holographic & $\left(S_A, S_B, S_{AB}\right)$ & Number of States & Subgraph\\
\hline
\hline
 Yes & $\textcolor{blue}{\bullet} (0,0,0)$ & $216$ & $g_{24}, g_{36}$\\
 \hline
 Yes & $\textcolor{red}{\bullet}(1,1,0)$ & $144$ & $g_{24}, g_{36}$\\
 \hline
 \hline
 Yes & $\textcolor{violet}{\bullet}(0,1,1)$ & $144$ & $g_{144}, g_{288}$\\
 \hline
 Yes & $\textcolor{orange}{\bullet}(1,0,1)$ & $144$ & $g_{144}, g_{288}$\\
 \hline
 Yes & $\textcolor{yellow}{\bullet}(1,1,1)$ & $432$ & $g_{144}, g_{288}$\\
 \hline
\end{tabular}
\end{center}
\caption{Table of 3-qubit entropy vectors. The last column lists which subgraph types exhibit each entropy vector.}
\label{tab:ThreeQubitEntropyVectors}
\end{table}

As discussed in section \ref{2qubitHCNOT}, phase actions on states in the $H_1,\, H_2,\, CNOT_{1,2},\, CNOT_{2,1}$ restricted graph are always either trivial, or move you to a different subgraph.  However, unlike in the 2-qubit case, in addition to the phase gates we are also not representing $H_3$, nor any CNOT gate involving the third qubit.  Thus,
not every subgraph can be reached by phase actions alone: applying a phase gate does not change the entropy vector, and two of the five entropy vectors appear only in $g_{24}, g_{36}$ while the other three appear in only $g_{144}, g_{288}$. We are also not representing $H_3$, nor any CNOT gate involving the third qubit.

In particular, the subgraphs of type $g_{144}$ and $g_{288}$ contain the entropy vectors $(0,1,1)$, $(1,0,1)$, and $(1,1,1)$, but the subgraphs of type $g_{24}$ and $g_{36}$ only contain the entropy vectors $(0,0,0)$ and $(1,1,0)$.  This separation occurs because CNOT gates involving the third qubit are required to change between the two sets of entropy vectors.
As reviewed in section \ref{sec:EntropyCone}, the $CNOT_{i,j}$ gate can only alter entropies $S_I$ where qubit $i\in I$ but qubit $j\in \bar{I}$.  Our entropy vectors are listed as $(S_A,S_B,S_{AB})$, where $A$ refers to qubit 1 and $B$ refers to qubit 2.  So the $CNOT_{1,2}$ or $CNOT_{2,1}$ gates can only affect $S_A$ and $S_B$, not $S_{AB}$.  Thus, entropy vectors with different $S_{AB}$ can only show up on separate subgraphs in our restricted graph.

As we will see in the next section, we can understand both the number of copies of each subgraph, as well as the shape of the subgraphs themselves, by seeing how the 2-qubit states and subgraphs can be embedded into the 3-qubit structures.

\subsection{Lifting States from Two Qubits to Three Qubits}\label{liftTwoToThree}

We apply the lifting procedure introduced in Section \ref{TwoQubitPandHSection} to lift 2-qubit states to three qubits. States in the 2-qubit $g_{24}$ subgraph all lift to states in the 3-qubit $g_{24}$ subgraphs by tensoring on a third qubit. For example, starting with the state $\ket{00}$ on qubits 1 and 2,  we can tensor on $\ket{0}$ on qubit 3 to find
    \begin{equation}\label{g24Lift}
        \ket{000} = \ket{00} \otimes \ket{0}.
    \end{equation}
There are $6$ possible states of the third qubit:\ $\{\ket{0},\, \ket{1}, \, \ket{\pm},\,\ket{\pm i}\}$. Tensoring on all $6$ of these states, as in \eqref{g24Lift}, to the $24$ states in the 2-qubit $g_{24}$ subgraph generates $144$ of the 3-qubit stabilizer states. These $144$ states make up the $6$ copies of $g_{24}$ at three qubits (shown in Figure \ref{ThreeQubitH1H2CNOT12CNOT21} of Appendix \ref{Two-Qubit Graphs}).

The next simplest set of lifts begins with states in the 2-qubit $g_{36}$ subgraph, and then tensors on a third qubit. We start with the 2-qubit state $\ket{i1}$, since the circuit  $\mathcal{C}$ \eqref{g36HamiltonianCircuit} defines a Hamiltonian path on $g_{36}$ starting at this state (Figure \ref{TwoQubitHamiltonianPath}). The process
    \begin{equation}\label{g36Lift}
        \ket{i1+} =  \ket{i1} \otimes \ket{+},
    \end{equation}
lifts $\ket{i1}$ to one of the six 3-qubit $g_{36}$ subgraphs. Since there are again 6 1-qubit stabilizer states available for the third qubit, we generate a further 216 states via this lift.  These 216 states are all located on one of the $6$ copies of $g_{36}$ at three qubits, completely covering those subgraphs. 

When we tensor on a third qubit, we extend the entropy vector of the original 2-qubit state.  For a 2-qubit state with entropy vector $(S_A)$, we have $S_A=S_B$ since it is a pure state. When we tensor on the third qubit, these two values stay the same. Additionally, the third qubit is just tensored on, so it is not entangled; thus $S_O=S_{AB}=0$.  Accordingly, states with the entropy vector $(1)$ lift to states with $(1,1,0)$ and those with entropy vector $(0)$ lift to $(0,0,0)$ under the tensoring lift of Equations \eqref{g24Lift} and \eqref{g36Lift}.

We have accounted so far for all of the states in the $g_{24}$ and $g_{36}$ subgraphs at three qubits.  As shown in Table \ref{tab:ThreeQubitEntropyVectors}, these are the only states with entropy vector $(0,0,0)$ or $(1,1,0)$.   To modify entanglement and reach new 3-qubit entropy arrangements, we must act with CNOT gates involving the third qubit, i.e.\ $CNOT_{1,3},\,CNOT_{3,1},\,CNOT_{2,3},$ or $CNOT_{3,2}$. For example, acting with $CNOT_{3,1}$ on the lifted state $\ket{i1+}$ in Equation \eqref{g36Lift} gives
    \begin{equation}\label{CNOT31OnLiftedState}
        CNOT_{3,1}\ket{i1+} = \ket{010} + i\ket{011}+ i\ket{110}  + \ket{111}.
    \end{equation}
This procedure lifts $\ket{i1}$ from the 2-qubit $g_{36}$ subgraph to a state on one $g_{144}$ subgraph at three qubits. Acting with $CNOT_{3,1}$ on $\ket{i1+}$ entangles qubits $1$ and $3$, resulting in a state with entropy vector $\vec{S} = (1,0,1)$, indicated with an orange vertex in Figure \ref{ThreeQubitH1H2CNOT12CNOT21Subgraphs}. Similarly replacing the third qubit in $\ket{i1+}$ with $\ket{-}, \, \ket{i},$ or $\ket{-i}$ instead also results in states on the same copy of $g_{144}$.  $\ket{i10}$ and $\ket{i11}$, however, return to the same copies of $g_{36}$ they started on under the action of $CNOT_{3,1}$.

Acting a different gate on a lifted version of $\ket{i1}$ can result in a lifted state on a different copy of $g_{144}$. For example, the gate $CNOT_{1,3}$ on state $\ket{i10} \equiv \ket{i1} \otimes \ket{0}$ gives
    \begin{equation}\label{CNOT13LiftedState}
        CNOT_{1,3}\ket{i10} = \ket{010} + i\ket{111},
    \end{equation}
which resides on a different copy of $g_{144}$ than the state lifted in Equation \eqref{CNOT31OnLiftedState}. As before, the third ket can be replaced with $\ket{1},\, \ket{i}$ or $\ket{-i}$ resulting in three other states on the same copy of $g_{144}$, but using $\ket{\pm}$ results in $CNOT_{1,3}$ mapping back to the same copies of $g_{36}$.

The third copy of $g_{144}$ is phase-separated from the previous two. Lifting 2-qubit starting states to this subgraph requires a phase gate. One possible such lift to this subgraph is
    \begin{equation}\label{CNOT13LiftedState}
        P_3(CNOT_{1,3}\ket{i10}) = \ket{010} - \ket{111}.
    \end{equation}
Again, the third qubit can be replaced with $\ket{1},\, \ket{i}$ or $\ket{-i}$ giving three further states on the last copy of $g_{144}$.

There are two ways that we can lift the 2-qubit state $\ket{i1}$ to the 3-qubit $g_{288}$ subgraph. The first begins by tensoring on a third qubit, then applying the appropriate CNOT gate to move the lifted state from $g_{36}$ directly to $g_{288}$. For example,
    \begin{equation}\label{CNOT32LiftTo288}
        CNOT_{3,2}\ket{i1+}) = \ket{010} + i\ket{011} + \ket{100} + i\ket{101}
    \end{equation}
resides on $g_{288}$. It has entropy vector $\vec{S} = (0,1,1)$, and is represented by a purple vertex in $g_{288}$ in Figure \ref{ThreeQubitH1H2CNOT12CNOT21Subgraphs}.  The same procedure works when we replace the third qubit by $\ket{-}, \,\ket{i},$ or $\ket{-i}$.

The second method first lifts state $\ket{i1}$ to $g_{144}$ as in Equation \eqref{CNOT31OnLiftedState}, entangling qubits $1$ and $3$. Applying a second CNOT gate then entangles qubit $2$ with the other qubits.  Explicitly, we have e.g.
    \begin{equation}\label{CNOT32CNOT31LiftTo288}
        CNOT_{3,2}(CNOT_{3,1}\ket{i1+}) = \ket{010} + i\ket{011} + i\ket{100} + \ket{101}.
    \end{equation}
This process results in a final state on $g_{288}$. Its entropy vector  is $(1,1,1)$, represented by a yellow vertex in $g_{288}$.  Three further states arise by replacing the third qubit $\ket{+}$ with $\ket{-}, \,\ket{i},$ or $\ket{-i}$.

In the next subsection, we will use the lifted states described here, combined with the 2-qubit Hamiltonian path $\mathcal{C}$, to reach the remaining states as well as to understand the new $g_{144}$ and $g_{288}$ subgraph structures that arise at three qubits.

\subsection{Lifting Paths from Two Qubits to Three Qubits}

At two qubits, the Hamiltonian path $\mathcal{C}$ \eqref{g36HamiltonianCircuit} starting from the state $\ket{i1}$ covered the $g_{36}$ subgraph.  Similarly, $\mathcal{C}$ starting from the lifted states described in the previous section, listed in Table \ref{tab:OneFullCoveringOfLiftedStates}, cover every vertex exactly once on each of the $g_{36}$, $g_{144}$, and $g_{288}$ subgraphs of the $H_1,\, H_2,\, CNOT_{1,2},\, CNOT_{2,1}$ restricted graph at three qubits.
\begin{table}[h]
\centering
\begin{tabular}{|c|c|c|}
 \hline
 Lift & Starting State & Subgraph  \\
 \hline
  $\ket{i1}\otimes\ket{0}$ & $\ket{i10}$   & $g^1_{36}$ \\
  \hline
  $\ket{i1}\otimes\ket{1}$ & $\ket{i11}$   & $g^2_{36}$ \\
  \hline
  $\ket{i1}\otimes\ket{+}$ & $\ket{i1+}$   & $g^3_{36}$ \\
  \hline
  $\ket{i1}\otimes\ket{-}$ & $\ket{i1-}$   & $g^4_{36}$ \\
  \hline
  $\ket{i1}\otimes\ket{i}$ & $\ket{i1i}$   & $g^5_{36}$ \\
  \hline
  $\ket{i1}\otimes\ket{-i}$ & $\ket{i1-i}$   & $g^6_{36}$ \\
  \hline
  \hline
  $CNOT_{3,1}(\ket{i1}\otimes\ket{+})$ & $\ket{i10}+i\ket{-i,11}$   & \multirow{4}{*}{$g_{144}^1$} \\
  \cline{1-2}
  $CNOT_{3,1}(\ket{i1}\otimes\ket{-})$ & $\ket{i10}-i\ket{-i,11}$   & \\
  \cline{1-2}
  $CNOT_{3,1}(\ket{i1}\otimes\ket{i})$ & $\ket{i10}-\ket{-i,11}$   &  \\
  \cline{1-2}
  $CNOT_{3,1}(\ket{i1}\otimes\ket{-i})$ & $\ket{i10}+\ket{-i,11}$   & \\
  \hline
  \hline
    $CNOT_{1,3}(\ket{i1}\otimes\ket{0})$ & $\ket{010}+i\ket{111}$   & \multirow{4}{*}{$g_{144}^2$} \\
  \cline{1-2}
  $CNOT_{1,3}(\ket{i1}\otimes\ket{1})$ & $\ket{011}+i\ket{110}$   & \\
  \cline{1-2}
  $CNOT_{1,3}(\ket{i1}\otimes\ket{i})$ & $\ket{01i}-\ket{11,-i}$   &  \\
  \cline{1-2}
  $CNOT_{1,3}(\ket{i1}\otimes\ket{-i})$ & $\ket{01,-i}+\ket{11i}$   & \\
  \hline
  \hline
    $P_3CNOT_{1,3}(\ket{i1}\otimes\ket{0})$ & $\ket{010}-\ket{111}$   & \multirow{4}{*}{$g_{144}^3$} \\
  \cline{1-2}
  $P_3CNOT_{1,3}(\ket{i1}\otimes\ket{1})$ & $\ket{011}+\ket{110}$   & \\
  \cline{1-2}
  $P_3CNOT_{1,3}(\ket{i1}\otimes\ket{i})$ & $\ket{01-}-\ket{11+}$   &  \\
  \cline{1-2}
  $P_3CNOT_{1,3}(\ket{i1}\otimes\ket{-i})$ & $\ket{01+}+\ket{11-}$   & \\
  \hline
  \hline
    $CNOT_{3,2}(\ket{i1}\otimes\ket{+})$ & $\ket{i10}+\ket{i01}$   & \multirow{8}{*}{$g_{288}$} \\
  \cline{1-2}
  $CNOT_{3,2}(\ket{i1}\otimes\ket{-})$ & $\ket{i10}-\ket{i01}$   & \\
  \cline{1-2}
  $CNOT_{3,2}(\ket{i1}\otimes\ket{i})$ & $\ket{i10}+i\ket{i01}$   &  \\
  \cline{1-2}
  $CNOT_{3,2}(\ket{i1}\otimes\ket{-i})$ & $\ket{i10}-\ket{i01}$   & \\
  \cline{1-2}
  $CNOT_{3,2}CNOT_{3,1}(\ket{i1}\otimes\ket{+})$ & $\ket{i10}+i\ket{-i,01}$   & \\
  \cline{1-2}
  $CNOT_{3,2}CNOT_{3,1}(\ket{i1}\otimes\ket{-})$ & $\ket{i10}-i\ket{-i,01}$   &  \\
  \cline{1-2}
  $CNOT_{3,2}CNOT_{3,1}(\ket{i1}\otimes\ket{i})$ & $\ket{i10}-\ket{-i,01}$   & \\
  \cline{1-2}
  $CNOT_{3,2}CNOT_{3,1}(\ket{i1}\otimes\ket{-i})$ & $\ket{i10}+\ket{-i,01}$   & \\
  \hline
\end{tabular}
\caption{Using the 26 starting states in this table, and the path $\mathcal{C}$ \eqref{g36HamiltonianCircuit}, we cover every vertex on the $g_{36}$, $g_{144}$, and $g_{288}$ subgraphs of the $H_1,\, H_2,\, CNOT_{1,2},\, CNOT_{2,1}$ restricted graph at 3 qubits.  The $g_{24}$ subgraphs cannot be covered by a single path, but all states in them can be generated from states in the 2-qubit $g_{24}$ subgraph via the lift described in \eqref{g36Lift}.}
\label{tab:OneFullCoveringOfLiftedStates}
\end{table}

For the $g_{36}$ subgraphs, the lift only involved a simple tensor product with the third qubit. Additionally, the lift actually worked on every state in the $g_{36}$ subgraph, so the structure of $g_{36}$ in each of the $6$ copies is completely preserved. All lifted states from the 2-qubit $g_{36}$ subgraph lift to the same relative position in the 3-qubit $g_{36}$ subgraphs. Therefore, applying $\mathcal{C}$ on each lift of $\ket{i1}$ on a 3-qubit copy $g_{36}$ still builds a Hamiltonian path on each subgraph. Accordingly, lifting $\mathcal{C}$ from two qubits to three qubits by lifting the starting state gives a complete vertex covering of each 3-qubit copy of $g_{36}$, just as in Figure \ref{TwoQubitHamiltonianPath}. This covering of higher qubit $g_{36}$ subgraphs by the 2-qubit $g_{36}$ structure persists to arbitrary qubit number.

\begin{figure}[h]
\begin{center}
\begin{overpic}[width=0.8\textwidth]{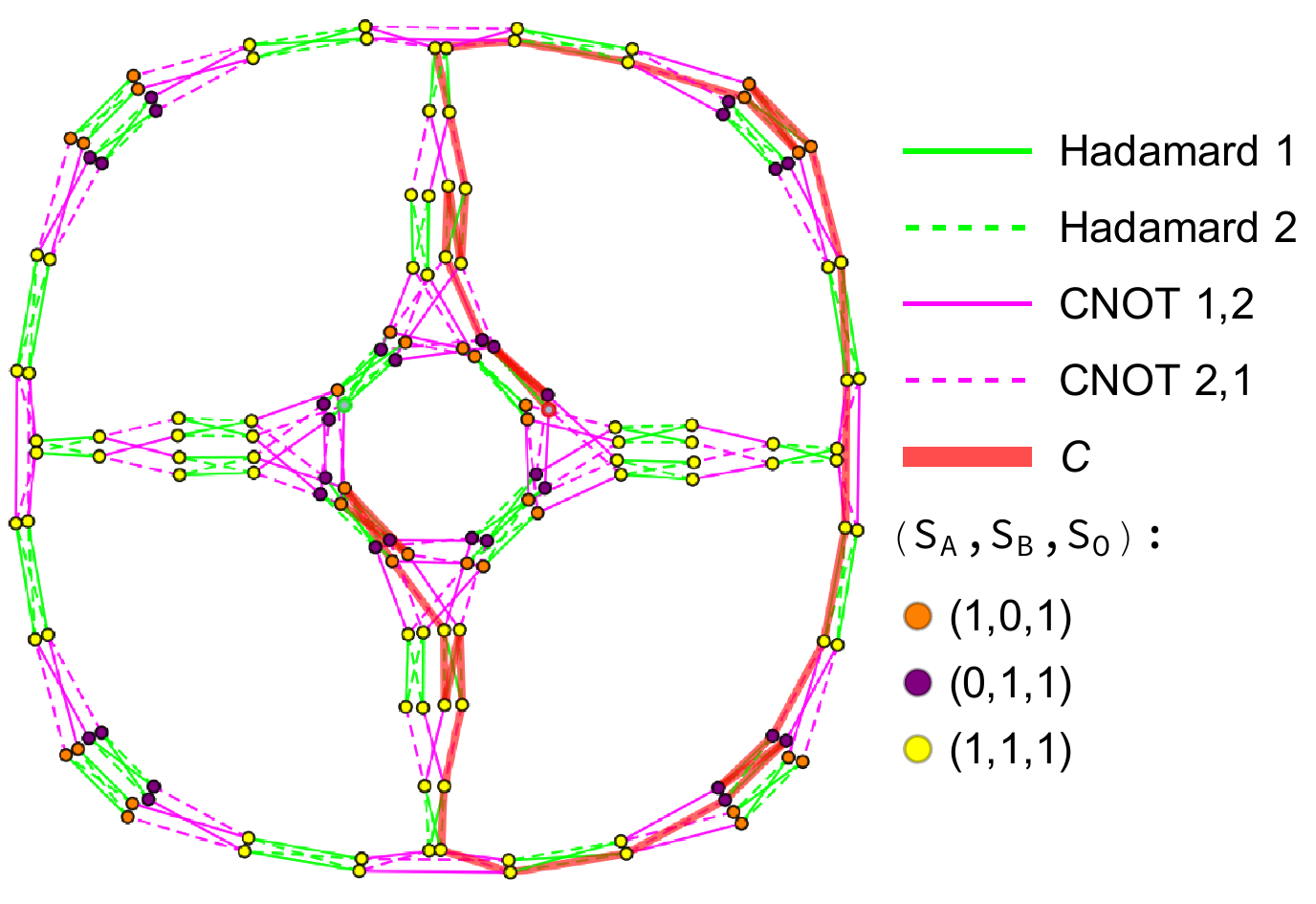}
		\put (27,37.3) {\footnotesize{$\leftarrow \ket{\psi}$}}
		\end{overpic}
\caption{Lift of Hamiltonian path $\mathcal{C}$ defined in Equation \eqref{g36HamiltonianCircuit}, from the 2-qubit $g_{36}$ subgraph to a copy of $g_{144}$ at three qubits. The path begins on $\ket{\psi} \equiv CNOT_{1,3}\ket{i10}$, the state lifted in Equation \eqref{CNOT13LiftedState}.}
\label{ThreeQubitLiftedHamiltonianPath}
\end{center}
\end{figure}

Lifting $\mathcal{C}$ to the three copies of $g_{144}$ and the single $g_{288}$ subgraph illustrates how the two-qubit subgraph $g_{36}$ is embedded into larger subgraphs at three qubits. Let us consider a specific example, beginning with the second $g_{144}$ subgraph, which we have termed $g_{144}^2$ in Table \ref{tab:OneFullCoveringOfLiftedStates}. As described in Equation \eqref{CNOT13LiftedState}, the lifted state $CNOT_{1,3}\ket{i10}$ is on this subgraph.  Applying the circuit $\mathcal{C}$ to this state produces a non-intersecting path on $g_{144}$, covering $1/4$ of its states as in Figure \ref{ThreeQubitLiftedHamiltonianPath}.

\begin{figure}[h]
\begin{center}
\begin{overpic}[width=0.8\textwidth]{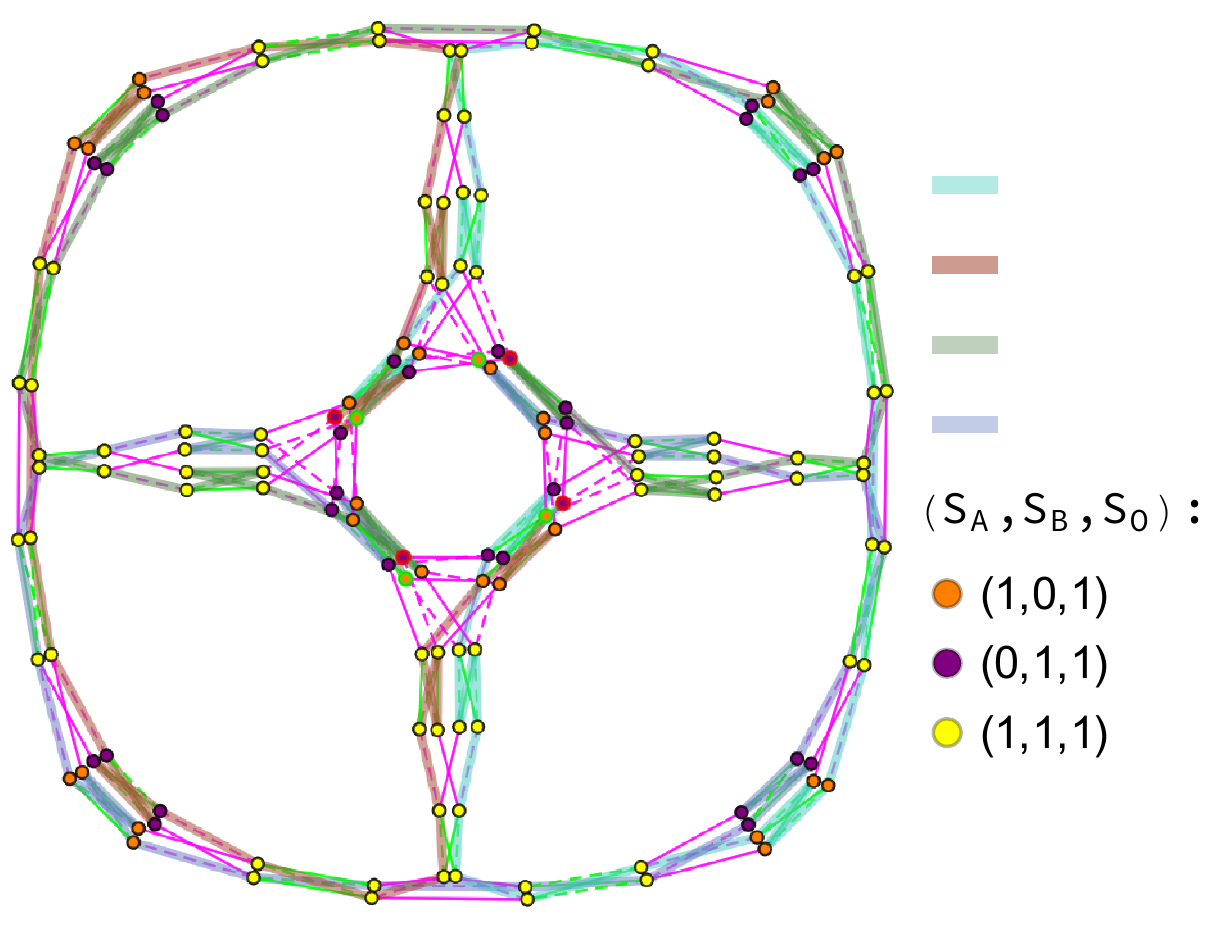}
		\put (84,60.5) {\footnotesize{$CNOT_{1,3}\ket{i,1,0}$}}
		\put (84,54) {\footnotesize{$CNOT_{1,3}\ket{i,1,1}$}}
		\put (84,47.5) {\footnotesize{$CNOT_{1,3}\ket{i,1,i}$}}
		\put (84,40.8) {\footnotesize{$CNOT_{1,3}\ket{i,1,-i}$}}
		\end{overpic}
\caption{Four copies of Hamiltonian path $\mathcal{C}$ that give a vertex cover of $g_{144}^2$. Each path is indicated by the lifted starting state where it begins.}
\label{CoveringG144}
\end{center}
\end{figure}

Each of the lifted starting states listed under $g_{144}^2$ in Table \ref{tab:OneFullCoveringOfLiftedStates} have entropy vector $\vec{S} = (1,0,1)$, and are located around the central octagonal structure of $g_{144}$. Applying $\mathcal{C}$ to each of the lifted states defines a separate path, covering $36$ vertices each, on $g_{144}$. The union of these $4$ lifts of $\mathcal{C}$ defines a vertex covering of $g_{144}$, displayed in Figure \ref{CoveringG144}. The other two $g_{144}$ subgraphs can be covered similarly, by four copies of $\mathcal{C}$ starting on each of the lifted starting states listed in the table.

A covering of the $g_{288}$ subgraph can also be constructed from lifts of $\mathcal{C}$.  First, we lift the 2-qubit $g_{36}$ state $\ket{i1}$ via Equations \eqref{CNOT32LiftTo288} and  \eqref{CNOT32CNOT31LiftTo288}, as listed in Table \ref{tab:OneFullCoveringOfLiftedStates}. We then apply $\mathcal{C}$ on these $8$ lifted states, covering every vertex of $g_{288}$ once.

We have now reached all of the 3-qubit states, and we have covered the six subgraphs $g_{36}$, three subgraphs $g_{144}$, and single subgraph $g_{288}$ with lifted versions of the 2-qubit Hamiltonian path $g_{36}$.

Just as we described at the end of section \ref{2qubitHCNOT} for the 2-qubit $g_{36}$ subgraph, these coverings are not unique.  We could have lifted the 2-qubit state $\ket{-i,1}$ instead, since $\mathcal{C}$ is also a Hamiltonian path from that starting point.  Alternatively, we could have used the path $\mathcal{C}^\top$ or $\mathcal{C}^{-1}$ instead, or any other Hamiltonian path from the 2-qubit $g_{36}$ subgraph.

Sticking to just $\mathcal{C}$, $\mathcal{C}^\top$, $\mathcal{C}^{-1}$, and $(\mathcal{C}^\top)^{-1}$, each path has two possible starting states on the 2-qubit $g_{36}$ subgraph.  Lifting these eight states to four starting states each on $g_{144}^2$, we hit all 32 purple and orange states in the central octagon in Figure \ref{CoveringG144}.  The same statement works for the other $g_{144}$ subgraphs, so every central octagon state is a lifted starting state for some 2-qubit Hamiltonian path, which together with 3 other states, fully covers the $g_{144}$ subgraph. A similar situation arises for the $g_{288}$ subgraph.

Rather than exploring these other possible coverings, we will instead move on to further structures which first arise at three qubits: Hamiltonian cycles and Eulerian paths.

\subsection{Hamiltonian cycles and Eulerian paths}\label{sec:3-qubitHamiltonianEulerian}

At one qubit, the complete reachability diagram did not have a Hamiltonian path.  At two qubits, the $H_1,\, H_2, \, CNOT_{1,2},\, CNOT_{2,1}$ restricted subgraph $g_{24}$ also did not have a Hamiltonian path, but $g_{36}$ did.
Beginning at three qubits, we are able to construct Hamiltonian cycles as well, on subgraphs $g_{144}$ and $g_{288}$.  Subgraph $g_{288}$ will additionally have an Eulerian cycle, which traverses every path exactly once (instead of visiting every vertex once).

Hamiltonian cycles cannot be built on the $g_{24}$ and $g_{36}$ subgraphs.  For $g_{24}$ this check is simple: it contains a vertex whose removal disconnects the graph, known as a cut-vertex; the $GHZ$ state in Figure \ref{TwoQubitH1H2CNOT12CNOT21} is one example. Subgraph $g_{36}$ contains no cut-vertex; however, it can be verified to have no Hamiltonian cycle by exhaustive search.%
\footnote{Any graph containing a Hamiltonian cycle must satisfy all of the following conditions: each vertex of degree $2$ must be included in the Hamiltonian cycle, once two edges incident to a vertex are included in the Hamiltonian cycle all other edges incident to that vertex must be removed from consideration, and the Hamiltonian cycle must contain no proper subcycles. These criteria are only sufficient to eliminate a graph's candidacy for having a Hamiltonian cycle. In general, the ``Hamiltonian path problem'' of proving that a given graph does admit a Hamiltonian path or cycle is NP-complete.}%

For the $g_{144}$ subgraph, the four copies of $\mathcal{C}$ in each graph do not directly connect into a Hamiltonian cycle simply by connecting their ends.  However, other Hamiltonian paths exist on the 2-qubit $g_{36}$ subgraph, and some of these paths, when their starting states are lifted to four states on $g_{144}$, can be directly connected into one large loop that builds a Hamiltonian cycle.

\begin{figure}[h]
\begin{center}
\includegraphics[scale=.9]{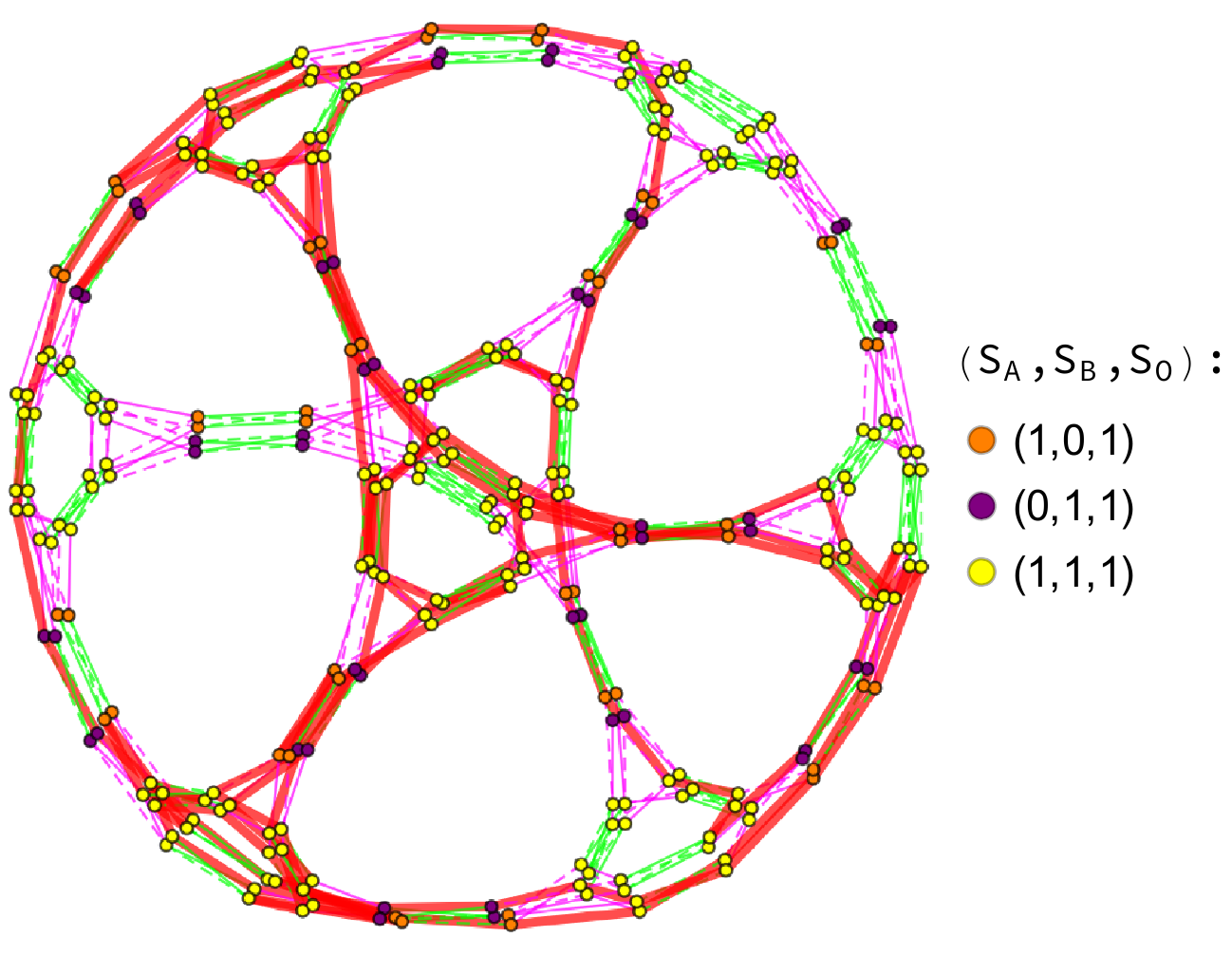}
\caption{The Hamiltonian cycle on $g_{144}$ can be mapped to a cycle on $g_{288}$ through a CNOT operation connecting the two subgraphs. Two copies of $g_{144}$ can be used to cover $g_{288}$ completely.}
\label{HamiltonianCycleMap}
\end{center}
\end{figure}

For the $g_{288}$ subgraph, we have three coverings.  First, as in the previous section, eight copies of $\cal{C}$ (or any other Hamiltonian path from $g_{36}$ ) completely cover the graph.  Next, the Hamiltonian cycle on $g_{144}$ can be mapped to a pair of cycles which each cover half of $g_{288}$, as shown in Figure \ref{HamiltonianCycleMap}.  And last, a new Hamiltonian cycle exists on $g_{288}$ itself.

We also note that Eulerian cycles, which traverse every edge, first exist on $g_{288}$.  The cycle is of length 576, and it must exist because all vertices have an even number of attached edges (namely, four).  Stated another way, all four gates in the restricted set $H_1,\, H_2, \, CNOT_{1,2},\, CNOT_{2,1}$ act nontrivially and differently from each other, on every state in the subgraph. In particular, this means that $g_{288}$ has no trivial loops.

Hamiltonian cycles can be embedded into to higher-qubit subgraphs in almost exactly the same way as Hamiltonian paths; for cycles we have to pick an arbitrary starting state to lift. Each Hamiltonian cycle embeds into every subgraph of greater or equal vertex number. We will also see that embeddings of Hamiltonian paths and cycles will continue to cover subgraphs at higher qubit number.

\section{Towards the $n$-qubit Stabilizer Graph}\label{sec:general}

We will now generalize our discussion to the cases of four and five qubits.  Two particularly important features arise first at four qubits. First, the last new subgraph shape in the $H_1,\, H_2, \, CNOT_{1,2}, \, CNOT_{2,1}$ restricted graph appears.  Second, some entropy vectors of stabilizer states will disobey monogamy of mutual information \eqref{eq:MonogamyofMutualInformation}.  These states thus lie outside the holographic entropy cone; they cannot have a classical gravitational representation.

\subsection{Four Qubits}

In order to explore four qubits, we have explicitly generated all of the 36720 4-qubit stabilizer states. The explicit form of these states is available in the GitHub repository \cite{github} along with associated reachability diagrams, entropy vector data, and a Mathematica package for simulating stabilizer circuits, generating sets of states, and constructing reachability diagrams.

\begin{figure}[h]
\begin{center}
\includegraphics[scale=0.75]{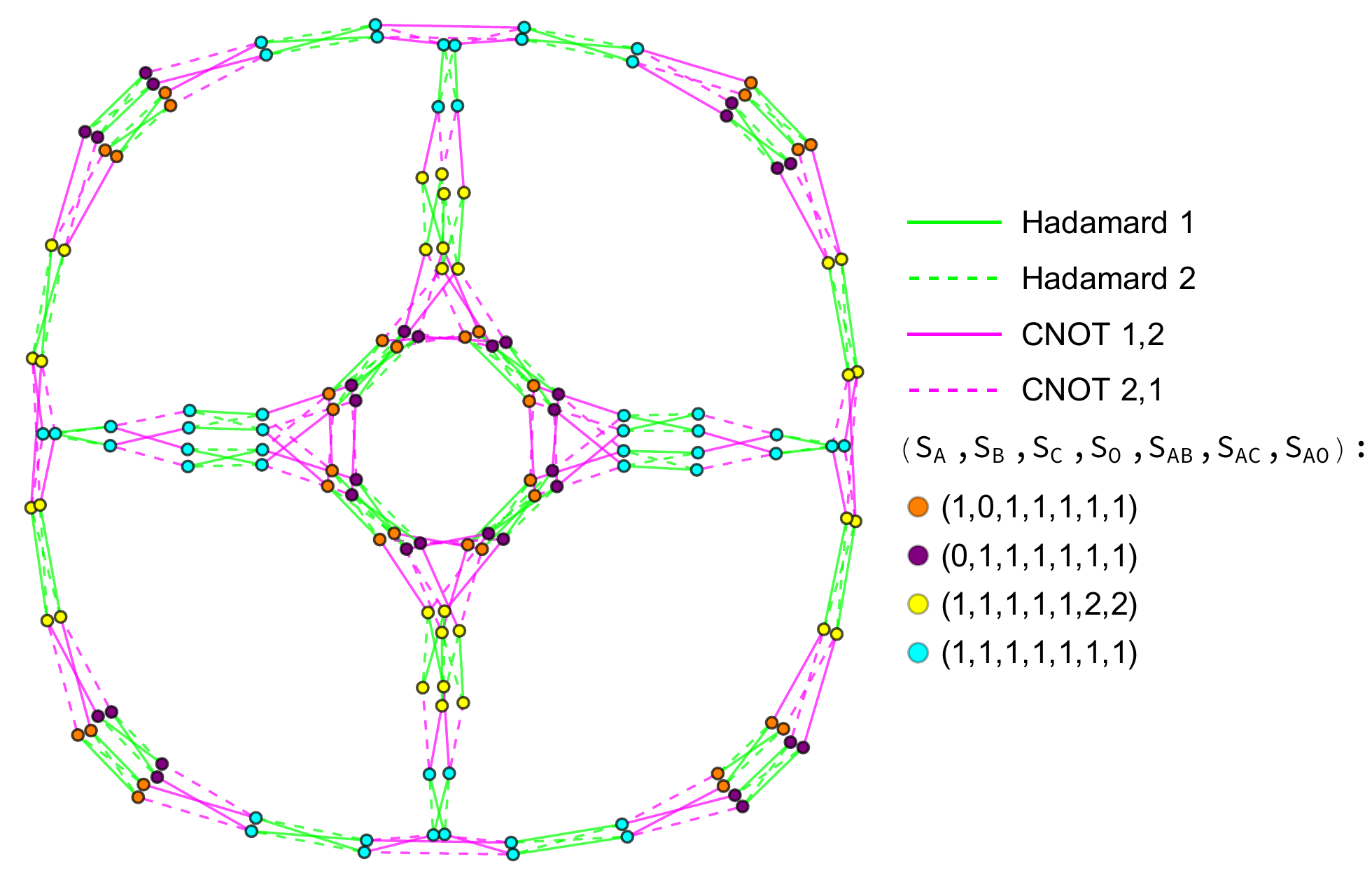}
\caption{At four qubits, the $g_{144}$ subgraph can have $4$ different entropy vectors among the vertices. The cyan vertices denote states with entropy vector $\vec{S} = (1,1,1,1,1,1,1)$, an entropic structure that violates monogamy of mutual information \eqref{eq:MonogamyofMutualInformation}. These 4-qubit stabilizer states, located on 4-qubit subgraphs $g_{144}$ and $g_{288}$, are the first instances of non-holographic states.}
\label{FourQubitG144}
\end{center}
\end{figure}

\begin{figure}[h]
\begin{center}
\includegraphics[scale=.9]{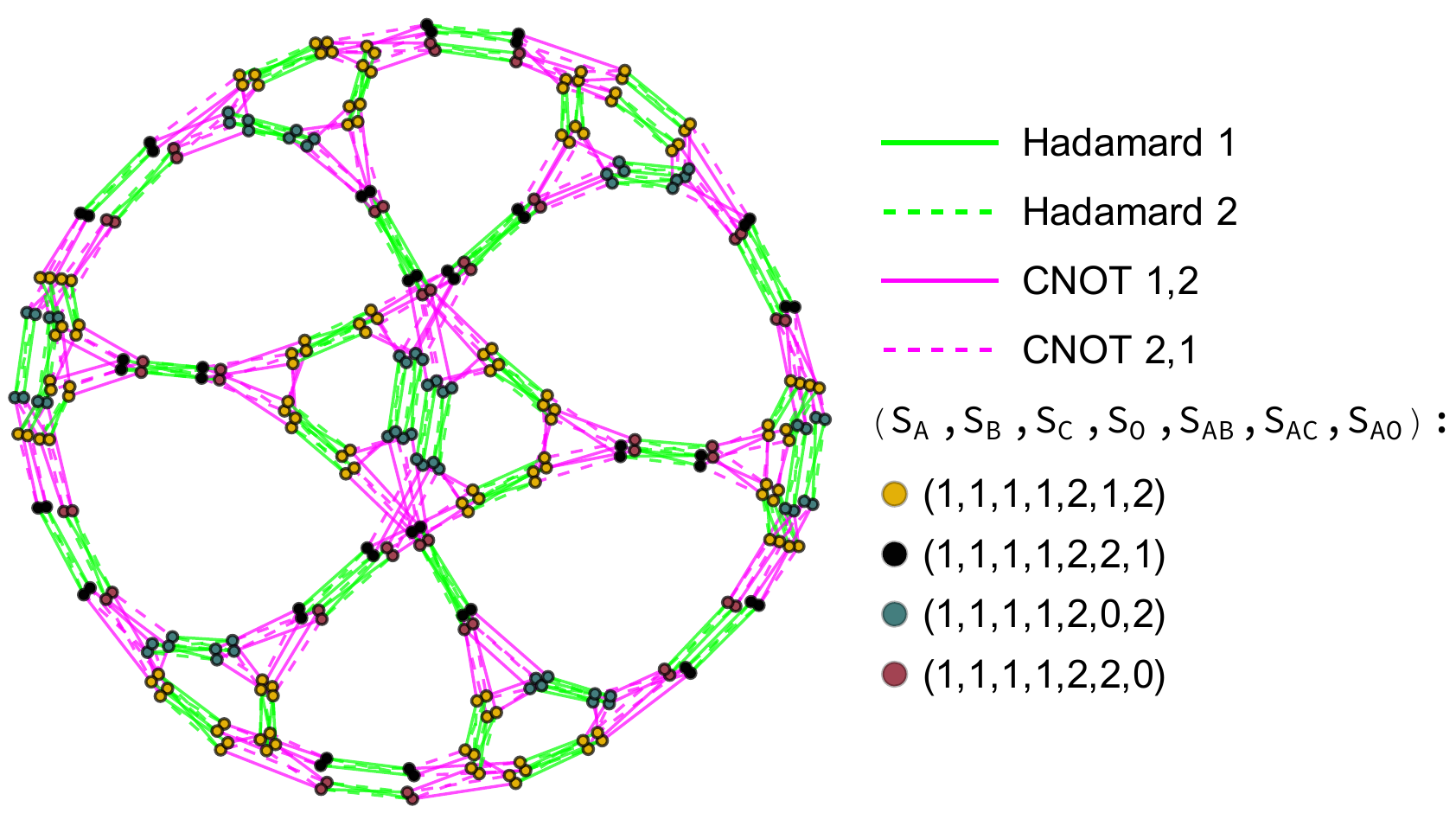}
\caption{At four qubits, we witness a variation to the $g_{288}$ subgraph first seen at three qubits. The structure is isomorphic to its 3-qubit counterpart; however, the number of entropy vectors present has increased to four, similar to what occurred in Figure \ref{FourQubitG144}.}
\label{FourQubitG288FourEntropies}
\end{center}
\end{figure}

Examining the $H_1,\, H_2, \, CNOT_{1,2}, \, CNOT_{2,1}$ restricted graphs, we find copies of the $g_{24}$ and $g_{36}$ subgraphs we have seen before. We also see again the $g_{144}$ subgraph with 3 different entropy vectors hereafter referred to as $g_{144}(3)$, and the $g_{288}$ with 3 entropy vectors, now termed $g_{288}(3)$.  We specify the number of entropy vectors because we also find $g_{144}(4)$ and $g_{288}(4)$, as in Figures \ref{FourQubitG144} and \ref{FourQubitG288FourEntropies}.

\begin{figure}[h]
\begin{center}
\includegraphics[scale=.87]{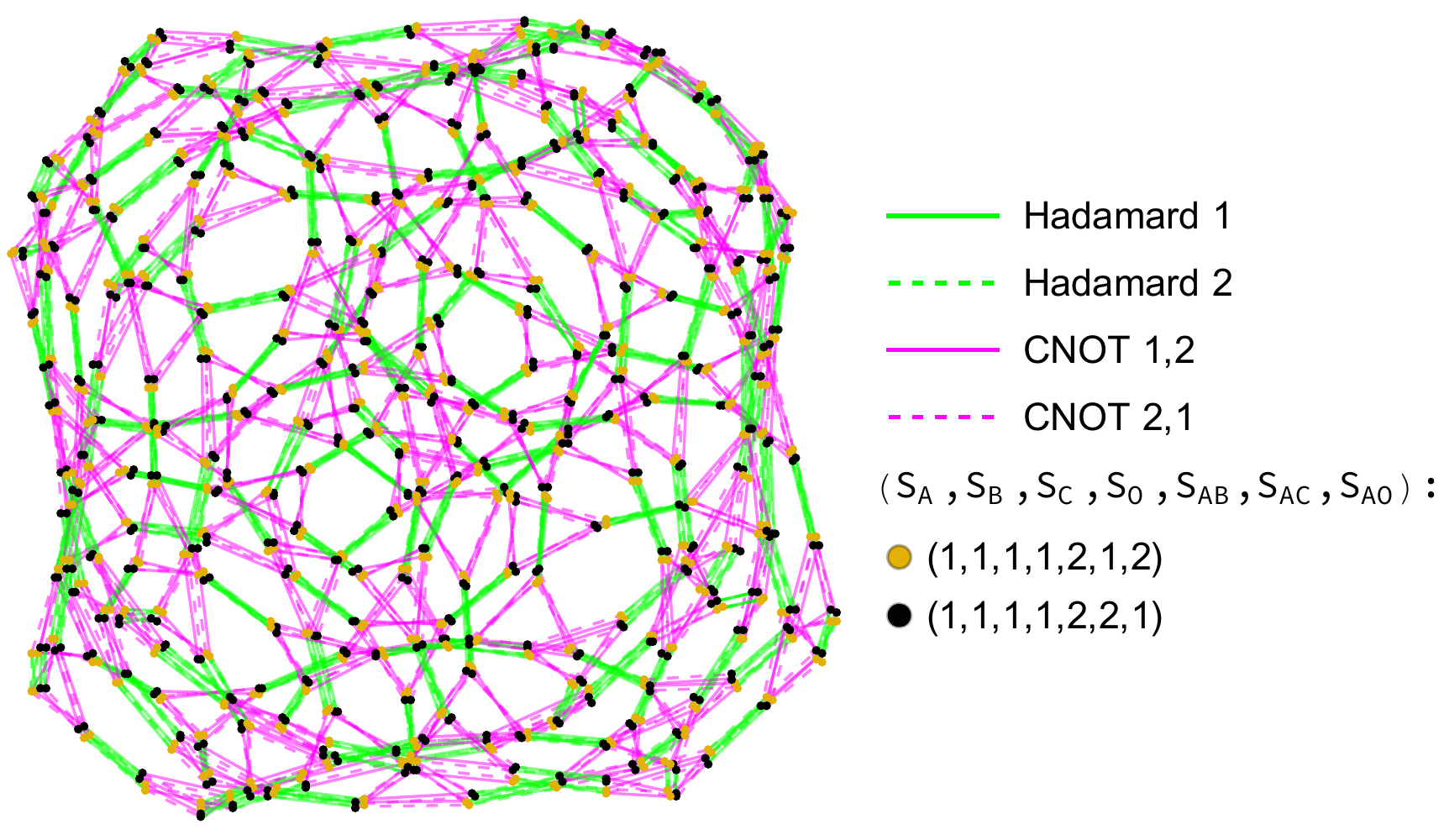}
\caption{One of the two $g_{1152}$ occurring at four qubits. This subgraph is isomorphic to the $g_{1152}$ subgraph in Figure \ref{FourQubitG1152FourEntropies}, but with only $2$ different entropy vectors among all the states in the subgraph.}
\label{FourQubitG1152TwoEntropies}
\end{center}
\end{figure}

\begin{figure}[h]
\begin{center}
\includegraphics[scale=0.9]{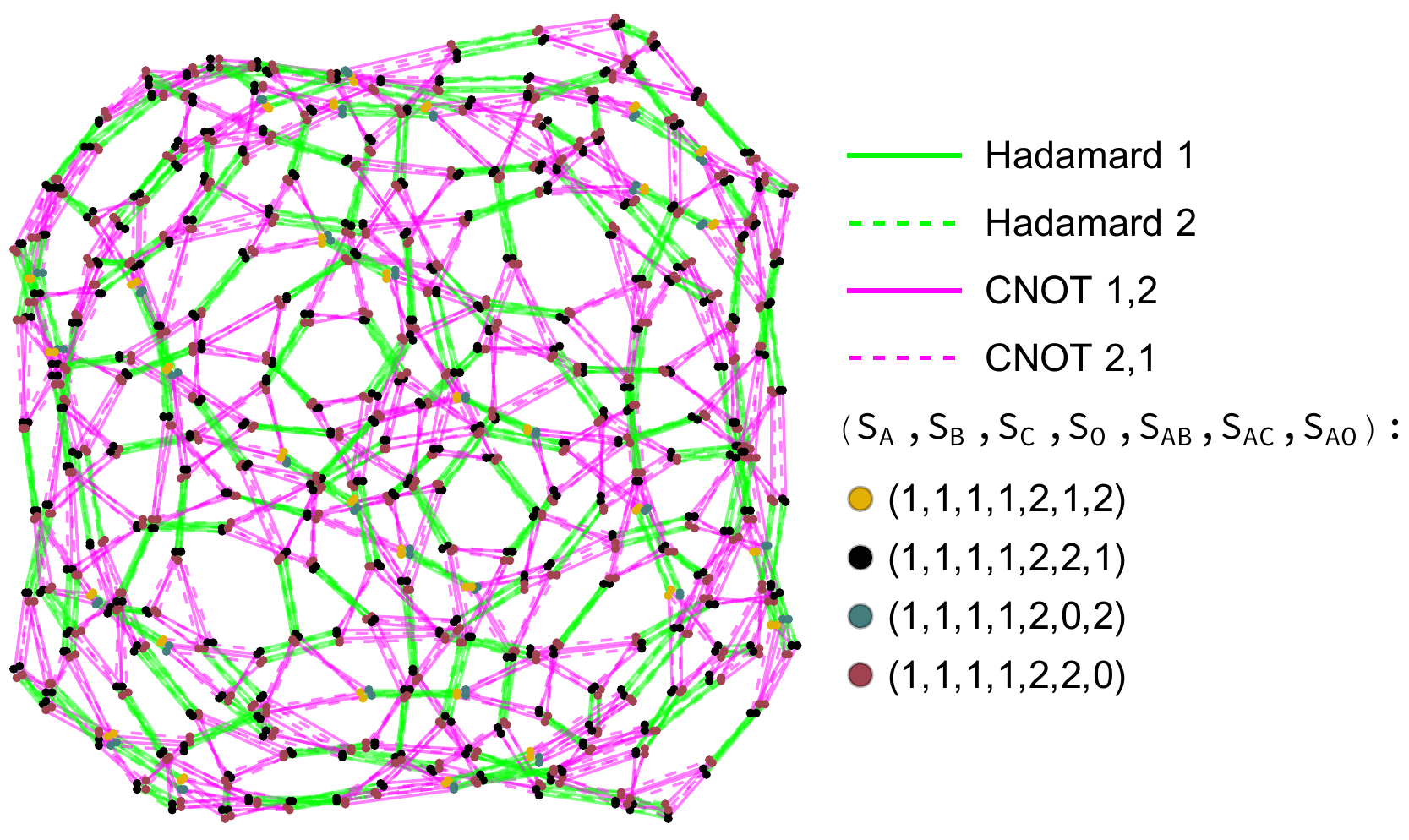}
\caption{Another copy of the $g_{1152}$ subgraph introduced at $4$ qubits. States located on this subgraph exhibit $4$ different entropy vectors, in contrast to the $g_{1152}$ subgraph in Figure \ref{FourQubitG1152TwoEntropies}.}
\label{FourQubitG1152FourEntropies}
\end{center}
\end{figure}

We also obtain our last new structure: $g_{1152}$.  At four qubits we find this subgraph with either two or four different entropy vectors, as in Figures \ref{FourQubitG1152TwoEntropies} and \ref{FourQubitG1152FourEntropies}. 

As in the 3-qubit case, lifting the $g_{24}$ and $g_{36}$ subgraphs from two qubits to four qubits proceeds in a simple manner. To lift the 2-qubit $g_{24}$ states to the 4-qubit $g_{24}$ subgraphs, we proceed as in \eqref{g24Lift}.  That is, we start with a state in the 2-qubit $g_{24}$ subgraph, and then tensor on any other 2-qubit state. Thus, both $\ket{0000}$ and $\ket{00i1}$ appear in copies of $g_{24}$ at four qubits. 

Similarly for $g_{36}$, we again tensor on 2-qubit states instead of the 1-qubit state added in \eqref{g36Lift}.  Thus $\ket{i100}$ and $\ket{i1i1}$ are in copies of $g_{24}$ at four qubits.  
Since there are 60 2-qubit stabilizer states, we get 60 copies each of $g_{24}$ and $g_{36}$.  For $g_{36}$ the Hamiltonian paths transfer just as before.

These Hamiltonian paths can be embedded in $g_{144}$ and $g_{288}$ by acting with the CNOT gates that involve qubits 3 and/or 4, or with phase gates when necessary.  The new structure, $g_{1152}$, behaves similarly, except 32 copies of $\mathcal{C}$ are needed for a full covering.

Not only the Hamiltonian paths lift; recall that Hamiltonian cycles exist for $g_{144}$ and $g_{288}$.  These cycles embed as well.  Choosing some starting state at 3 qubits, we tensor on a 1-qubit stabilizer state, act with a CNOT to entangle it and arrive on one of the ten copies of $g_{1152}$, and then enact the cycle in the same gate order as the Hamiltonian cycle at three qubits.  In order to cover one full $g_{1152}$ subgraph, four such cycles are needed.

Lastly, the $g_{1152}$ structure, like the $g_{288}$ subgraph, contains an Eulerian cycle of length 2304.  Just as for $g_{288}$, as discussed at the end of section \ref{sec:3-qubitHamiltonianEulerian}, this cycle exists because all four of the gates $H_1,\, H_2, \, CNOT_{1,2}, \, CNOT_{2,1}$ act nontrivially and differently from each other, on every state in the subgraph; no degeneracies or loops remain.

\subsubsection{Entropy Vector Analysis}

In Table \ref{tab:FourQubitEntropyVectors}, we list the 18 different entropy vectors which stabilizer states exhibit at four qubits.  In the table, the vectors are grouped together if they appear in the same subgraphs in the $H_1,\, H_2, \, CNOT_{1,2}, \, CNOT_{2,1}$ restricted graph.  These groupings arise because $S_C, \, S_O, $ and $S_{AB}$ cannot be altered by the Hadamards and CNOTs on qubits 1 and 2 only.
\begin{table}[h]
    \centering
    \small{
    \begin{tabular}{|c||c|c|c|} 
 \hline
Holographic & $\left(S_A,S_B,S_C,S_O,S_{AB},S_{AC},S_{AO}\right)$ & Number of States & Subgraph\\
\hline
\hline
 Yes & $(0,0,0,0,0,0,0)$ & $1296$ & $g_{24}$, $g_{36}$\\
 \hline
 Yes & $(1,1,0,0,0,1,1)$ & $864$ & $g_{24}$, $g_{36}$\\
 \hline
 \hline
 Yes & $(0,0,1,1,0,1,1)$ & $864$ & $g_{24}$, $g_{36}$\\
 \hline
 Yes & $(1,1,1,1,0,2,2)$ & $576$ & $g_{24}$, $g_{36}$\\
 \hline
 \hline
 Yes & $(0,1,1,0,1,1,0)$ & $864$ & $g_{144}, g_{288}$\\
 \hline
 Yes & $(1,0,1,0,1,0,1)$ & $864$ & $g_{144}, g_{288}$\\
 \hline
 Yes & $(1,1,1,0,1,1,1)$ & $2592$ & $g_{144}, g_{288}$\\
 \hline
 \hline
 Yes & $(0,1,0,1,1,0,1)$ & $864$ & $g_{144}, g_{288}$\\
 \hline
 Yes & $(1,0,0,1,1,1,0)$ & $864$ & $g_{144}, g_{288}$\\
 \hline
  Yes & $(1,1,0,1,1,1,1)$ & $2592$ & $g_{144}, g_{288}$\\
 \hline
 \hline
 Yes & $\textcolor{orange}{\bullet}(1,0,1,1,1,1,1)$ & $2592$ & $g_{144}, g_{288}$\\
 \hline
 Yes & $\textcolor{violet}{\bullet}(0,1,1,1,1,1,1)$ & $2592$ & $g_{144}, g_{288}$\\
 \hline
  Yes & $\textcolor{yellow}{\bullet}(1,1,1,1,1,2,2)$ & $5184$ & $g_{144}, g_{288}$\\
 \hline
 \color{red} No & $\textcolor{cyan}{\bullet}(1,1,1,1,1,1,1)$ & $2592$ & $g_{144}, g_{288}$\\
 \hline
 \hline
 Yes & $\textcolor{lime}{\bullet}(1,1,1,1,2,1,2)$ & $5184$ & $g_{1152}$ (2,4)\\
 \hline
 Yes & $\textcolor{black}{\bullet}(1,1,1,1,2,2,1)$ & $5184$ & $g_{1152}$ (2,4)\\
 \hline
 Yes & $\textcolor{teal}{\bullet}(1,1,1,1,2,0,2)$ & $576$ & $g_{1152}$ (4)\\
 \hline
 Yes & $\textcolor{purple}{\bullet}(1,1,1,1,2,2,0)$ & $576$ & $g_{1152}$ (4)\\
 \hline
\end{tabular}}
    \caption{Entropy vectors for the set of four-qubit stabilizer states. One entropy vector disobeys the holographic inequalities. Graph $g_{1152}$ comes in two varieties: one with only two different entropy vectors, and one with four.  The last two entropy vectors only appear on subgraphs with four different entropy vectors, while the previous two appear on subgraphs with either two or four entropy vectors. }
    \label{tab:FourQubitEntropyVectors}
\end{table}

Importantly, notice that our first non-holographic state, which disobeys the holographic entropy cone relation \eqref{eq:MonogamyofMutualInformation}, arises here.  CNOT gates are the only actions which could move a state from within the entropy cone to outside of it, so by studying the restricted graphs where this non-holographic entropy vector lives, we can define how to reach such states. In particular, notice that in $g_{144}(4)$, shown in Figure \ref{HamiltonianCycleMap}, the non-holographic states divide the subgraph into multiple parts; removing them entirely would separate the inner octagonal structure from the outside, and divide the outer ring into several pieces.

\subsection{Five Qubits and Beyond}

At five qubits, no new structures arise. The subgraph types may begin to have more entropy vectors on a given subgraph, and each subgraph has more copies, but the shapes themselves do not increase in size. Again, we have generated the full set of stabilizer states as well as the complete reachability diagram, and this data is fully accessible at \cite{github}. The full five qubit list of entropy vectors is in Table \ref{tab:5QubitEntropiesPart1} of the appendix.

Again, we note 12 entropy vectors arise which disobey the holographic entropy cone relations; as usual, examining the gate actions which move to and from those entropy vectors will help us to learn how states move in and out of the holographic entropy cone.

For now, we note that the $g_{24}$ and $g_{36}$ subgraphs each have 1080 copies, as expected since they are built by tensoring 2-qubit stabilizer states with 3-qubit stabilizer states; there are 1080 3-qubit stabilizer states.  We also note that the number of $g_{1152}$ copies scales faster than for the smaller index number graphs, as shown in Table \ref{tab:NumberOfSubgraphs}.
\begin{table}[h]
    \centering
    \begin{tabular}{|c||c|c|c|c|c|}
    \hline
    Qubit \# & $g_{24}$ & $g_{36}$ & $g_{144}$ & $g_{288}$ & $g_{1152}$\\
    \hline
    \hline
    Two & $1$ & $1$ & - & - & -\\
    \hline
    Three & $6$ & $6$ & $3$ & $1$ & -\\
    \hline
    Four & $60$ & $60$ & $90$ & $30$ & $10$\\
    \hline
    Five & $1080$ & $1080$ & $3780$ & $1260$ & $1260$\\
    \hline
    \end{tabular}
\caption{The number of occurrences of each subgraph structure at two, three, four, and five qubits.  The number of copies of $g_{1152}$ exhibits the most dramatic growth with qubit number.}
\label{tab:NumberOfSubgraphs}
\end{table}

We believe no further new structures arise at higher qubits beyond the $g_{1152}$ first found at four qubits.%
\footnote{\label{NoHigherStructures} This claim has since been verified explicitly by directly computing the full set of unique Clifford group elements generated by $H_1, H_2, CNOT_{1,2},$ and $CNOT_{2,1}$. A simple justification can be observed by allowing each leg of the $g_{1152}$ subgraph to represent a unique operation generated by $H_1, H_2, CNOT_{1,2}$ and $CNOT_{2,1}$. The proof of this result will be presented in forthcoming work.} %
However, individual subgraphs will start to contain larger numbers of different entropy vectors, just as we found when increasing to four and then five qubits.

\section{Discussion}\label{sec:discussion}

Our goal in the body of the paper has been to understand the $n$-qubit stabilizer states by first constructing their reachability graphs, then passing to the restricted graphs formed using only the Hadamards and CNOTs on two qubits. We were able to understand how the simpler structures at low qubit number ($n=1$ and $n=2$) assemble, via lifts, to build the more complicated structures observed at higher qubit number ($n=3$ and $n=4$). We have argued that the four graph structures shown already in the Introduction are ``all there is:'' passing to higher qubit number merely proliferates larger numbers of the four basic structures, perhaps with more complicated entropic arrangements. We noted the presence of non-holographic states beginning at four qubits, and their confinement to particular groups of subgraphs, as demanded by their having entropy vectors that violate holographic inequalities. In this Discussion, we collect some other ways of thinking about the reachability graphs we have constructed, along with potential generalizations and questions for further research.

We first note that, to our knowledge, already at two qubits this is the first time that the complete, explicit reachability graphs have been presented in the literature. We have constructed the full reachability graphs up through five qubits, and have made them available in our GitHub repository \cite{github}.

Having the explicit form of the graphs allows for direct computation of some interesting quantities. For example, the minimum distance on the graph from one state to another is, explicitly, the gate complexity of the stabilizer circuit which maps the first state to the second. The state with largest minimum distance relative to a reference state thus is the ``most complicated'' state by this measure. We have explicitly computed this minimum distance relative to the all-zero state for $n=1\ldots4$, and find that it is circuit distance $4,7,10,13$, respectively. Since the asymptotic complexity of $n$-qubit stabilizer states is known to go as $n^2/\mathrm{log}(n)$ \cite{aaronson2004improved}, this is perhaps a surprising result---it indicates that at small, finite qubit number the main contribution to the complexity is not the CNOT gate applications which lead to the asymptotic result. Of course, we note that, even with the explicit reachability graph, finding a state with maximal relative distance is a highly nontrivial problem requiring an exhaustive numerical search.

We recall also that the group-theoretic identity in \eqref{eq:orbit_stabilizer} equates the number of elements in a given subgraph to the orbit length of an element in that subgraph under the action of the subgroup of the Clifford group generated by the subset of the Clifford gates we are considering (see Table \ref{tab:OneQubitSubgroupOrbits}). Hence we can interpret $24, 36, 144, 288, 1152$ as the orbit lengths of various stabilizer states under the action of the subgroup generated by $\{H_1,H_2,CNOT_{1,2},CNOT_{2,1}\}$. It would be interesting to find these orbits directly using a group-theoretic approach, and confirm or disprove our conjecture that no additional structures appear beyond $g_{1152}$.

We emphasize that we chose to focus primarily on Clifford gates acting on the first two qubits because the stabilizer states exhibit only bipartite entanglement, constructed explicitly through CNOT gate applications. Hence our restricted graphs are sufficient to analyze how the entropy vector of a state changes as it is acted upon by various stabilizer circuits. Of course, adding additional gates to our restricted gate will result in more complicated connected subgraphs, culminating in the full reachability graph at each qubit number.

In the holographic literature, we typically think of each subsystem as some portion of a spatial boundary, with the exception of an additional ``purifier'' subsystem which does not necessarily have an interpretation as a spatial subsystem. The holographic interpretation of our results will therefore differ depending on whether we take the purifier to be one of the first two qubits, or instead some other qubit $>=$ 3. In the body of the paper we have taken the purifying system $O$ to be the last qubit.

In what ways can our reachability graph be generalized? First consider staying within the setting of $n$-qubit states. The existence of a graph structure is most useful when we have a gate set that can be used to pick out only a finite number of states, rather than a continuous subspace of $\mathbb{C}^{2n}$ or the entire Hilbert space itself. Hence we do not expect choosing a modification of the Clifford gates which yields a universal gate set to provide interesting results. We could instead consider the stabilizers of some other finite group of operators besides the Pauli group. It is an interesting question whether there are any such finite groups with a compelling physical interpretation. We could also consider allowing a small number of discrete applications of non-Clifford gates, i.e. allowing a small amount of ``magic'' \cite{bravyi2005universal,White:2020zoz} as a resource. Or, we could consider restricting ourselves to stabilizer states, but allowing some applications of operators which are in the Clifford group but not Clifford generators themselves, which could be used to ``fast-forward'' a circuit. Finally, it would be very interesting to understand whether there is a set of gates that produce a discrete set of states but allow, for example, for tripartite entanglement.

In more general Hilbert spaces which are not isomorphic to tensor products of qubits, we expect that a similar story should hold with a different group than the Pauli group. There are a number of approaches in the literature to generalizing Pauli matrices to larger discrete systems. One approach, well-suited to qudit systems, is to use Gell-Mann matrices and their generalizations. Another, quite different, approach better suited to approaching the continuum limit of a lattice system is to use the ``clock'' and ``shift'' matrices that generate a ``generalized Clifford algebra'' \cite{Jagannathan:2010sb,Pollack:2018yum}.

Finally, we recall our motivation in the Introduction, where we described both entanglement entropies in a factorized system, and stabilizer states relative to a preferred group of algebra, as two different ways of imposing structure on the set of states in Hilbert space. We can further generalize this picture by noting that the von Neumann entropy of a reduced density matrix of a subsystem can be identified with the \emph{algebraic} entropy associated with the algebra of operators which act as the identity everywhere outside that subsystem. Hence entropies are defined relative to an \emph{algebra} of operators while stabilizer states are defined relative to a \emph{group} of operators. In particular, recall that the algebra generated by the Pauli operators acting on a qubit is in fact the full algebra of linear operators on that qubit. 

When we consider more general von Neumann algebras, instead of the Hilbert space decomposing as a product of tensor factors, we get a more general \emph{Wedderburn decomposition} \cite{wedderburn1934lectures,Harlow:2016vwg,Kabernik:2019jko}, which decomposes the Hilbert space as a \emph{sum} of products of tensor factors, where each term in the sum represents a superselection sector which has its own associated entropy. Thus we can envision a very abstract version of our picture in which we fix a \emph{set} of operators, and then find stabilizer states by using this set to generate a \emph{group} of operators and entropies by using the same set to generate an \emph{algebra} of operators. It would be interesting to see how far this picture can be taken in generic cases. A first step would be to understand how to generate not just a single entropy for a given reduced density matrix on a state, but rather a larger entropy vector, which would pick out some \emph{sequence} of algebras, the equivalent of the operators acting on the first, second, and in general $n^{th}$ qubit.

\paragraph{Data Repository} The full set of states, reachability diagrams, and entropy vectors for stabilizer states at $n \leq 5$ qubits can be accessed via the GitHub repository \cite{github}. The repository additionally includes Mathematica notebooks used to generate the data for this paper, as well as a Stabilizer State package designed to generate reachability graphs and analyze stabilizer state structure.

\acknowledgments

The authors thank Scott Aaronson, Ning Bao, Charles Cao, and Sergio Hernandez-Cuenca for helpful discussions. J.P. is supported by the Simons Foundation through \emph{It from Qubit: Simons Collaboration on Quantum Fields, Gravity, and Information}. C.K. and W.M. are supported by the U.S. Department of Energy under grant number DE-SC0019470 and by the Heising-Simons Foundation “Observational Signatures
of Quantum Gravity” collaboration grant 2021-2818.

\begin{appendices}
\addtocontents{toc}{\protect\setcounter{tocdepth}{1}}

\section{Additional Graphs}\label{Two-Qubit Graphs}

Contained below are some graphs not featured in the body of this paper. Futher graphs are available in the repository \cite{github}.

\subsection{Two Qubit Graphs}
 
 The 2-qubit complete reachability graph and some 2-qubit restricted graphs were discussed in section \ref{sec:two}. Here we compile additional restricted graphs to further illustrate the relation between states under action $C_2$ subgroup action. Figure \ref{TwoQubitP1P2} restricts to only phase operations, while Figure \ref{TwoQubitCNOT12CNOT21} shows the restricted graph for only CNOT operation.
 
\subsubsection{Two Qubit $P_1,P_2$ Restricted Graph}

\begin{figure}[h]
\begin{center}
\includegraphics[scale=0.82]{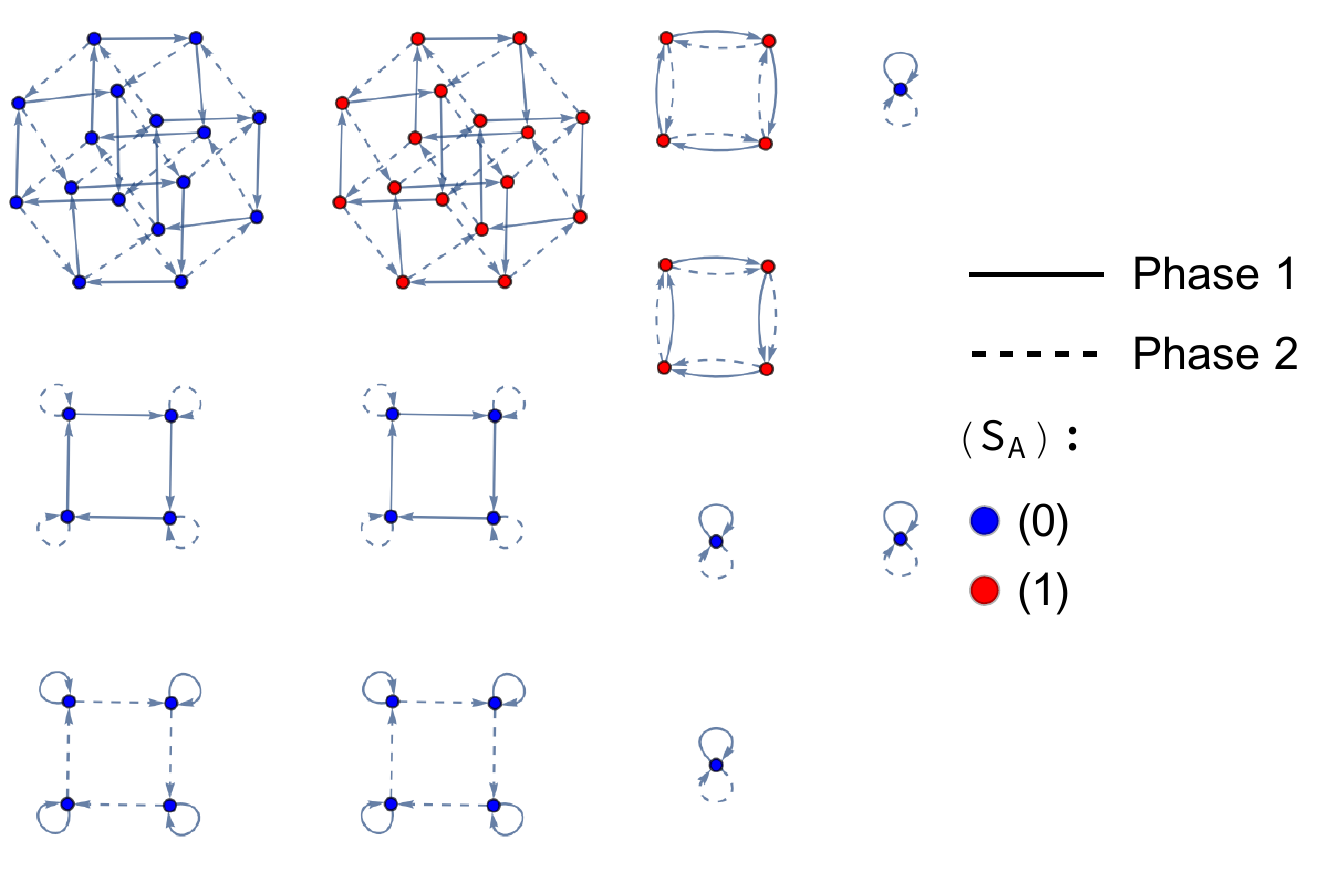}
\caption{The 2-qubit $P_1,P_2$ restricted graph contains $5$ unique substructures. The $4$ isolated points are states on which both $P_1$ and $P_2$ act trivially. The box-like structures come in two varieties. The box of unentangled states has a trivial loop at each corner, while the boxes of entangled states witness degenerate action instead. There exist two largest structures of states (top-left) on which both phase gates act non-trivially and non-degenerately.}
\label{TwoQubitP1P2}
\end{center}
\end{figure}

Figure \ref{TwoQubitP1P2} shows the 2-qubit graph restricted to only phase operations. The graph consists of $5$ distinct and disconnected substructures.

\subsubsection{Two Qubit $CNOT_{1,2},CNOT_{2,1}$ Restricted Graph}

\begin{figure}[h]
\begin{center}
\includegraphics[scale=0.5]{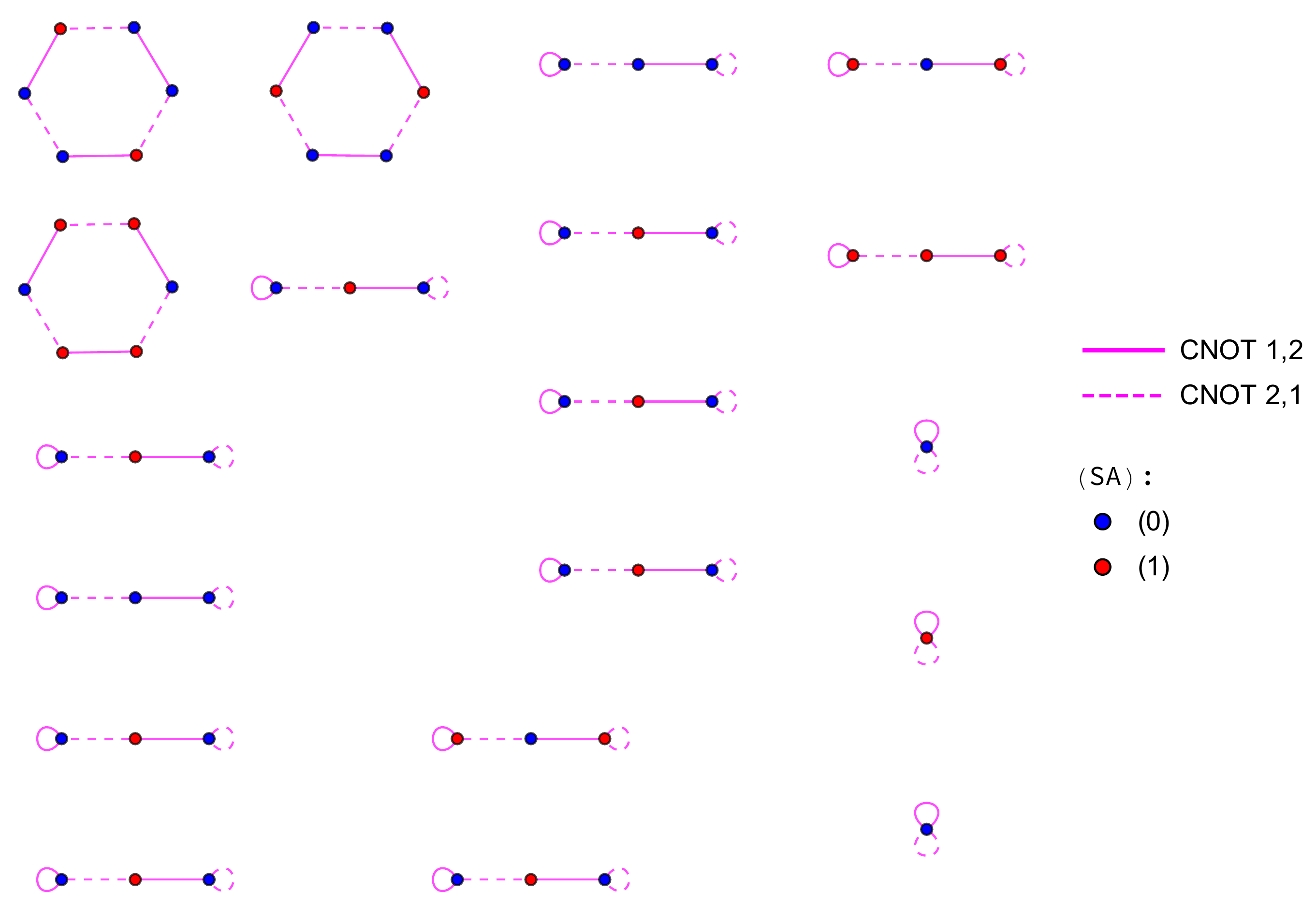}
\caption{The subgroup generated by $CNOT_{1,2}$ and $CNOT_{2,1}$ partitions the set of 2-qubit states into $3$ graph substructures. The isolated points are states on which both CNOT gates act trivially. The linear triplets are built of one state, on which both CNOT gates act non-trivially, connected to two states on which opposing one CNOT gate acts trivially. These triplets can occur with states of similar or different entropic structure. The largest structure is a hexagon which illustrates the maximum cycle of the subgroup generated by $CNOT_{1,2}$ and $CNOT_{2,1}$.}
\label{TwoQubitCNOT12CNOT21}
\end{center}
\end{figure}

Figure \ref{TwoQubitCNOT12CNOT21} shows all interactions between 2-qubit states under only CNOT operations. The CNOT gate can modify entropy structure, therefore we witness the first occurrences of states with different entropy vectors lying in the same substructures. Alternating action of $CNOT_{1,2}$ and $CNOT_{2,1}$ has a maximum cycle of $6$, seen in the hexagonal structures top-left.

\newpage

\subsection{Three Qubit Graphs}

 The 3-qubit reachability diagrams were discussed in section \ref{sec:three} with a focus on the $H_1,H_2,CNOT_{1,2},CNOT_{2,1}$ restricted graph (Figure \ref{ThreeQubitH1H2CNOT12CNOT21}). Here we provide additional graphs of potential interest, including the complete reachability graph on three qubits (Figure \ref{ThreeQubitCompleteGraph}). Figure \ref{ThreeQubitP1P2P3Subgraphs} displays the action of all phase gates on three qubits, generalizing the cycles and structures seen at two qubits (Figure \ref{TwoQubitP1P2}) to three qubits. 
 
\subsubsection{Three Qubit Complete Reachability Graph}

\begin{figure}[h]
\begin{center}
\includegraphics[scale=0.6]{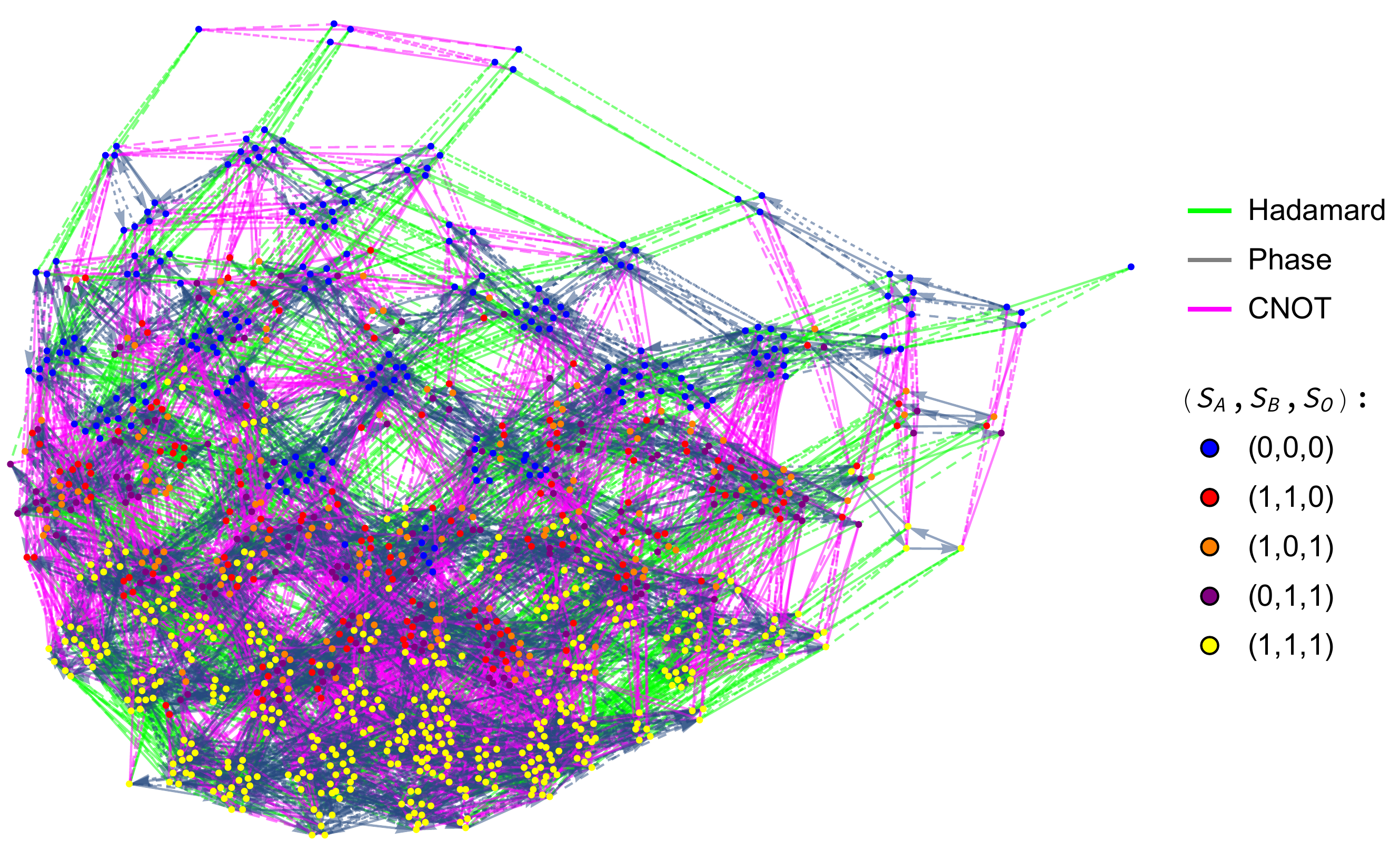}
\caption{Complete reachability diagram on three qubits with trivial loops removed. The Hadamard and phase gates act individually on all three qubits, while the CNOT gate acts on any pair of qubits. Line texture indicates the particular action, e.g. a solid line for $H_1$ and medium dashed line for $H_2$.}
\label{ThreeQubitCompleteGraph}
\end{center}
\end{figure}

Figure \ref{ThreeQubitCompleteGraph} displays the full reachability graph for three qubits (trivial loops removed). This graph contains all non-trivial information about 3-qubit interaction under operations of the Clifford group ($C_3$).

\newpage

\subsubsection{Three Qubit $P_1,P_2,P_3$ Restricted Graph Subgraphs}

\begin{figure}[h]
\begin{center}
\includegraphics[scale=0.58]{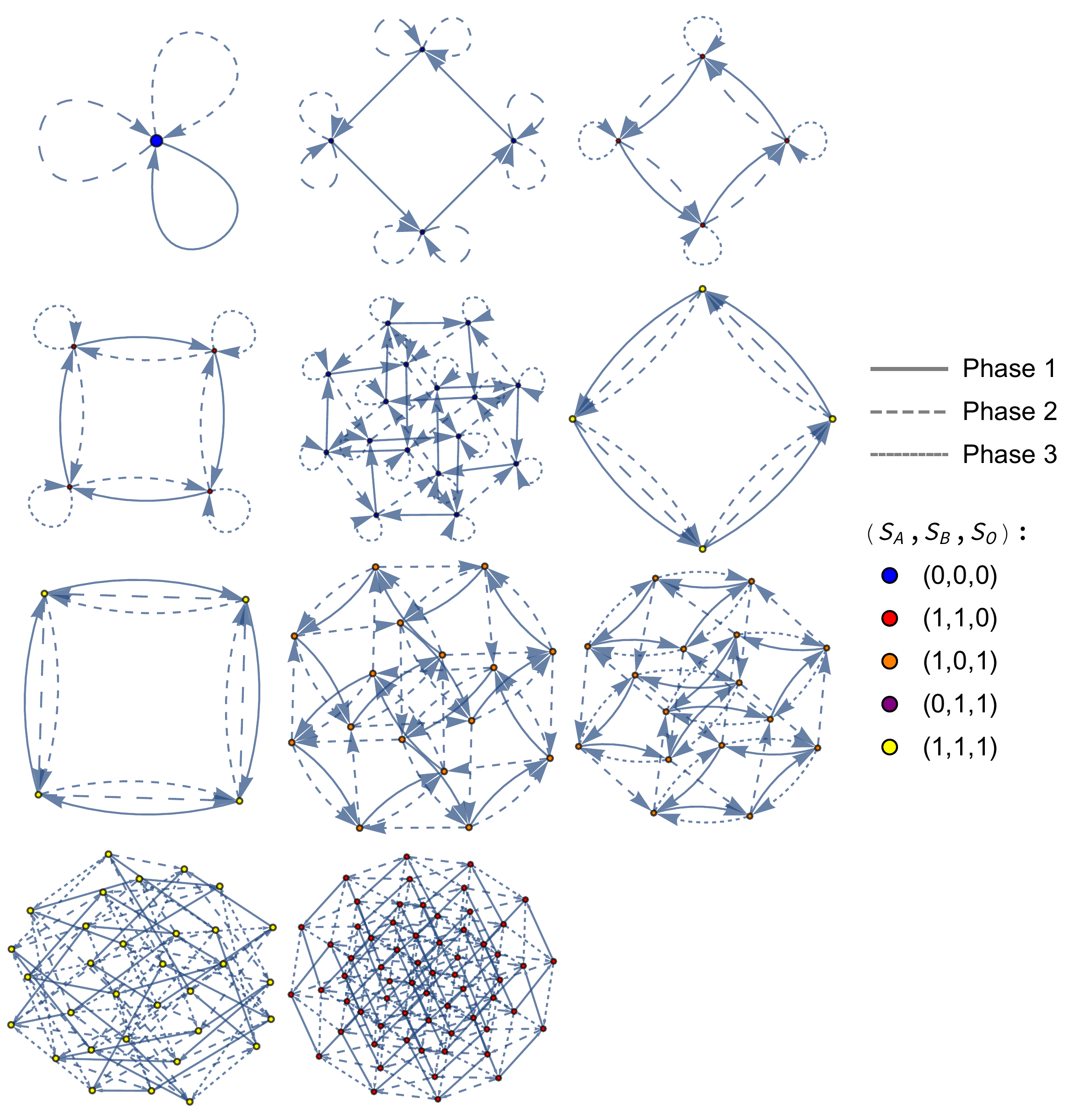}
\caption{There are $11$ unique subgraphs whose copies build the 3-qubit $P_1,P_2,P_3$ restricted graph. These subgraphs segregate according to which sequences of $P_1,P_2,$ and $P_3$ are equivalent.}
\label{ThreeQubitP1P2P3Subgraphs}
\end{center}
\end{figure}

In Figure \ref{ThreeQubitP1P2P3Subgraphs} is the 3-qubit graph restricted to only phase gates. The addition of $P_3$ to the generating set allows for longer cycles, resulting in more complex structures than were witness at lower qubit number.

\newpage

\subsubsection{Three Qubit $H_1,H_2,CNOT_{1,2},CNOT_{2,1}$ Restricted Graph}

\begin{figure}[h]
\begin{center}
\includegraphics[scale=0.54]{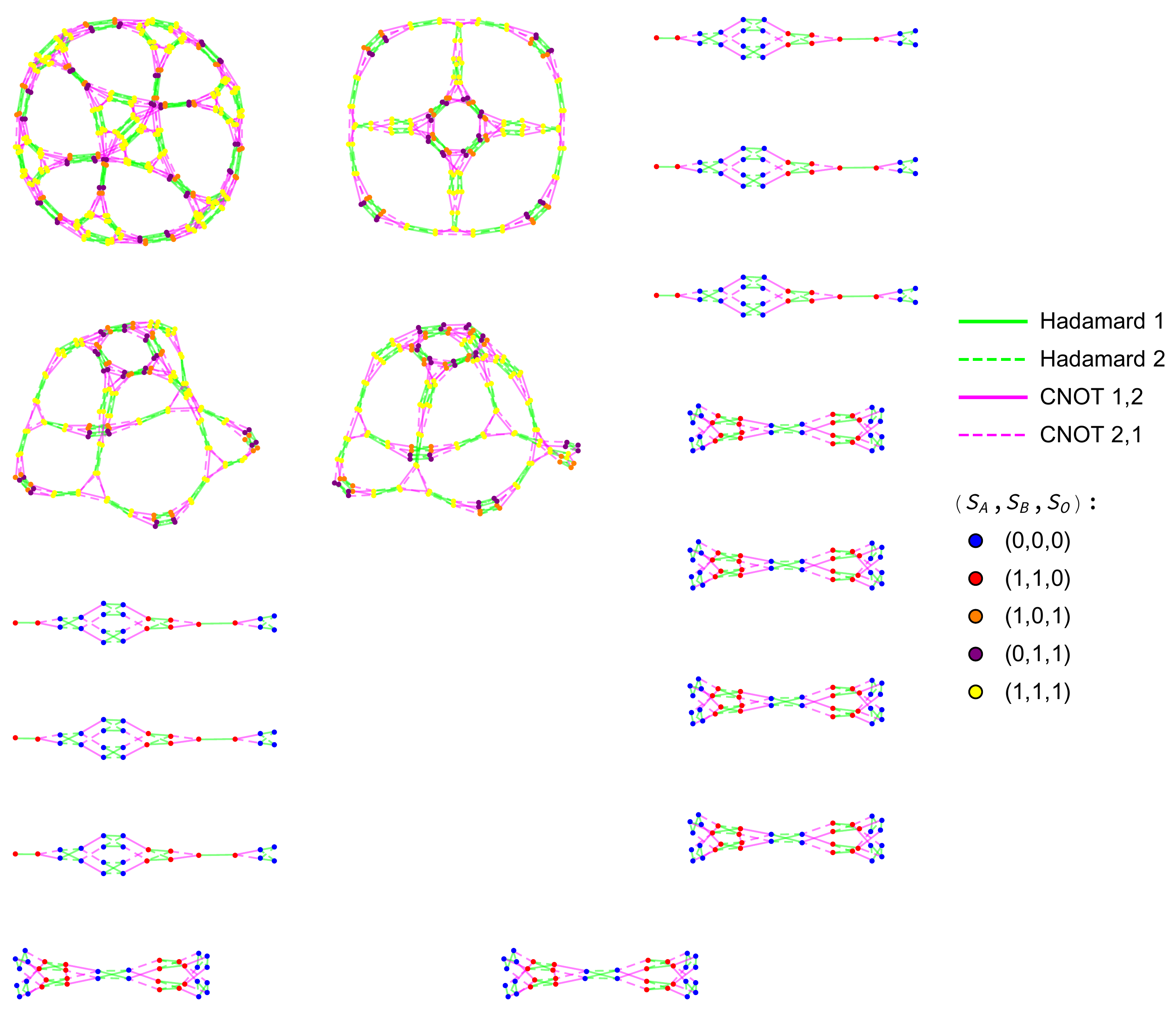}
\caption{The graph contains $6$ copies of $g_{24}$ and $g_{36}$, $3$ copies of $g_{144}$, and a single copy of $g_{288}$. The $3$ $g_{144}$ subgraphs are isomorphic, but were generated with slightly different layouts by the software used to build this graph.}
\label{ThreeQubitH1H2CNOT12CNOT21}
\end{center}
\end{figure}

Figure \ref{ThreeQubitH1H2CNOT12CNOT21} shows the 3-qubit Three-qubit restricted graph displaying only Hadamard and CNOT operations on the first two qubits of in the system. The graph has repeated copies of structures found at two qubits, as well as the addition of two new subgraphs $g_{144}$ and $g_{288}$.

\section{Five-Qubit Entropy Vectors}\label{sec:entropy_vectors}

Below we present the complete set of entropy vectors for the 5-qubit stabilizer set. There are $2423520$ stabilizer states at five qubits, with $93$ different entropic arrangements. There are $16$ of these $93$ entropy vectors which violate the monogamy of mutual information (Equation \eqref{eq:MonogamyofMutualInformation}), and therefore correspond to non-holographic states. %The ordering of entropies used in the table, following the conventions of Section \ref{sec:EntropyCone} above, is $(S_A,S_B,S_C,S_D,S_O,S_{AB},S_{AC},S_{AD},S_{AO},S_{BC},S_{BD},S_{BO},S_{CD},S_{CO},S_{DO})$
\begin{table}[h]
    \centering
    \small{
    \begin{tabular}{|c||c|c|} 
 \hline
Holographic & Entropy Vector & Subgraph\\
\hline
\hline 
Yes & $(0,0,1,0,1,0,1,0,1,1,0,1,1,0,1)$ & $g_{24}$, $g_{36}$\\
 \hline
 Yes & $(1, 1, 1, 0, 1, 0, 2, 1, 2, 2, 1, 2, 1, 0, 1)$ & $g_{24}$, $g_{36}$\\
\hline
 \hline
 Yes & $(0, 0, 1, 1, 0, 0, 1, 1, 0, 1, 1, 0, 0, 1, 1)$ & $g_{24}$, $g_{36}$\\
 \hline
 Yes & $(1, 1, 1, 1, 0, 0, 2, 2, 1, 2, 2, 1, 0, 1, 1)$ & $g_{24}$, $g_{36}$\\
\hline
 \hline
 Yes & $(0, 0, 0, 1, 1, 0, 0, 1, 1, 0, 1, 1, 1, 1, 0)$ & $g_{24}$, $g_{36}$\\
 \hline
 Yes & $(1, 1, 0, 1, 1, 0, 1, 2, 2, 1, 2, 2, 1, 1, 0)$ & $g_{24}$, $g_{36}$\\
\hline
 \hline
 Yes & $(0, 0, 0, 0, 0, 0, 0, 0, 0, 0, 0, 0, 0,0, 0)$ & $g_{24}$, $g_{36}$\\
 \hline
 Yes & $(1, 1, 0, 0, 0, 0, 1, 1, 1, 1, 1, 1, 0,0, 0)$ & $g_{24}$, $g_{36}$\\
\hline
 \hline
 Yes & $(0, 0, 1, 1, 1, 0, 1, 1, 1, 1, 1, 1, 1,1, 1)$ & $g_{24}$, $g_{36}$\\
 \hline
 Yes & $(1, 1, 1, 1, 1, 0, 2, 2, 2, 2, 2, 2, 1,1, 1)$ & $g_{24}$, $g_{36}$\\
 \hline
 \hline
 Yes & $(0, 1, 1, 1, 1, 1, 1, 1, 1, 0, 2, 2, 2, 2, 0)$ & $g_{144}, g_{288}$\\
 \hline
 Yes & $ (1, 0, 1, 1, 1, 1, 0, 2, 2, 1, 1, 1, 2, 2, 0),$ & $g_{144}, g_{288}$\\
  \hline
 Yes & $(1, 1, 1, 1, 1, 1, 1, 2, 2, 1, 2, 2, 2, 2, 0)$ & $g_{144}, g_{288}$\\
  \hline
  \hline
 Yes & $(0, 1, 1, 1, 1, 1, 1, 1, 1, 2, 2, 0, 0, 2, 2)$ & $g_{144}, g_{288}$\\
 \hline
 Yes & $(1, 0, 1, 1, 1, 1, 2, 2, 0, 1, 1, 1, 0, 2, 2)$ & $g_{144}, g_{288}$\\
  \hline
 Yes & $(1, 1, 1, 1, 1, 1, 2, 2, 1, 2, 2, 1, 0, 2, 2)$ & $g_{144}, g_{288}$\\
\hline
\hline
 Yes & $(0, 1, 0, 1, 0, 1, 0, 1, 0, 1, 0, 1, 1, 0, 1)$ & $g_{144}, g_{288}$\\
 \hline
 Yes & $(1, 0, 0, 1, 0, 1, 1, 0, 1, 0, 1, 0, 1, 0, 1)$ & $g_{144}, g_{288}$\\
  \hline
 Yes & $(1, 1, 0, 1, 0, 1, 1, 1, 1, 1, 1, 1, 1, 0, 1)$ & $g_{144}, g_{288}$\\
 \hline
\hline
 Yes & $(0, 1, 0, 0, 1, 1, 0, 0, 1, 1, 1, 0, 0, 1, 1)$ & $g_{144}, g_{288}$\\
 \hline
 Yes & $(1, 0, 0, 0, 1, 1, 1, 1, 0, 0, 0, 1, 0, 1, 1)$ & $g_{144}, g_{288}$\\
  \hline
 Yes & $(1, 1, 0, 0, 1, 1, 1, 1, 1, 1, 1, 1, 0, 1, 1)$ & $g_{144}, g_{288}$\\
\hline
\hline
 Yes & $(0, 1, 1, 0, 0, 1, 1, 0, 0, 0, 1, 1, 1, 1, 0)$ & $g_{144}, g_{288}$\\
 \hline
 Yes & $(1, 0, 1, 0, 0, 1, 0, 1, 1, 1, 0, 0, 1, 1, 0)$ & $g_{144}, g_{288}$\\
  \hline
 Yes & $ (1, 1, 1, 0, 0, 1, 1, 1, 1, 1, 1, 1, 1, 1, 0)$ & $g_{144}, g_{288}$\\
\hline
\end{tabular}}
    \caption{At five qubits, there are $93$ stabilizer state entropy vectors (listed here and on the next two pages). Of these, $16$ correspond to non-holographic states. All non-holographic states are located on subgraphs with $4$ different entropy vectors.}
    \label{tab:5QubitEntropiesPart1}
\end{table}

\newpage

\begin{table}[h]
    \centering
    \small{
    \begin{tabular}{|c||c|c|c|} 
 \hline
Holographic & Entropy Vector & Subgraph\\
\hline
\hline
 Yes & $(0, 1, 1, 1, 1, 1, 1, 1, 1, 2, 0, 2, 2, 0, 2)$ & $g_{144}, g_{288}$\\
 \hline
 Yes & $(1, 0, 1, 1, 1, 1, 2, 0, 2, 1, 1, 1, 2, 0, 2)$ & $g_{144}, g_{288}$\\
  \hline
 Yes & $(1, 1, 1, 1, 1, 1, 2, 1, 2, 2, 1, 2, 2, 0, 2)$ & $g_{144}, g_{288}$\\
\hline
\hline
  \color{red} No & $(0, 1, 1, 1, 1, 1, 1, 1, 1, 1, 1, 1, 1, 1, 1)$ & $g_{144}, g_{288}$\\
 \hline
  \color{red} No & $ (1, 0, 1, 1, 1, 1, 1, 1, 1, 1, 1, 1, 1, 1, 1)$ & $g_{144}, g_{288}$\\
  \hline
  \color{red} No & $(1, 1, 1, 1, 1, 1, 1, 1, 1, 1, 1, 1, 1, 1, 1)$ & $g_{144}, g_{288}$\\
   \hline
  \color{red} No & $(1, 1, 1, 1, 1, 1, 2, 2, 2, 2, 2, 2, 1, 1, 1)$ & $g_{144}, g_{288}$\\
  \hline
\hline
  Yes & $(0, 1, 1, 1, 1, 1, 1, 1, 1, 2, 2, 1, 1, 2, 2)$ & $g_{144}, g_{288}$\\
 \hline
 Yes & $(1, 0, 1, 1, 1, 1, 2, 2, 1, 1, 1, 1, 1, 2, 2)$ & $g_{144}, g_{288}$\\
  \hline
  \color{red} No & $(1, 1, 1, 1, 1, 1, 2, 2, 1, 2, 2, 1, 1, 2, 2)$ & $g_{144}, g_{288}$\\
   \hline
 Yes & $(1, 1, 1, 1, 1, 1, 2, 2, 2, 2, 2, 2, 1, 2, 2)$ & $g_{144}, g_{288}$\\
\hline
\hline
Yes & $(0, 1, 1, 1, 0, 1, 1, 1, 0, 1, 1, 1, 1, 1, 1)$ & $g_{144}, g_{288}$\\
 \hline
 Yes & $(1, 0, 1, 1, 0, 1, 1, 1, 1, 1, 1, 0, 1, 1, 1)$ & $g_{144}, g_{288}$\\
  \hline
  \color{red} No & $(1, 1, 1, 1, 0, 1, 1, 1, 1, 1, 1, 1, 1, 1, 1)$ & $g_{144}, g_{288}$\\
   \hline
 Yes & $ (1, 1, 1, 1, 0, 1, 2, 2, 1, 2, 2, 1, 1, 1, 1)$ & $g_{144}, g_{288}$\\
\hline
\hline
  Yes & $(0, 1, 0, 1, 1, 1, 0, 1, 1, 1, 1, 1, 1, 1, 1)$ & $g_{144}, g_{288}$\\
 \hline
 Yes & $(1, 0, 0, 1, 1, 1, 1, 1, 1, 0, 1, 1, 1, 1, 1)$ & $g_{144}, g_{288}$\\
  \hline
  \color{red} No & $ (1, 1, 0, 1, 1, 1, 1, 1, 1, 1, 1, 1, 1, 1, 1)$ & $g_{144}, g_{288}$\\
   \hline
 Yes & $(1, 1, 0, 1, 1, 1, 1, 2, 2, 1, 2, 2, 1, 1, 1)$ & $g_{144}, g_{288}$\\
\hline
\hline
  Yes & $(0, 1, 1, 1, 1, 1, 1, 1, 1, 1, 2, 2, 2, 2, 1)$ & $g_{144}, g_{288}$\\
 \hline
 Yes & $ (1, 0, 1, 1, 1, 1, 1, 2, 2, 1, 1, 1, 2, 2, 1)$ & $g_{144}, g_{288}$\\
  \hline
  \color{red} No & $(1, 1, 1, 1, 1, 1, 1, 2, 2, 1, 2, 2, 2, 2, 1)$ & $g_{144}, g_{288}$\\
   \hline
 Yes & $(1, 1, 1, 1, 1, 1, 2, 2, 2, 2, 2, 2, 2, 2, 1)$ & $g_{144}, g_{288}$\\
\hline
\hline
  Yes & $(0, 1, 1, 0, 1, 1, 1, 0, 1, 1, 1, 1, 1, 1, 1)$ & $g_{144}, g_{288}$\\
 \hline
 Yes & $(1, 0, 1, 0, 1, 1, 1, 1, 1, 1, 0, 1, 1, 1, 1)$ & $g_{144}, g_{288}$\\
  \hline
  \color{red} No & $(1, 1, 1, 0, 1, 1, 1, 1, 1, 1, 1, 1, 1, 1, 1)$ & $g_{144}, g_{288}$\\
   \hline
 Yes & $(1, 1, 1, 0, 1, 1, 2, 1, 2, 2, 1, 2, 1, 1, 1)$ & $g_{144}, g_{288}$\\
\hline
\hline
  Yes & $(0, 1, 1, 1, 1, 1, 1, 1, 1, 2, 1, 2, 2, 1, 2)$ & $g_{144}, g_{288}$\\
 \hline
 Yes & $(1, 0, 1, 1, 1, 1, 2, 1, 2, 1, 1, 1, 2, 1, 2)$ & $g_{144}, g_{288}$\\
  \hline
  \color{red} No & $(1, 1, 1, 1, 1, 1, 2, 1, 2, 2, 1, 2, 2, 1, 2)$ & $g_{144}, g_{288}$\\
   \hline
 Yes & $(1, 1, 1, 1, 1, 1, 2, 2, 2, 2, 2, 2, 2, 1, 2)$ & $g_{144}, g_{288}$\\
\hline
\hline
   Yes & $(1, 1, 1, 0, 1, 2, 2, 1, 1, 1, 1, 2, 1, 2, 1)$ & $g_{1152}$ (2,4)\\
  \hline
   Yes & $(1, 1, 1, 0, 1, 2, 1, 1, 2, 2, 1, 1, 1, 2, 1)$ & $g_{1152}$ (2,4)\\
  \hline
   Yes & $(1, 1, 1, 0, 1, 2, 0, 1, 2, 2, 1, 0, 1, 2, 1)$ & $g_{1152}$ (4)\\
 \hline
   Yes & $ (1, 1, 1, 0, 1, 2, 2, 1, 0, 0, 1, 2, 1, 2, 1)$ & $g_{1152}$ (4)\\
 \hline
\end{tabular}}
    \label{tab:5QubitEntropiesPart2}
\end{table}

\newpage

\begin{table}[h]
    \centering
    \small{
    \begin{tabular}{|c||c|c|c|} 
 \hline
Holographic & Entropy Vector & Subgraph\\
\hline
  \hline
   Yes & $(1, 1, 0, 1, 1, 2, 1, 2, 1, 1, 1, 2, 1, 1, 2)$ & $g_{1152}$ (2,4)\\
  \hline
   Yes & $(1, 1, 0, 1, 1, 2, 1, 1, 2, 1, 2, 1, 1, 1, 2)$ & $g_{1152}$ (2,4)\\
  \hline
   Yes & $(1, 1, 0, 1, 1, 2, 1, 0, 2, 1, 2, 0, 1, 1, 2)$ & $g_{1152}$ (4)\\
 \hline
   Yes & $ (1, 1, 0, 1, 1, 2, 1, 2, 0, 1, 0, 2, 1, 1, 2)$ & $g_{1152}$ (4)\\
 \hline
  \hline
   Yes & $(1, 1, 1, 1, 0, 2, 2, 1, 1, 1, 2, 1, 2, 1, 1)$ & $g_{1152}$ (2,4)\\
  \hline
   Yes & $(1, 1, 1, 1, 0, 2, 1, 2, 1, 2, 1, 1, 2, 1, 1)$ & $g_{1152}$ (2,4)\\
  \hline
   Yes & $(1, 1, 1, 1, 0, 2, 0, 2, 1, 2, 0, 1, 2, 1, 1)$ & $g_{1152}$ (4)\\
 \hline
   Yes & $(1, 1, 1, 1, 0, 2, 2, 0, 1, 0, 2, 1, 2, 1, 1)$ & $g_{1152}$ (4)\\
 \hline
\hline
 \color{red} No & $(1, 1, 1, 1, 1, 2, 1, 2, 2, 2, 1, 1, 2, 2, 1)$ & $g_{1152}$ (4,6)\\
 \hline
  \color{red} No & $ (1, 1, 1, 1, 1, 2, 2, 1, 1, 1, 2, 2, 2, 2, 1)$ & $g_{1152}$ (4,6)\\
 \hline
 Yes & $(1, 1, 1, 1, 1, 2, 1, 2, 2, 2, 2, 2, 2, 2, 1)$ & $g_{1152}$ (4,6)\\
  \hline
 Yes & $(1, 1, 1, 1, 1, 2, 2, 2, 2, 1, 2, 2, 2, 2, 1)$ & $g_{1152}$ (4,6)\\
  \hline
   Yes & $(1, 1, 1, 1, 1, 2, 0, 2, 2, 2, 1, 1, 2, 2, 1)$ & $g_{1152}$ (6)\\
  \hline
     Yes & $(1, 1, 1, 1, 1, 2, 2, 1, 1, 0, 2, 2, 2, 2, 1)$ & $g_{1152}$ (6)\\
\hline
\hline
      \color{red} No & $(1, 1, 1, 1, 1, 2, 1, 1, 2, 2, 2, 1, 1, 2, 2)$ & $g_{1152}$ (4,6)\\
\hline
  \color{red} No & $(1, 1, 1, 1, 1, 2, 2, 2, 1, 1, 1, 2, 1, 2, 2)$ & $g_{1152}$ (4,6)\\
\hline
   Yes & $(1, 1, 1, 1, 1, 2, 2, 2, 1, 2, 2, 2, 1, 2, 2)$ & $g_{1152}$ (4,6)\\
   \hline
 Yes & $(1, 1, 1, 1, 1, 2, 2, 2, 2, 2, 2, 1, 1, 2, 2)$ & $g_{1152}$ (4,6)\\
  \hline
     Yes & $(1, 1, 1, 1, 1, 2, 2, 2, 0, 1, 1, 2, 1, 2, 2)$ & $g_{1152}$ (6)\\
  \hline
     Yes & $(1, 1, 1, 1, 1, 2, 1, 1, 2, 2, 2, 0, 1, 2, 2)$ & $g_{1152}$ (6)\\
\hline
\hline
  \color{red} No & $(1, 1, 1, 1, 1, 2, 1, 2, 1, 2, 1, 2, 2, 1, 2)$ & $g_{1152}$ (4,6)\\
 \hline
  \color{red} No & $(1, 1, 1, 1, 1, 2, 2, 1, 2, 1, 2, 1, 2, 1, 2)$ & $g_{1152}$ (4,6)\\
  \hline
   Yes & $(1, 1, 1, 1, 1, 2, 2, 1, 2, 2, 2, 2, 2, 1, 2)$ & $g_{1152}$ (4,6)\\
 \hline
 Yes & $ (1, 1, 1, 1, 1, 2, 2, 2, 2, 2, 1, 2, 2, 1, 2)$ & $g_{1152}$ (4,6)\\
  \hline
     Yes & $(1, 1, 1, 1, 1, 2, 1, 2, 1, 2, 0, 2, 2, 1, 2)$ & $g_{1152}$ (6)\\
  \hline
  Yes & $(1, 1, 1, 1, 1, 2, 2, 0, 2, 1, 2, 1, 2, 1, 2)$ & $g_{1152}$ (6)\\
  \hline
  \hline
 Yes & $(1, 1, 1, 1, 1, 2, 1, 2, 2, 2, 1, 2, 2, 2, 2)$ & $g_{1152}$ (7)\\
\hline
Yes & $(1, 1, 1, 1, 1, 2, 1, 2, 2, 2, 2, 1, 2, 2, 2)$ & $g_{1152}$ (7)\\
\hline
Yes & $(1, 1, 1, 1, 1, 2, 2, 1, 2, 1, 2, 2, 2, 2, 2)$ & $g_{1152}$ (7)\\
\hline
Yes & $ (1, 1, 1, 1, 1, 2, 2, 1, 2, 2, 2, 1, 2, 2, 2)$ & $g_{1152}$ (7)\\
\hline
Yes & $(1, 1, 1, 1, 1, 2, 2, 2, 1, 1, 2, 2, 2, 2, 2)$ & $g_{1152}$ (7)\\
\hline
Yes & $(1, 1, 1, 1, 1, 2, 2, 2, 1, 2, 1, 2, 2, 2, 2)$ & $g_{1152}$ (7)\\
\hline
Yes & $(1, 1, 1, 1, 1, 2, 2, 2, 2, 2, 2, 2, 2, 2, 2)$ & $g_{1152}$ (7)\\
\hline
\end{tabular}}
    \label{tab:5QubitEntropiesPart3}
\end{table}

\end{appendices}

\bibliographystyle{JHEP}
\bibliography{paper}

\end{document}